\documentclass[preprint,12pt]{elsarticle}

\usepackage{amssymb}
\usepackage{amsmath}
\usepackage[version=4]{mhchem}
\usepackage{xcolor}
\usepackage{float, subfig}
\usepackage{multirow} 

\newcommand{\degC}{$^\circ$C}
\newcommand{\degF}{$^\circ$F}

\journal{Acta Astronautica}

\begin{document}

\begin{frontmatter}



\title{Titan's Resources and their Utilization}


\author[nasa]{Conor A. Nixon} 

\address[nasa]{Solar System Exploration Division, NASA Goddard Space Flight Center, 8800 Greenbelt Road, Greenbelt, MD 20771, USA}


\author[wpi]{Ye Lu} 
\address[wpi]{Department of Aerospace Engineering, Worcester Polytechnic Institute, 100 Institute Road, Worcester, MA 01609, USA}


\author[ufl]{Jennifer E. Ruliffson} 
\address[ufl]{Department of Materials Science and Engineering, University of Florida, 100 Rhines Hall, 549 Gale Lemerand Drive, Gainesville, FL 32611, USA} 

            
\begin{abstract}
Saturn's moon Titan is a unique environment in the solar system. It is the only moon with an atmosphere, composed primarily of the gases \ce{N2} and \ce{CH4}. It is also the only world to have abundant surface hydrocarbons C$_x$H$_y$, which are found as both liquids (seas, lakes) and solids (dunes). Meanwhile, oxygen is also readily accessible in the form of crustal water. This combination of abundant reduced carbon, along with available nitrogen and oxygen makes Titan an enticing world rich in resources that can be readily used to make food, fuel, building materials and more - potentially mission-enabling for long-duration voyages or habitats in the outer solar system. At the same time Titan, as an icy moon, is likely to be depleted at the surface in heavier elements including metals, which must therefore be found and brought from elsewhere. In this article we describe both the available resources on Titan, and also their potential uses. We compare and contrast the resource availability and potential in-situ utilization (ISRU) with other destinations suggested for human habitation such as the Moon and Mars. We conclude by discussing what future work will be important to further characterize Titan's resources, and to develop technologies for their utilization.

\end{abstract}

\begin{graphicalabstract}
\vspace*{1cm} \\
\includegraphics[width=1.0\linewidth]{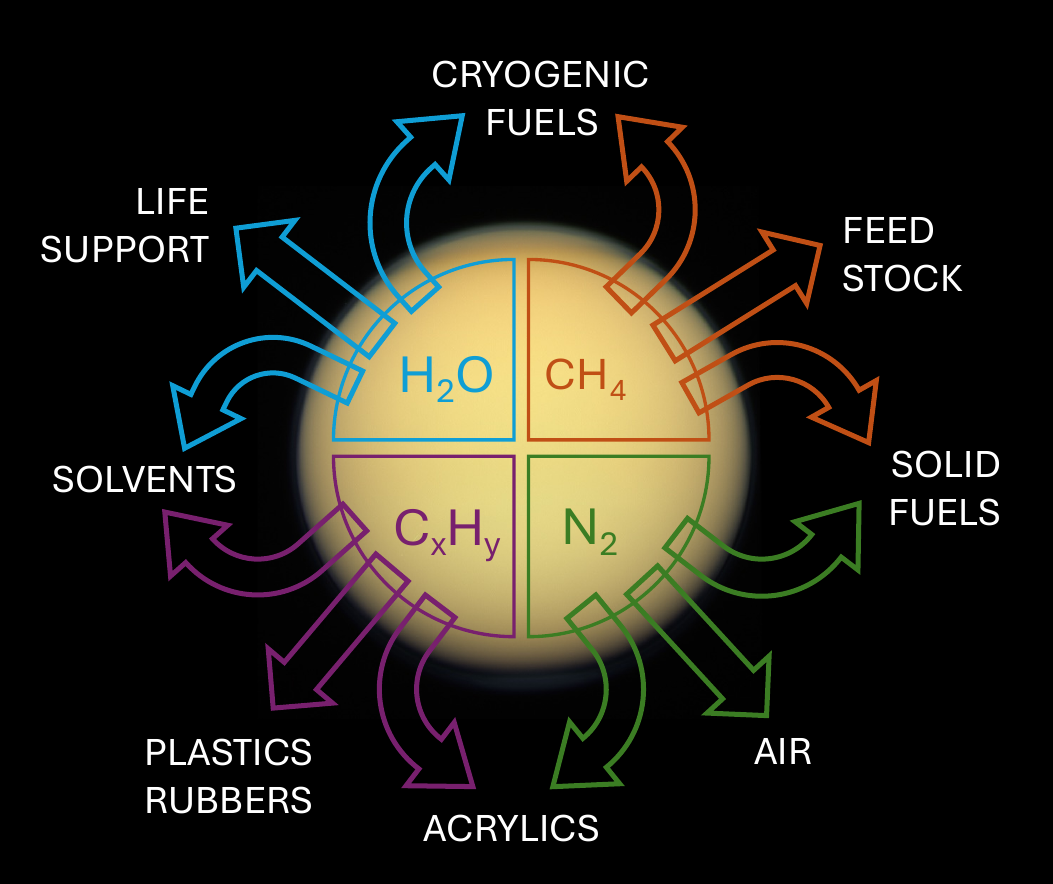}
\end{graphicalabstract}

\begin{highlights}
\item Saturn's moon Titan is abundant in useful resources. 
\item Methane and oxygen for fuel are easily obtained from air and bedrock.
\item Simple CHON chemicals can be used to produce more complex chemicals. 
\item Titan's resources can enable long-duration missions to the outer solar system.
\end{highlights}

\begin{keyword}

Titan \sep Saturn \sep ISRU \sep hydrocarbons \sep space manufacturing \sep 3D printing



\end{keyword}

\end{frontmatter}




\section{Introduction}
\label{sect:intro}

Titan is a unique world in the solar system. As a moon it stands apart as the only natural satellite with a dense atmosphere \cite{horst2017titan}. As a ‘terrestrial planet’ – meaning a body with both an atmosphere and solid surface, like Venus, Earth and Mars – it is the only such world to have a largely anoxic atmosphere \cite{nixon2024composition}. Titan’s atmosphere, while mostly composed of nitrogen gas (\ce{N2}), also contains a substantial amount of methane (2-5\%), which, through photochemistry, leads to the production of a wide range of organic molecules. These include especially hydrocarbons (C$_{\rm x}$H$_{\rm y}$) and nitriles (C$_{\rm x}$H$_{\rm y}$[CN]$_{\rm z}$), although the atmosphere has small amounts of oxygen in the form of CO, \ce{CO2} and \ce{H2O}. As these chemicals ‘stew’ in the atmosphere they react many times, reaching large molecular sizes, eventually clumping together as microscopic haze particles akin to Earth’s smog or soot \cite{waite2007process}. 
Together and separately these photochemical products (gases, haze particles) condense, sediment and rain out onto Titan’s solid icy crust, collecting in vast dune fields of solid grains \cite{lorenz2006sand}, and in polar hydrocarbon lakes and seas \cite{stofan2007lakes}.

Titan therefore has both large atmospheric reservoirs and also surface deposits of the types of hydrocarbons used on Earth for fuel (e.g. methane-ethane heating/cooking gas, household propane, butane lighters, acetylene for welding) and as industrial feedstock for making plastics, rubbers and other synthetic materials \cite{brydson1999plastics}. Other than on Earth - where crustal hydrocarbons are the result of geochemical processing of biological materials \cite{sephton2013origins} - large reservoirs of complex organics are not found elsewhere in the solar system. Moreover, Titan’s hydrocarbons and other organics are found in much larger quantities than on Earth \cite{lorenz2008titan}.

These unique deposits make Titan potentially an important and unique resource for resupply of fuel and raw materials in the outer solar system, {e.g.} for an exploration vessel many years of travel from Earth, or for the construction and maintenance of a longer-term habitat. Fuel is available in the form of methane, acetylene etc, while raw materials such as hydrocarbon monomers can be used in 3D printing to make a wide range of useful products. Water is also available from Titan’s crust which can be electrolyzed to make \ce{H2} and \ce{O2}, as is the case for other icy moons in the outer solar systems, {e.g.} Europa, Enceladus etc \cite{davenport1991space}.

The use of resources on other worlds to support (typically) human missions is known as ISRU (In-Situ Resource Utilization, \cite{linne2017overview}). ISRU has become an area of active study, especially for the Moon and Mars that are the most likely near-term destinations for human landings and prolonged stays \cite{glaze2025moon}. For the remainder of this section we briefly review the topic of ISRU at the Moon and Mars, before proceeding to discuss Titan's resources and their usage in the remainder of this paper.

\subsection{Lunar ISRU}

\begin{figure}
    \centering
    \includegraphics[width=1.0\linewidth]{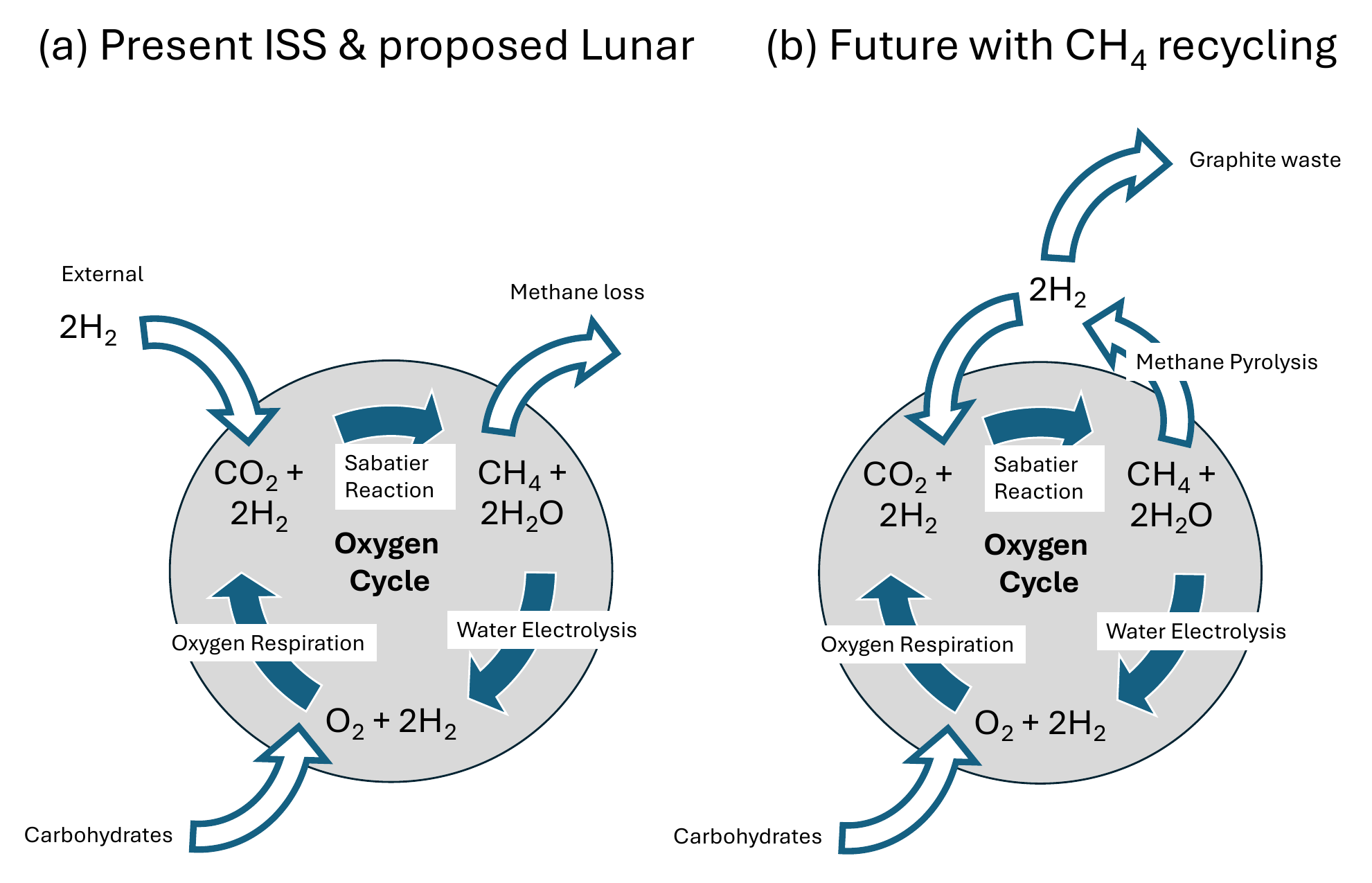}
    \caption{Chemical recycling of oxygen on the ISS and similar closed, resource-poor environments such as lunar bases. Oxygen is generated from electrolysis of water, and after use in respiration, \ce{CO2} may be recycled to water by the Sabatier technique, requiring \ce{H2}. Currently \ce{H2} is resupplied from the Earth (a), but in future it may be recycled from methane (b).}
    \label{fig:oxygen-cycle}
\end{figure}

The Moon, lacking in an atmosphere, consequently has its surface exposed directly to the solar wind. It also experiences large diurnal temperature swings during the 708.7 hour long lunar day (29.5 terrestrial days) that have left the regolith largely devoid of light element volatiles \cite{albarede2013asteroidal}. There is evidence for water trapped inside grains \cite{honniball2021molecular} and possibly exposed on the surface in the permanently shadowed regions (PSRs) of polar craters \cite{lawrence2017tale}, although evidence remains inconclusive \cite{li2026searching}. Lunar ISRU studies have therefore focused on electrolysis of water to generate \ce{O2} and \ce{H2} \cite{kleinhenz2020case}, which provides both fuel and breathable air - in addition to direct use of water itself. This begins the oxygen cycle, a series of reactions which can in principle become a closed loop preserving oxygen and even hydrogen. This allows recycling of scarce light elements in resource-poor environments, such as lunar colonies and space stations (Fig.~\ref{fig:oxygen-cycle}) \cite{sakurai2013fundamental}. 

More speculatively the lunar crust contains smaller amounts of implanted volatiles of other light species, such as ices of C and N, as well as $^3$He \cite{fegley1993lunar}. $^3$He can potentially be used as a fuel for fusion reactors, if it can be concentrated sufficiently \cite{wittenberg1992review}. For a wider review of other lunar ISRU possibilities see \cite{anand2012brief}. A major open question at the present time is the extent and distribution of useful ores on the Moon, which will greatly affect its utility as a mining destination \cite{crawford2015lunar, safronov2023theory}.

\subsection{Martian ISRU}

For Mars, comparatively rich in resources compared to the Moon, a recent overview study \cite{pazar2020resource} concluded that seven specific types of resources can be extracted directly or derived from basic existing resources in the Martian atmosphere, surface and surface: water, air, power, fuel, food, plastics and metals. 


Water (\ce{H2O}) is available at the polar caps and in permafrost or condensed from air, which can then be purified for drinking/food production and electrolyzed to produce \ce{H2} and \ce{O2}, as for the the Moon. \ce{H2} and \ce{O2} can be stored and later recombined in a fuel cell to regenerate electrical power \cite{iacomini2005electrolyzer}, or used for other purposes:

\begin{align}
    \ce{2H2O + 4e^-}  &\rightarrow \ce{2H2 + O2} & 
   (\mathrm{electrolysis:} \Delta H = 286 \: \mathrm{kJ/mol})
    \label{eq:electrolysis} \\
        \ce{2H2 +O2} &\rightarrow \ce{2H2O +4e^-} & \text{(fuel cell)} 
        \label{eq:fuelcell}
\end{align}

\noindent
\ce{H2} may then be combined with atmospheric \ce{CO2} (or from polar ice) in the Sabatier reaction to generate methane (\ce{CH4}) and \ce{O2} \cite{clark1997situ}:

\begin{flalign}
\label{eq:sabatier}
 &&  \ce{CO2 + 4H2} \rightarrow \ce{CH4 + 2H2O} && 
 (\mathrm{Sabatier:} \Delta H = -165 \: \mathrm{kJ/mol})
\end{flalign}

\noindent
and the resultant oxygen can be used to sustain plant and human respiration.

\ce{CO2} can be reduced to more useful CO in several ways, including by solid oxide electrolysis (SOEC):

\begin{flalign}
    && \ce{2CO2} \rightarrow \ce{2CO + O2} && 
     (\mathrm{SOEC:} \Delta H = 280 \: \mathrm{kJ/mol})
\end{flalign}

\noindent
which also produces \ce{O2}, as demonstrated recently by the MOXIE experiment on the Perseverance rover \cite{hoffman2022mars}, or with graphite in the Boudouard reaction \cite{zeitlin2015self}:

\begin{flalign}
    && \ce{CO2 + C} \rightarrow \ce{2CO} && (\mathrm{Boudouard:} \Delta H = 172 \: \mathrm{kJ/mol})
\end{flalign}

\noindent
Either way, the resulting CO can then be combined with \ce{H2} in the Fischer-Tropsch process to yield \ce{C2H4} \cite{deliismail2024mini}:

\begin{flalign}
    && \ce{4H2 + 2CO} \rightarrow \ce{C2H4 + 2H2O} && (\mathrm{Fischer}-\mathrm{Tropsch:} \Delta H = -165 \: \mathrm{kJ/mol \: CO})
\end{flalign}

\noindent
providing a feedstock for plastics \cite{greenblatt2023design} - and water. 

Another variant adds water along with \ce{CO2} in the SOEC, creating both parts of syngas (CO + \ce{H2}) in one step, as well as oxygen:

\begin{flalign}
    && \ce{CO2 + H2O} \rightarrow \ce{CO + H2 + O2} && 
    \mathrm{Syngas:} \Delta H = 525 \: \mathrm{kJ/mol})
\end{flalign}

\begin{figure}
    \centering
    \includegraphics[width=0.95\linewidth]{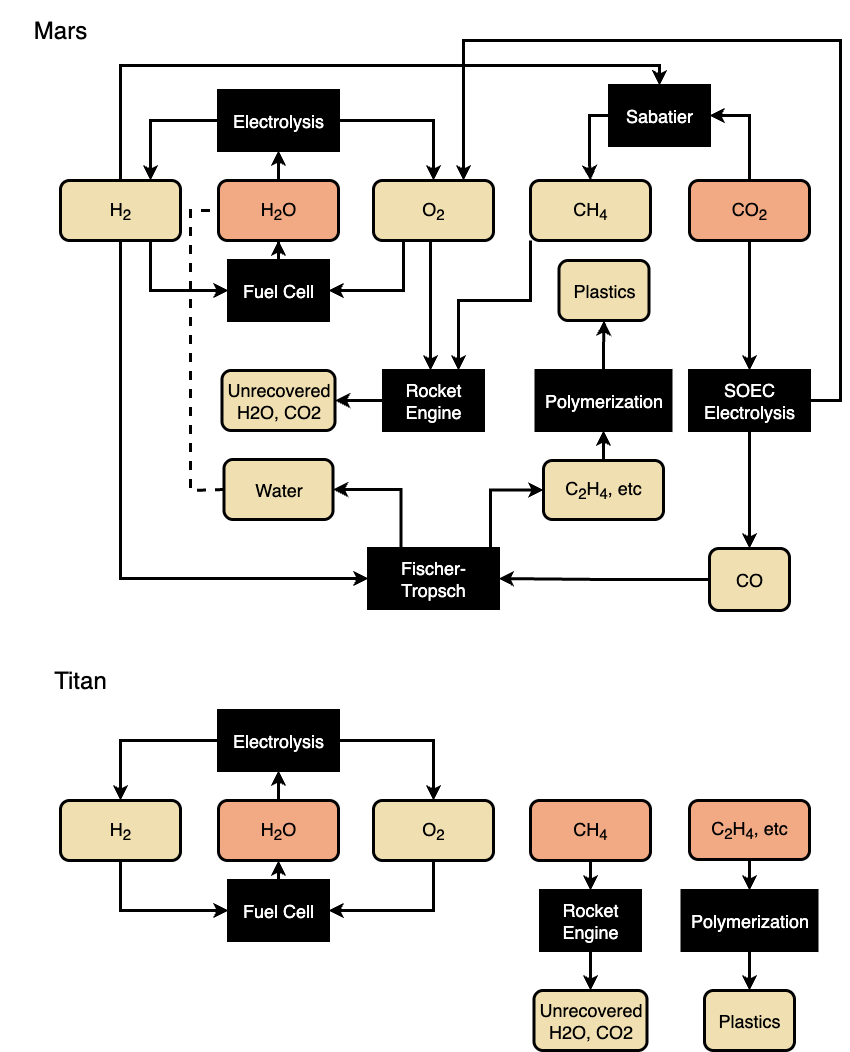}
    \caption{Simple C-H-O chemical production pathways on Mars (top) and Titan (bottom). Orange cells indicate abundant resources to be used as raw materials, while yellow boxes are derived resources. Black boxes indicate processes. The availability of hydrocarbons on Titan obviates the need for chemically reducing processes such as Sabatier and Fischer-Tropsch that are needed on Mars to remove oxygen from carbon. }
    \label{fig:cho-cycle}
\end{figure}

Figure~\ref{fig:cho-cycle} (top) summarizes these processes.

Nitrogen (\ce{N2}) is available in the ambient Mars air (2.7\%), and can be used both to create human breathable air, and to create other needed chemicals such as fertilizers (nitrates, NOx). These can  be produced directly from Martian air using energetic dissociation of \ce{CO2} and \ce{N2} \cite{kelly2022producing}, or by fixing of atmospheric nitrogen via cyanobacteria \cite{shen2019nitrates}. Nitrates are furthermore found in Martian soil \cite{stern2017nitrate}, as on deserts on Earth \cite{shen2019nitrates}. 

Finally, metals such as Fe, Ca, Al, Mg and Si are abundant in the Martian regolith, although reducing their ores from highly oxidized states is energy intensive. A feasible route appears to be carbothermal reduction (heating with graphite, which can be generated from atmospheric \ce{CO2}) \cite{nababan2025metals}. The reaction proceeds via intermediary CO, which reduces the metal oxide (MOx) and is turn reduced by graphite:

\begin{align}
    \ce{MOx + xCO}  &\rightarrow \ce{M + xCO2} & \text{(oxide reduction)} \\
        \ce{CO2 + C} &\rightarrow \ce{2CO} & \text{(CO regeneration)} 
\end{align}

\subsection{Asteroids and Comets}

Near-earth asteroids (NEAs or NEOs) represent a source of resources nearby to the Earth \cite{sercel2018practical} whose diversity is only beginning to be realized. Their orbits in general are relatively unstable, and therefore NEAs persist in their positions for only a few million years before being ejected from the solar system or colliding with a planet or the Sun, due to gravitational perturbations \cite{shustov2023dynamic}. In turn, this implies that a constant source of new NEAs must exist, which is thought to be from the main asteroid belt \cite{deelia2007collisional}. NEAs are therefore likely to have a similar range of compositions to main belt asteroids (MBAs) \cite{dunn2013mineralogies}. 

Several missions to NEAs have been completed, including the NEAR-Shoemaker mission that landed on asteroid 433 Eros in 2001 \cite{prockter2002near}; the Chang'e 2 mission that flew by 4179 Toutatis in December 2012 \cite{zou2014preliminary}; along with three completed sample return missions:

\begin{itemize}
    \item {\em Hayabusa:} the first asteroid sample return mission that brought back material from 25143 Itokawa to Earth in 2010 \cite{yoshikawa2021hayabusa}.
    \item {\em Hayabusa-2:} a second JAXA mission that returned samples from 162173 Ryugu in 2020 \cite{tsuda2013system}.
    \item {\em OSIRIS-REx:} NASA's first asteroid sample return mission that collected regolith from the surface of 101955 Bennu and returned this to Earth in 2023 \cite{lauretta2017osiris}.   
\end{itemize}

These samples continue to be analyzed to better understand the composition and diversity of asteroids, and a full review of findings is beyond the scope of this paper. However, at the high level preliminary findings show that NEAs have similar diversity to MBAs, including stony types (S-types, Eros and Itokawa for example) and darker, more carbon-rich types (C-type) including Ryugu and Bennu \cite{mo2025taxonomic}. 

For asteroid ISRU this implies that plentiful water and organic materials can be found on C-type objects while S-type objects may prove to be useful sources of metals. Asteroid ISRU is an active area of research with hopeful commercial entities being created to mine asteroids \cite{zacny2013asteroid, hein2020techno}. While such exploits remain ambitious with current technologies, in the more distant future asteroid ISRU both of NEAs and MBAs will likely prove highly enabling for human expansion and colonization of the solar system. 

For missions to reach the outer solar system and Titan therefore, asteroid ISRU may prove to be an important stepping stone, and later an important supplementary source of materials difficult to find on icy moons, such as metals \cite{cannon2023precious, kargel1994metalliferous, rios2024platinum}. We will return to this topic in Section \ref{sect:aerospaceneeds}.

\subsection{ISRU at Titan}

ISRU at Titan has received relatively less attention, with the notable exception of a recent sample return  mission study (Titan ISRU Sample Return - TISR, \cite{landis2022mission, oleson2022titan} and is hence the topic of the current paper. We first quantify some of the resources that are available on Titan for use by future missions, especially the ready availability of reduced hydrocarbons which greatly simplifies the production of some complex materials (see Fig.~\ref{fig:cho-cycle} - bottom), and mention some that are not (especially metals). We then outline some possible mission architectures, followed by methods of resource extraction, refining and utilization.  Finally, we present our summary and conclusions.



\section{Titan’s resources and their potential uses}
\label{resources}

Titan's atmosphere is composed mostly of \ce{N2} ($\sim$95\% at surface) and \ce{CH4} ($\sim$5\% at surface) \cite{niemann2010composition}, with the next most abundant gases being \ce{H2} (0.1\%) \cite{courtin2012abundance} and CO \cite{serigano2016isotopic}. Atmospheric photochemistry then creates from these a wide variety of other substances \cite{nixon2024composition}, primarily hydrocarbons (C$_x$H$_y$) and organic cyanides (nitriles: C$_x$H$_y$(CN)$_z$)). These provide a diverse array of available substances that are potentially useful to support exploration and life-support. Table~\ref{tab:resources} provides estimates for the prevalence of some of the more abundant chemicals in Titan's atmosphere and seas/lakes \cite{vuitton2019simulating, cordier2009estimate}.

\begin{table}[ht]
    \begin{centering}
            \caption{Relative amounts of different hydrocarbons and other light molecules available in Titan's atmosphere and in surface lakes/seas. 
            } 
\begin{tabular}{lrr}
 \label{tab:resources}
    \\
\hline
Substance &	Mole Fraction & Atmos. Col. Abund.   \\
 &  Seas (\%)$^a$ & (molecules/cm$^{-2}$)$^b$  \\
		 	\hline		
                & & \\
            \multicolumn{3}{l}{\em Hydrocarbons:} \\
Methane &       9.69 &		1.10E+25  \\
Ethane &        76.40 &		2.65E+19  \\
Acetylene &     1.15 &		5.48E+18  \\
Propane &       7.42 &		1.55E+18  \\
Butane &        1.21 &		1.06E+17  \\
Butene &        1.39 &		2.20E+16  \\
Benzene &       0.02 &		1.54E+14  \\
 &  & \\
            \multicolumn{3}{l}{\em Other species:} \\
Nitrogen &      	0.49 &	2.76E+26  \\
Hydrogen &      	0.00 &	2.75E+23  \\
Argon &	        	0.00 &	1.04E+22  \\
Carbon Monoxide &  	0.00 &	1.47E+22  \\
Hydrogen Cyanide &  2.09 &	3.00E+17  \\
Acetonitrile &		0.10 &	2.21E+16  \\
Carbon Dioxide &    0.03 &	7.84E+15  \\
\hline
   \end{tabular}
       \end{centering}
    \\   \\ {  Notes: \\ 
$^a$ From \cite{cordier2009estimate}. \\
$^b$ Typical Titan column profile from \cite{vuitton2019simulating}. }
\normalsize
\end{table}


\subsection{Hydrocarbons}
\label{sect:hydrocarbons}

\subsubsection{Alkanes}
\label{sect:alkanes}

Fully saturated hydrocarbons (alkanes: C$_{\rm n}$H$_{\rm 2(n+1)}$) are valuable fuels due their high H/C ratio. The simplest alkane, methane (\ce{CH4}) is already ubiquitous on Titan in both gaseous and liquid form \cite{niemann2010composition, cordier2009estimate}, thereby negating the need for the Fischer-Tropsch reduction of carbon needed on Mars (Fig.~\ref{fig:cho-cycle}, \cite{pazar2020resource}). The next simplest hydrocarbons - ethane (\ce{C2H6}) and propane (\ce{C3H8}) - have also been detected in Titan's atmosphere \cite{ lombardo2019ethane, nixon2009titan} and ethane has been detected in several Titan seas \cite{brown2008identification, mastrogiuseppe2014bathymetry} and in the crust \cite{niemann2010composition}, where it is expected to widespread in form of the clathrate \cite{mousis2008sequestration}. 

\subsubsection{Alkenes}
\label{sect:alkenes}

Alkenes are \emph{unsaturated} hydrocarbons with one or more carbon-carbon double bonds and chemical formula: C$_{\rm n}$H$_{\rm 2(n+1-m)}$, where $n$ and $m$ are the number of carbon atoms and the number of double bonds respectively. The simplest two mono-alkenes ethene (\ce{C2H4}) and  propene (\ce{C3H6}) have been identified in Titan's atmosphere \cite{hanel1982infrared, nixon2013detection} along with the simplest di-alkene, propadiene (\ce{CH2CCH2}, \cite{lombardo2019detection}).

\begin{figure}
    \centering
    \includegraphics[width=1.0\linewidth]{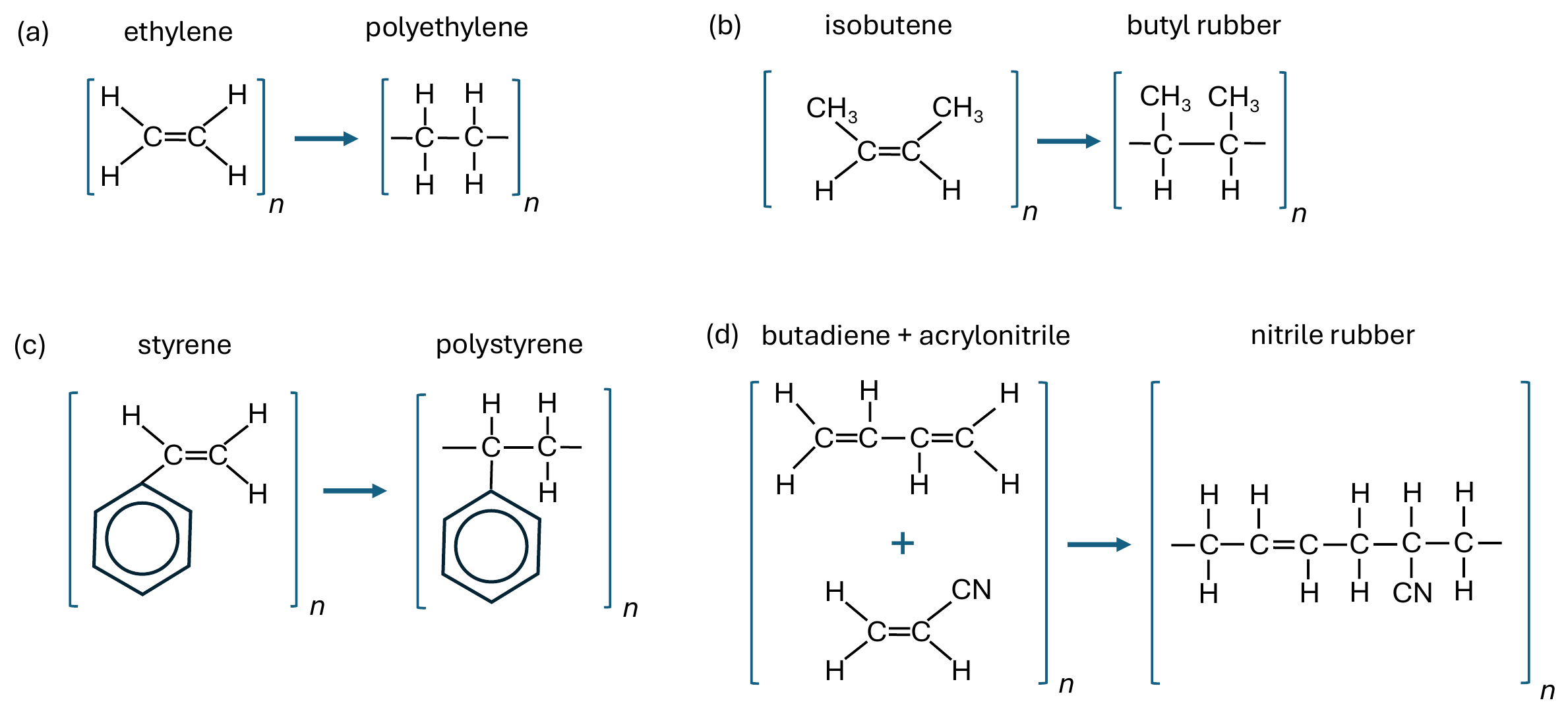}
    \caption{Examples of plastic and rubber-forming through polymerization of C=C double bonds: (a) polyethylene; (b) polysytrene; (c) butyl rubber; (d) nitrile rubber. The monomers required to form these polymers are either detected (ethylene, acrylonitrile) or likely to exist (styrene, isobutene, butadiene) on Titan.}
    \label{fig:polymerization}
\end{figure}

Alkenes are valuable feedstocks for production of plastics and rubbers. The industrial importance of alkenes lies in their ability to polymerize – a process where the double bond is partially broken (under sufficient conditions of temperature/pressure and possible catalysis) \cite{schmerling1950mechanism, bochmann2006kinetic} , leaving a single bond within the original structural unit and a new bond created to an adjacent molecule. The process is illustrated in Fig.~\ref{fig:polymerization}(a).

Polymerization of ethene leads to polythene (polyethylene, PE) \cite{zhong2017polyethylene}, a light-weight plastic used for  storage bags, milk jugs, pipes and, at higher densities, bearings and fittings used in industry and medicine, such as artificial joints \cite{chen2021polyethylene}. Polyethylene is the most widely produced industrial plastic, followed by polypropylene (PP) which is less dense but harder. PP is used in plastic boxes (e.g. as Tupperware), tubes and medical materials such as hernia repair meshes. Both PE and PP come in different types depending on how the polymers are packed, branched etc, and may be co-polymerized with each other and other alkenes \cite{karian2003polypropylene, maddah2016polypropylene}. 

Heavier alkenes are used to make synthetic rubbers. Polymerization of isobutene ((\ce{CH3})$_3$C-CH=\ce{CH2}) leads to butyl rubber \cite{higgins1990butyl} (Fig.~\ref{fig:polymerization}(b)), which being impermeable to air is used in all manners of caulks, sealants, adhesives, gloves, cling wrap etc. Butadiene is often used to make co-polymers (polymers from more than one alkene) such as SBR (styrene-butadiene rubber), commonly found in car tires \cite{zubov2012styrene}. 

\subsubsection{Alkynes}
\label{sect:alkynes}

Alkynes, like the alkenes, are unsaturated hydrocarbons but possess stronger (and more energetic) triple bonds and formula: C$_{\rm n}$H$_{\rm 2(n-m)}$. The simplest mono-alkynes ethyne (\ce{C2H2}, commonly called acetylene) and propyne (\ce{CH3C2H}) have been detected in Titan's atmosphere, along with the lightest di-alkyne - di-acetylene (\ce{C4H2}) \cite{teanby2009titan}. On Earth, alkynes find uses in energy-dense fuel applications such as welding and also formerly for lighting (e.g. lighthouses) and in the production of various chemicals including organic semiconductors \cite{voronin2018acetylene}.

\subsubsection{Benzene, styrene and cyclic molecules}
Benzene, the archetypal six-sided carbon ring molecule (\ce{C6H6}) was first detected on Titan using the Infrared Space Observatory (ISO, \cite{coustenis2003titan}). Benzene is frequently used as the basis for modification into other industrial feedstocks, especially to form styrene (\ce{C6H5}-\ce{C2H3}) by addition of the vinyl group \cite{matsumoto2002direct}. The double bond in the side group permits polymerization into the ubiquitous low-density packing material polystyrene \cite{lynwood2014polystyrene} (Fig.~\ref{fig:polymerization}(c)). Benzene is also used a fuel additive, especially when modified by the addition of the methyl moiety to form toluene (\ce{C6H5CH3}). Benzene can also be chemically transformed to the saturated ring cyclohexane (\ce{C6H12}) used as common solvent, e.g. in correction fluid.


\subsection{Nitrogen species}
\label{sect:nitrogen}

Nitrogen in the form of \ce{N2} is readily available on Titan, being the major component of the atmosphere \cite{niemann2010composition}. The atmosphere also contains significant amounts of nitriles (organic cyanides) such as HCN with the general formula X-CN \cite{molina2002nitriles, gautier2011nitrile}. Nitriles, especially vinyl cyanide (acrylonitrile, \ce{C2H3CN}) which also possess the alkene C=C double bond, can be co-polymerized with other hydrocarbon monomers to form a variety of plastics and rubbers with different properties (see section...). Plastics formed largely from acrylonitrile (>85\%) are referred to as `acrylics' and have widespread use in modern manufacturing, from clothing to tennis racquets, tents and aerospace components \cite{mosley2017acrylic}. The co-polymer of acrylonitrile with butadiene (\ce{H2}C=C-C=\ce{CH2}) leads to nitrile rubber (Fig.~\ref{fig:polymerization}(d)), used in medical gloves, footwear, adhesives, sponges, mats and more \cite{sruthi2020overview, mackey2000nitrile} . Finally, nitrogen can potentially be used to manufacture types of rocket fuels such nitrogen tetroxide (\ce{N2O4}) and various methylated hydrazines ((CH$_3$)$_\mathrm{n}$N$_2$H$_{(\mathrm{4-n})}$ - see Section~\ref{sect:fuelproduction}).


\subsection{Oxygen}
\label{sect:oxygen}

\begin{figure}
    \centering
    \includegraphics[width=1.0\linewidth]{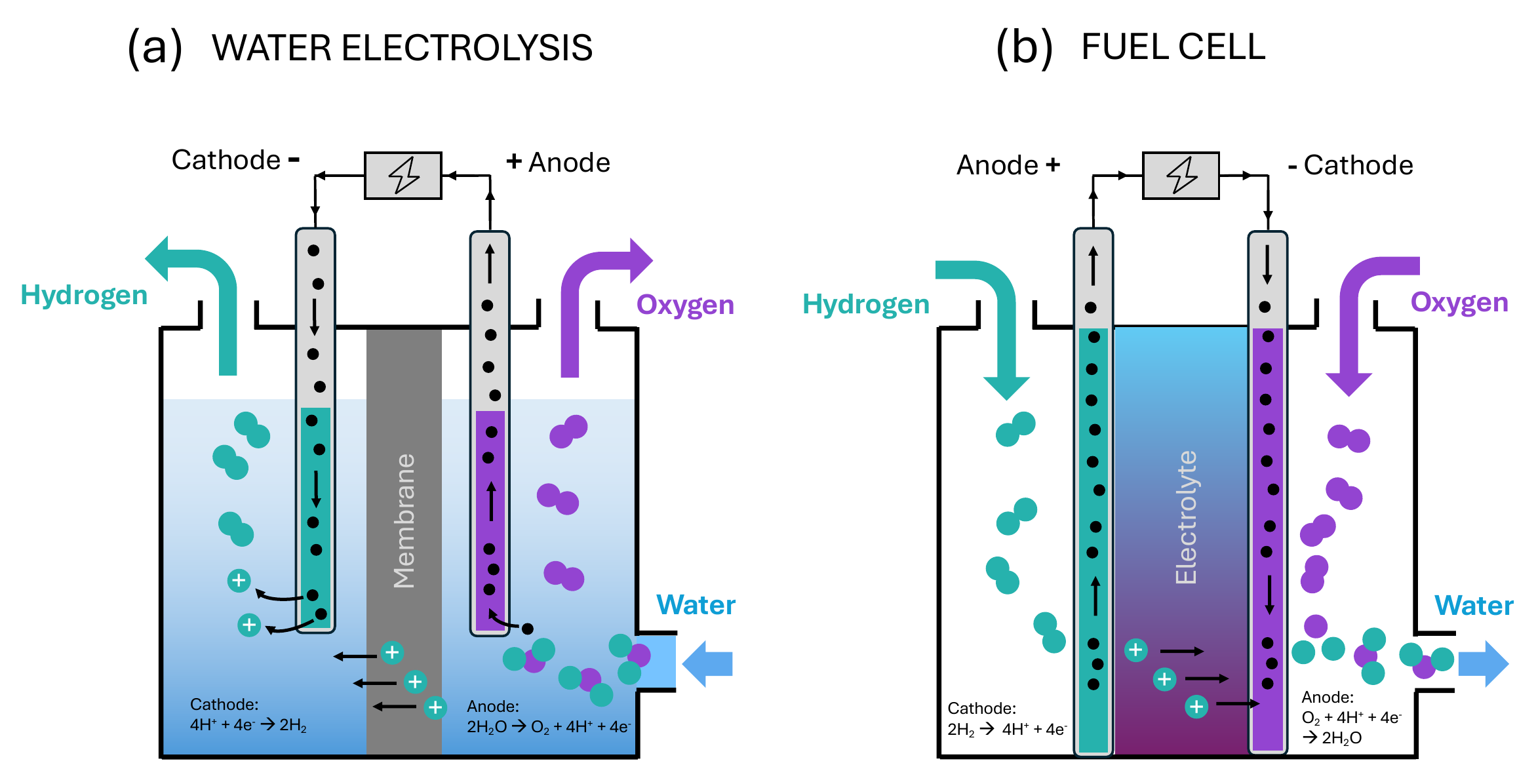}
 \caption{(a) Electrolysis of water via Proton Exchange Membrane (PEM) method. (b) PEM Fuel cell for electricity generation.}
    \label{fig:fuel-cell}
\end{figure}

Atmospheric oxygen on Titan is scarce, being found mostly in the form of CO (50 ppm, \cite{serigano2016isotopic}), with trace amounts of \ce{CO2} and \ce{H2O}. On the other hand, water is readily available on Titan's surface, being the principal ingredient of its crust \cite{tobie2014origin}. Crustal ice will require melting and purifying before usage for drinking, washing, plant irrigation etc. Water ice is also the best and most convenient source of oxygen, which may be produced via electrolysis using several different technologies (e.g. Proton Exchange Membrane - PEM - Fig.~\ref{fig:fuel-cell}(a) \cite{liu2023recent}) and is the method by which most oxygen is generated on the ISS \cite{takada2024status}.

Besides direct use for life support and combustion as rocket fuel, fuel cell electricity generation (Fig.~\ref{fig:fuel-cell}(b)), welding, heating etc - oxygen is also an important and necessary component for making various chemicals such as alcohols, aldehydes and ketones \cite{lawrence1998aldehydes} (see Section~\ref{sect:solventsglues}). While such chemicals may exist in trace amounts in the atmosphere, their naturally occurring abundances may be too low for efficient extraction, necessitating their manufacture instead via chemical engineering methods \cite{crawford1994aldehydes}.

\subsection{Resources Missing}
\label{sect:missing}

Due its original composition as a 50\% rock/50\% ice body and subsequent differentiation \cite{tobie2014origin}, Titan’s crust at the present day is thought to be composed mostly of water ice, with some amount of methane and ammonia in clathrate form \cite{kalousova2020insulating}. The crust is blanketed in solid refractory hydrocarbons, a product of atmospheric photochemistry \cite{lorenz2006sand, radebaugh2008dunes}.

A consequence is that heavy materials such as rock and metals are likely to be extremely scarce at Titan’s surface and would pose a limitation and need for any long-lasted surface colonies. Such materials would need to be brought from outside (e.g. mined and refined from other rocky moons, asteroids, or inner planets) and likewise would constitute resources that could not be refueled from Titan to any passing spacecraft.

Sulfur and phosphorus are also important elements whose abundance on Titan is currently unknown; however unlike metals they are more potentially present in trace amounts in the atmosphere or crust \cite{nixon2013upper, pasek2011phosphorus, hickson2014evolution, fortes2007ammonium}.

\subsection{Power generation}
\label{sect:power}

Available power on Titan is in short supply. Atmospheric and surface hydrocarbons can be oxidized, but require \ce{O2}, which is lacking. Generation of \ce{O2} has been discussed earlier in the this section, but itself requires power to split from water or \ce{CO2}. Solar power is also scarce at Titan's surface, with perhaps only 0.1\% of the photons that reach the surface of the Earth \cite{tomasko2008heat}. Wind power in Titan's sluggish lower atmosphere is also not likely to provide much energy \cite{tokano2015wind}.

The most logical power source is therefore nuclear, requiring a small reactor to be brought from Earth. Small nuclear reactors have been studied for in-space propulsion \cite{velidi2020nuclear, koroteev2015nuclear, mason2022nuclear, loeb2015realistic} as well as to power lunar \cite{belov2022some, nikitaeva2022power} and martian settlements \cite{bushman2004martian, balint2004nuclear, fan2024conceptual, yuan2024design}. A nuclear reactor powering a spacecraft could be repurposed for use on Titan's surface, providing a long-lived power source. Of course, replacement nuclear fuel would need to be sourced from elsewhere, most likely the inner solar system.




\section{Environmental Challenges at Titan}
\label{sect:challenges}

\subsection{Gravity}
\label{sect:gravity}

Gravity is an essential and often ignored part of materials processing and manufacturing on Earth, simplifying many processes.
Modification of processes for Titan's surface and orbital gravities will require an interdisciplinary team of material scientists, manufacturing experts,  engineers and scientists. Surface-based operations can leverage many traditional approaches, even in the reduced Titan gravity of 1.35 $ \mathrm{m/s}^2$ (about 1/7 Earth's gravity) \cite{jacobson_gravity_2006}. This is close to the lunar gravity, so any research to adjust Earth-based manufacturing for lunar ISRU will be applicable to Titan \cite{Araghi2022LunarISRU}. That said, care must be taken to unravel requirements for low-gravity versus for lunar pressures (high vacuum). 


\begin{figure}[ht]
    \centering
    \includegraphics[width=1.0\linewidth]{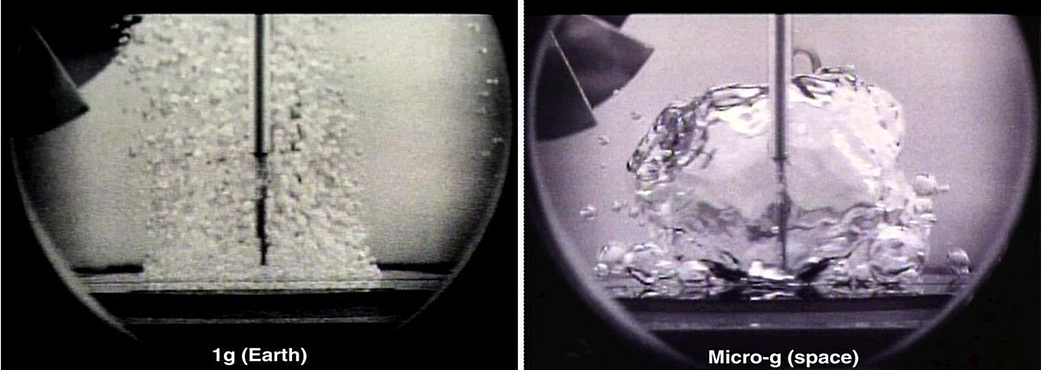}
    \caption{Liquid heated to boiling point in 1-g, generating vapor bubbles that rise. Right: Liquid heated to boiling point in microgravity, which occurs in the absence of natural convection or buoyant flows. Under these conditions, bubbles generated at the heater coalesce into a single large bubble. Photo Credit: NASA}
    \label{fig:poolboiling}
\end{figure}

On-orbit activities fall into the microgravity regime, and experience gained in weightless environments on Earth-orbiting space stations such as MIR, the ISS, and Tiangong have illuminated many of the problems of microgravity and suggest some remedies. For example, fractional distillation (discussed later in Section~\ref{sect:collectingrefining}, which separates substances based on boiling points, generates bubbles at surfaces that do not evolve (Fig.~\ref{fig:poolboiling}), \cite{Warrier2015NPBXMicrogravity}, requiring unusual approaches, such as specialized surface engineering of hardware \cite{Wen2018CapillaryFilmBoiling}. 

\begin{figure}[ht]
    \centering
    \includegraphics[width=0.75\linewidth]{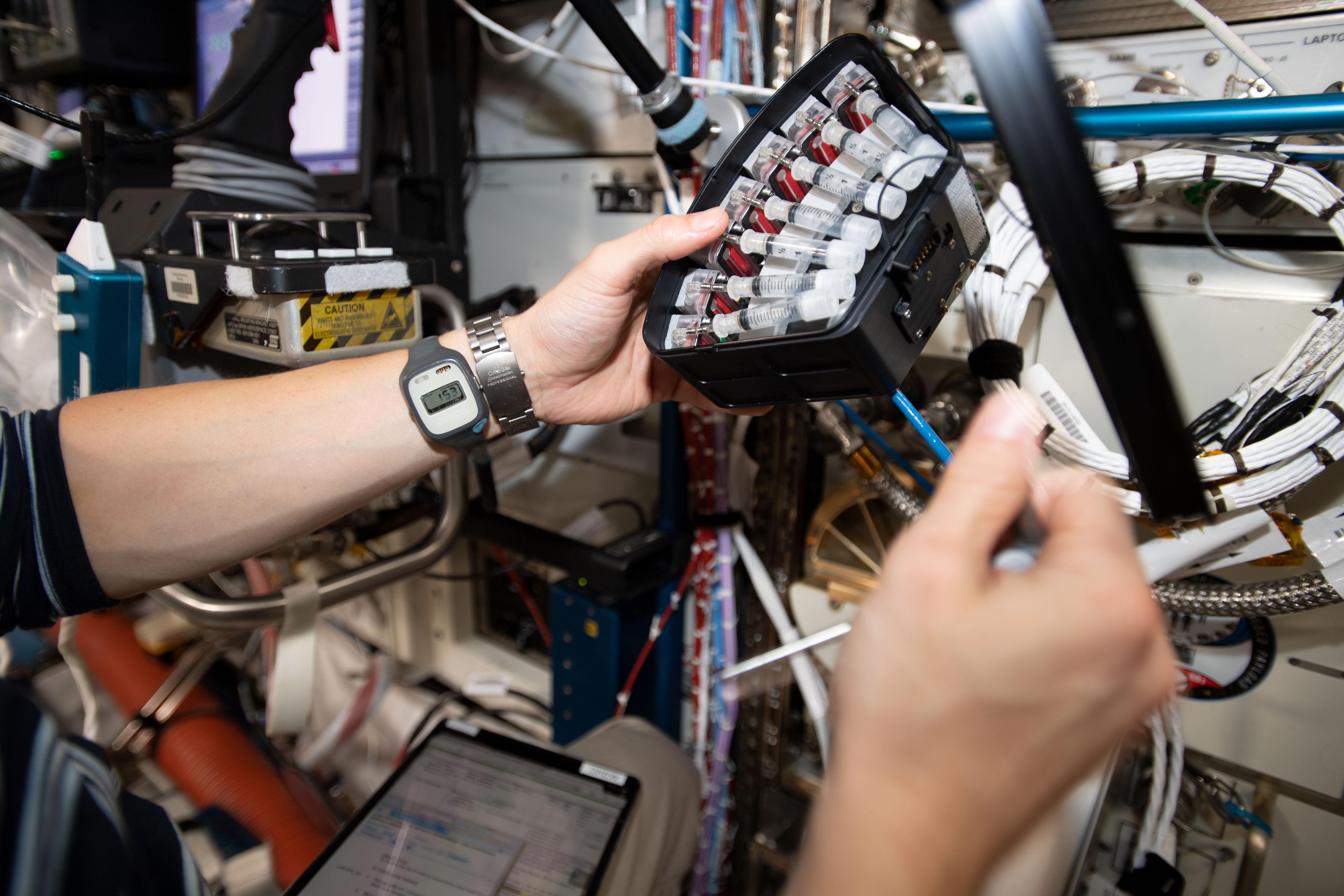}
    \caption{The Multi-use Variable-g Platform (MVP) is a centrifuge-based ISS research platform capable of generating artificial gravity in 0.1 g increments up to 2.0 g, enabling controlled gravity comparisons between Earth, Moon, Mars, and potentially Titan. Credit: NASA}
    \label{fig:mvp}
\end{figure}

Particle sedimentation on-orbit will require artificial gravity, such as in a centrifuge. This was demonstrated on the ISS at up to two times Earth's gravity using the Multi-use Variable-gravity Platform (MVP), a space biology research platform that can produce up to 2g of artificial gravity. (Fig.~\ref{fig:mvp}) \cite{mvp_hardware}. 

Common manufacturing techniques that may require adapting to on-orbit operations (microgravity), are mixing, gluing, soldering, and welding – where gravity is commonly employed, but not intrinsically required. Free-floating drips could pose a problem, without the simplifying `clean up' of gravity, although a fan may serve a similar purpose to direct drips in one direction.  In addition, on-orbit welding has been an established activity in space since the 1960s \cite{naden2020review}. 

Techniques that do not depend on gravity include machining, riveting, drilling, nailing, screwing – so long as a force can be applied in some way (which requires careful consideration of Newton’s third law). For example, drilling a hole would be entirely possible in low or zero gravity, provided that the drill is braced or attached to a large, fixed mass (e.g. a tabletop or wall). We note that metals typically needed for machining tools are not widely available on Titan, and hence will need to be brought from elsewhere, used sparingly and if possible, recycled. 

\subsection{Pressure and Temperature Considerations}
\label{sect:presstemp}

\begin{figure}[ht]
    \centering
    \includegraphics[width=1\linewidth]{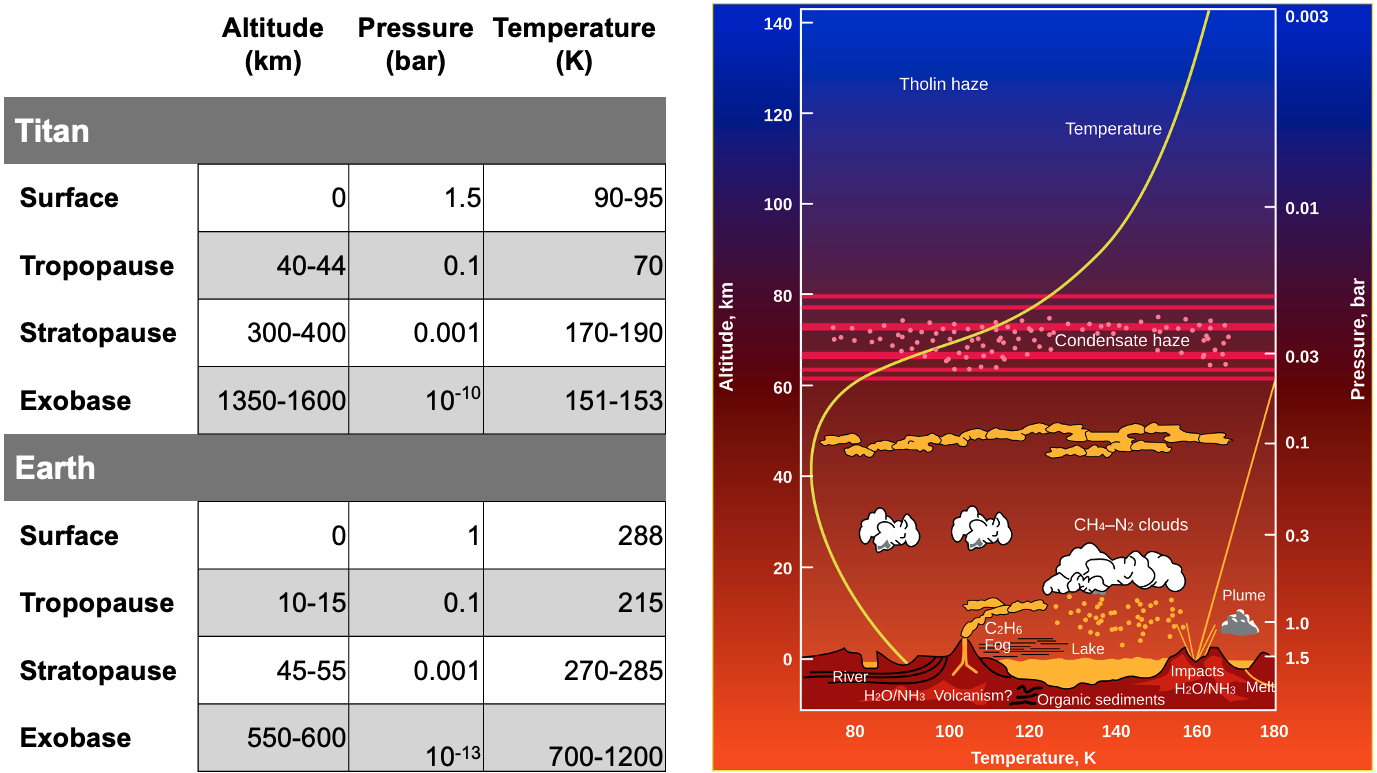}
     \caption{Right: Vertical cross-section diagram of Titan's atmosphere. Image: ESA. Left: Comparison of atmospheric profiles between Earth and Titan.}
    \label{fig:profile}
\end{figure}

Titan's surface has a dense, cold atmosphere (Fig.~\ref{fig:profile}), with near-Earth pressures and frigid temperatures. The Huygens probe, the first probe to land on a body in the outer solar system, collected data on Titan's surface in 2005 and measured about 1.5 bar and 94 K at its landing site \cite{Fulchignoni2005TitanHuygens}, and global analysis shows minimal seasonal and spatial variation \cite{horst2017titan}. The low temperatures seen on the surface cause ethane and methane to exist as liquids \cite{stofan2007lakes}, and permit methane phase cycling \cite{turtle2011seasonal}, which shapes the landscape in a similar manner to water on Earth \cite{jaumann2008fluvial}. 

Figure \ref{fig:profile} compares Titan and Earth's atmospheres and highlights significant differences in atmospheric layer heights and thicknesses. In addition, Titan's atmosphere is roughly 35\% of Titan's terrestrial radius, while Earth's atmosphere is only 8\%. Mission hardware must operate at cryogenic temperatures, however the increased surface pressures relative to Earth can enable unique spacecraft design such as rotocraft and aerial platforms.

Given the pressure and temperature differences on Titan's surface and atmosphere compared to Earth, boiling points for known species are different for any manufacturing process. Manufacturing inside pressurized surface structures can mimic some conditions of Earth, and so developing manufacturing processes should take this into consideration when developing engineering requirements.

\subsection{Surface Atmospheric Chemical Exposure}
As noted in earlier sections, Titan's surface and surface atmosphere are rich in chemicals, including \ce{H2}, that are known to affect structural materials. Candidate flight hardware should therefore be evaluated against relevant chemical mixtures. NASA Kennedy Space Center's Swamp Works has developed expertise and standards for ISRU development for the Moon and Mars \cite{NASASwampWorksKSC}, providing research facilities tailored to these unique environments. Similar environmental simulation and analysis programs are needed for Titan's surface environment. These include atmospheric chambers, specialized materials testing equipment, and campaigns to develop and maintain a reference database of flight hardware performance for Titan.

\subsection{Thermal management}
\label{sect:powerthermal}


In general, processes selected for manufacturing benefit from maximizing temperature-tolerance for the location of interest. In orbit, spacecraft are in vacuum and so management of heat generated by on-board processes is limited to radiative heat transfer. Therefore, despite ambient temperatures being low, heat can build up in sensitive components. These craft should include heat conductive pathways to shift generated heat to areas which need heating. This thermal differential could be leveraged for power generation using thermoelectric materials along the thermal gradient (Seebeck effect). 

On Titan's surface, heat transfer is open to radiative, conduction, and convection processes. Therefore, the extremely low temperatures will be a major challenge for maintaining high enough operations temperatures as heat is quickly dissipated. Thermally insulated structures may be required for surface operations. Heat dissipation from surface structures and equipment may interfere with measurement sensors \cite{Lorenz2016HeatRejectionTitan}, so minimizing thermal loss from structures, or strategically recycling waste heat should be considered. 


\section{Mission Architectures}
\label{sect:architectures}

Several previous studies have proposed mission concepts for sample return from Titan surface,  assuming either that all propellant is launched from Earth \cite{Donahue2009,Marlin2022}, or an ISRU operation is used for propellant production \cite{oleson2022titan,wpi_titansamplereturn}. The mission architectures for Titan sample return are largely similar, with launch from Earth launch, Titan entry and surface ascent. Sample return missions have notably smaller payload masses (on the order of 10 kg) while human-class missions have payload masses on the order of $10^1-10^2$ tons. In this section, we provide an overview of several options for end-to-end mission architecture, from Earth launch, interplanetary transfer to and from Saturn, to flight elements supporting human presence on Titan, and Titan orbit. Such endeavors will leverage from knowledge and infrastructure of previous human missions to the Moon \cite{Green2023-vq} and Mars \cite{Drake2010}. 

\subsection{Mission Trajectories}
\label{sect:trajectories}

\subsubsection{Earth Launch and Interplanetary Transfer to Saturn}
Interplanetary transfers generally have two parameters for high-level trade-off --- time-of-flight and total fuel mass required ($\Delta V$). A mission with a shorter time-of-flight will typically have a higher $\Delta V$ requirement, resulting in lower usable mass for payload; whereas a mission with a longer time-of-flight will have a lower $\Delta V$ requirement, resulting in higher usable mass for payload. 

\begin{figure}[ht]
    \centering
    \includegraphics[width=1\linewidth]{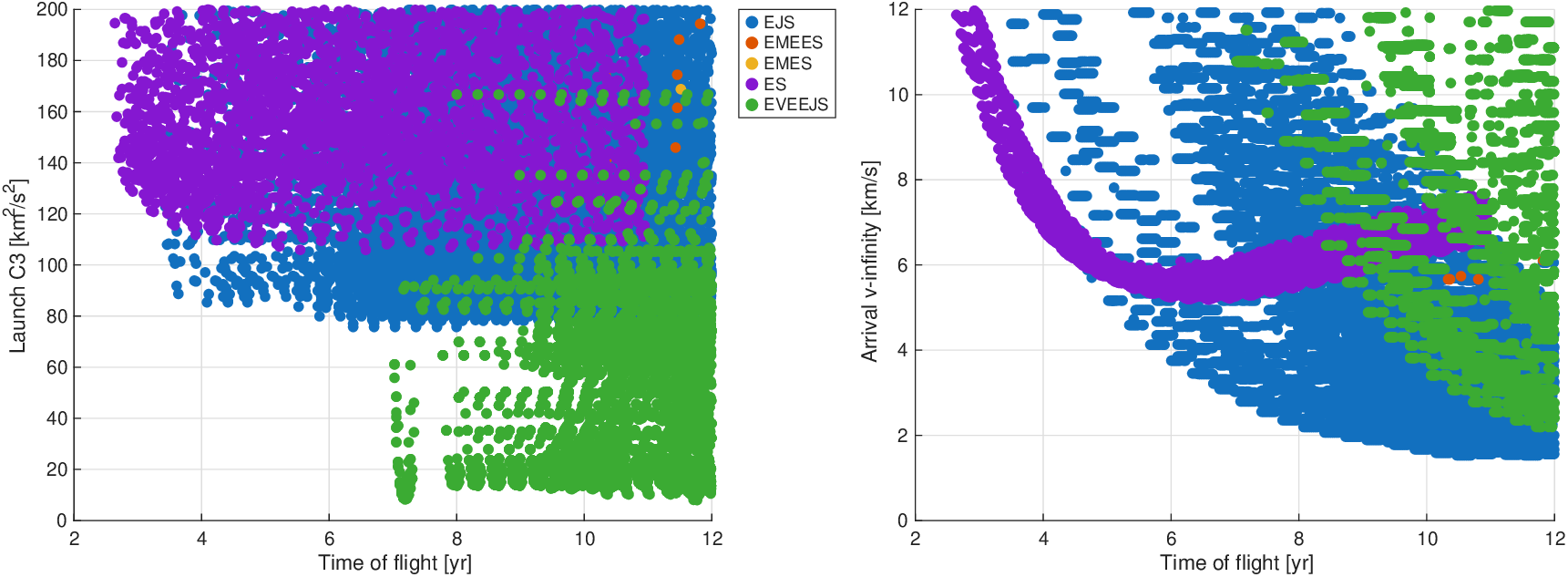}
    \caption{Launch C3 and Arrival $V_\infty$ vs time-of-flight for Earth-Saturn transfer during 2030-2060}
    \label{fig:c3-vinf-vs-tof}
\end{figure}

As shown in Fig. \ref{fig:c3-vinf-vs-tof}, there exists a pareto-optimal front between launch C3 and arrival $V_\infty$ between Earth and Saturn. Transfers denoted with cyan are direct Earth-Saturn transfer, which come with yearly-launch opportunities (Earth-Saturn synodic period is 378 days). Transfers denoted with red use one Jupiter gravity-assist after launch, the EJS sequence (Earth-Jupiter-Saturn), which have sporadic launch opportunities depending on the phasing of Jupiter relative to Saturn. 

Longer time-of-flight transfer can generally reduce the total $\Delta V$ requirement, but those transfers require gravity-assist at Mars, Venus, and Earth. For example, the dark green trajectories on the bottom right of the left figure in Fig. \ref{fig:c3-vinf-vs-tof} are EVEEJS gravity-assist sequence, Earth-Venus-Earth-Earth-Jupiter-Saturn, that has significantly lower launch C3 than ES and EJS options, but longer time-of-flight. 

Human missions will typically require multiple launches, while some flights may be robotic explorers or cargo only, resulting in different requirements for transfers. Transfers with humans on board will likely prioritize shorter time-of-flight for crew health and safety (i.e., ES and EJS trajectories with short TOF), while transfers for cargo-only missions can use more efficient transfer opportunities using multiple gravity-assists, i.e., EVEEJS sequence. The options shown here are for ballistic transfers and these general trends are still valid even if a mission uses a low-thrust propulsion system for post-launch transfers. 

\subsubsection{Arriving at Saturn and Titan}
Establishing a human presence at Titan will require orbit insertion. Conventional methods have relied on propulsive orbit insertion. For human-class vehicles, aerocapture may be a promising orbit insertion method. NASA's Human Mars Mission DRA 5.0 \cite{Drake2010} has assumed aerocapture orbit insertion for cargo missions, but opted for a propulsive maneuver for crewed missions due to the perceived risk of aerocapture at Mars during the DRA study in 2009. Since then, there has been extensive literature and analysis on aerocapture orbit insertion \cite{Lu2018-jc,Girija2020-kj,Girija2022-pb,Deshmukh2024-or}. 

Studies have shown that Titan's thick atmosphere is extremely benign for aerocapture orbit insertion \cite{bailey2003titan, lockwood2003titan, spilker2005significant, nixon2016aerocapture}. Aerocapture at Titan does not have the inherent risks and technical issues that were observed for aerocapture at Mars and ice giants (Uranus and Neptune). In addition, Titan aerogravity-assist is another similar maneuver that uses aerobraking from Titan's atmosphere to become captured into an orbit around Saturn \cite{Lu2020}. 

\subsubsection{Departing Titan/Saturn and Return Transfer }
Inbound missions returning from Titan/Saturn to Earth follow similar trends for time-of-flight vs. departure and arrival velocities as discussed previously for outbound Earth-Saturn transfer. Direct transfer from Saturn to Earth will have yearly launch opportunities aligned with the synodic period. Transfers with gravity-assists, especially Jupiter, may reduce time-of-flight and the total $\Delta V$ requirements. In contrast to outbound Earth-to-Saturn transfer, returning from Titan to Earth will likely be all-crew missions, prioritizing shorter time-of-flight and implying chemical propulsion systems. Orbital transfer techniques are known that can reduce  $\Delta V$ requirements by using an Oberth maneuver, V-infinity leveraging, and Titan gravity-assist maneuver(s) \cite{Pekosh2022}. 

\begin{figure}[h!]
    \centering
    \includegraphics[width=1\linewidth]{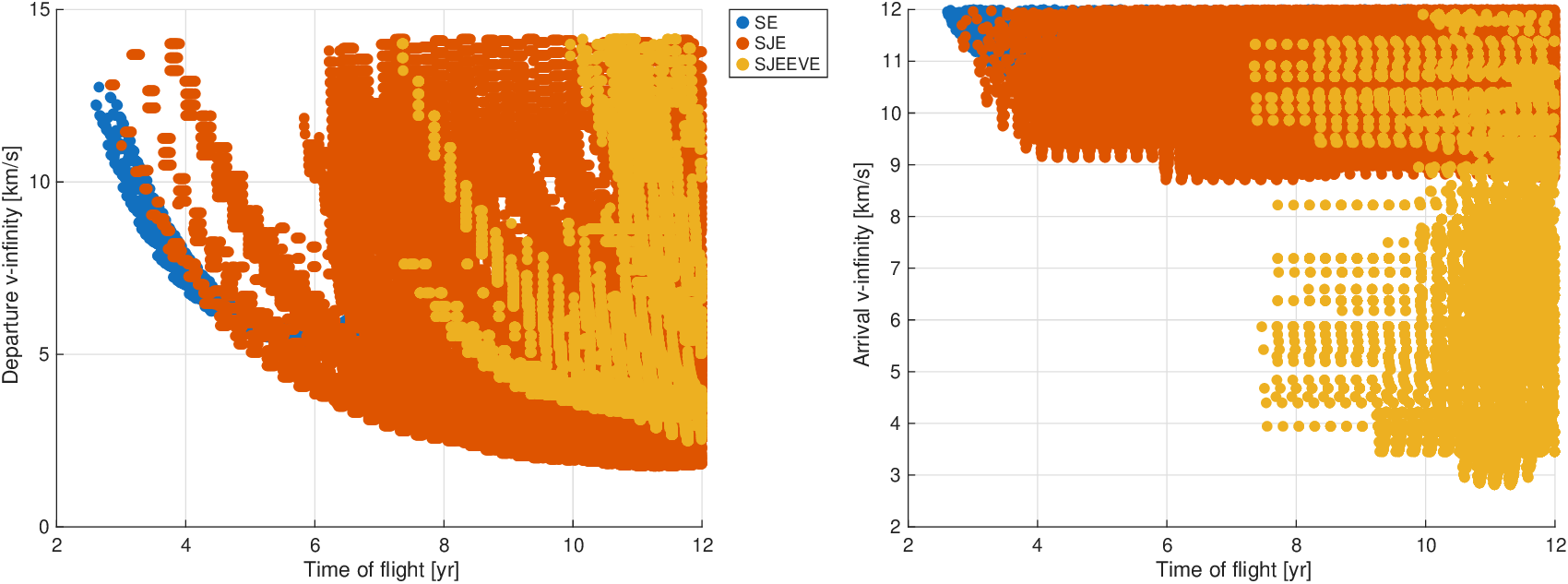}
    \caption{Saturn departure $V_\infty$ and Earth arrival $V_\infty$ for return transfer options.}
    \label{fig:placeholder}
\end{figure}

Another option for transfer to and from Saturn is the cycler trajectory \cite{Byrnes1993} as advocated by Aldrin and colleagues for Human to Mars missions \cite{chen2002preliminary, chen2005powered}. A cycle trajectory is where a transfer stage encounters Earth and Mars repeatedly (cyclically) using gravity-assists from both Earth and Mars. The transfer stage hosts all life support systems, thus providing a more efficient mission architecture. This then permits smaller launches for each subsequent human mission using the pre-established infrastructure; this is similar to how International Space Station (ISS) operates. A drawback of the cycler transfer stage is the restriction of launch opportunities. A workaround is to have multiple transfer stages in the mission architecture, but these incur cost and risk considerations that are beyond the scope of this paper \cite{rogers2012preliminary, rogers2015establishing, pelle2019earth}.

\subsubsection{Trajectory to other Moons and Asteroids}

Titan's environment provides easy / direct access to hydrogen, carbon, nitrogen, and oxygen, however access to metallic elements will likely be difficult. Metallic asteroids (e.g. Psyche) may be a necessary waypoint to acquire needed metals for long-duration, self-sufficient missions.

\subsection{Saturn In-System Mission Architectures}
\label{sect:saturnsystem}

There are several options for Saturn in-system mission architectures. With a targeted human presence on Titan's surface, there will likely be support vehicles in orbit. Some options include a dedicated Titan orbiter, an orbiter leveraging a periodic orbit at the Saturn-Titan libration points, and a Saturn orbiter that is in a resonance transfer with Titan. 

This section will primarily focus on four distinct types of refueling and provisioning (see Table~\ref{tab:architectures} and Figs.~\ref{fig:flyby}-\ref{fig:depots}) that could be envisaged for a passing spacecraft at Titan. While these present four quite distinct, high-level mission architectures, many more types are possible, including hybrids between these models. 

\begin{table}[h!]
    \centering \footnotesize
    \begin{tabular}{lllll}
    \hline
  Code	& Mission Type	& Resource Collection	& Refining 	& Manufacturing \\
   & & & \\
    \hline
FB	& Flyby	& Atmospheric scoop	& Small scale, 	& Small scale,  \\
 & & & in spacecraft & in space \\
 
OL	& Orbit with 	& Surface retrieval from   	& On surface	& Small scale, \\
 & Lander & lakes/seas or atmosphere; & & in space  \\
  & & materials refined & & \\
& & before launch. & & \\

RS	& Rendezvous  	& Passing ship collects refined  	& On surface	& Large scale,  \\
& with Orbital & fuel and manufactured  & & in orbit \\
 & Station & materials from orbiting & &  (on station) \\
 & & station, which in & & \\
 & & turn collects and refines raw & & \\
 & & materials on surface. & & \\

SS	& Surface Station	& Shuttle retrieves fuel and  and   	& On surface	& Large scale,  \\
 & & manufactured materials & & on surface \\
 & & from surface station and  & & \\
 & & returns to orbiting & & \\
 & & spacecraft; surface station & & \\
 & & extracts raw materials & & \\
 & & from surface & & \\
\hline
    \end{tabular}
    \caption{Architecture-specific differences in resource collection, refining and manufacture.}
    \label{tab:architectures}
    \normalsize
\end{table}



\subsubsection{Atmospheric Flybys}
\label{sect:flybys}

By the term ‘flyby’ in this context, we consider together three distinct types of ‘flybys’ – (i) a flyby of Titan from outside of Saturn orbit (i.e. heliocentric orbit, or hyperbolic orbit escaping the solar system entirely); (ii) a Saturn orbit with Titan encounters; (iii) an elliptical Titan orbit with low altitude ‘scooping’ periapsis. A circular, low-altitude orbit encountering the atmosphere in any significant way seems ruled out, since the spacecraft would rapidly lose energy and re-enter, unless constantly burning fuel to maintain station. 

\begin{figure}
    \centering
    \includegraphics[width=0.8\linewidth]{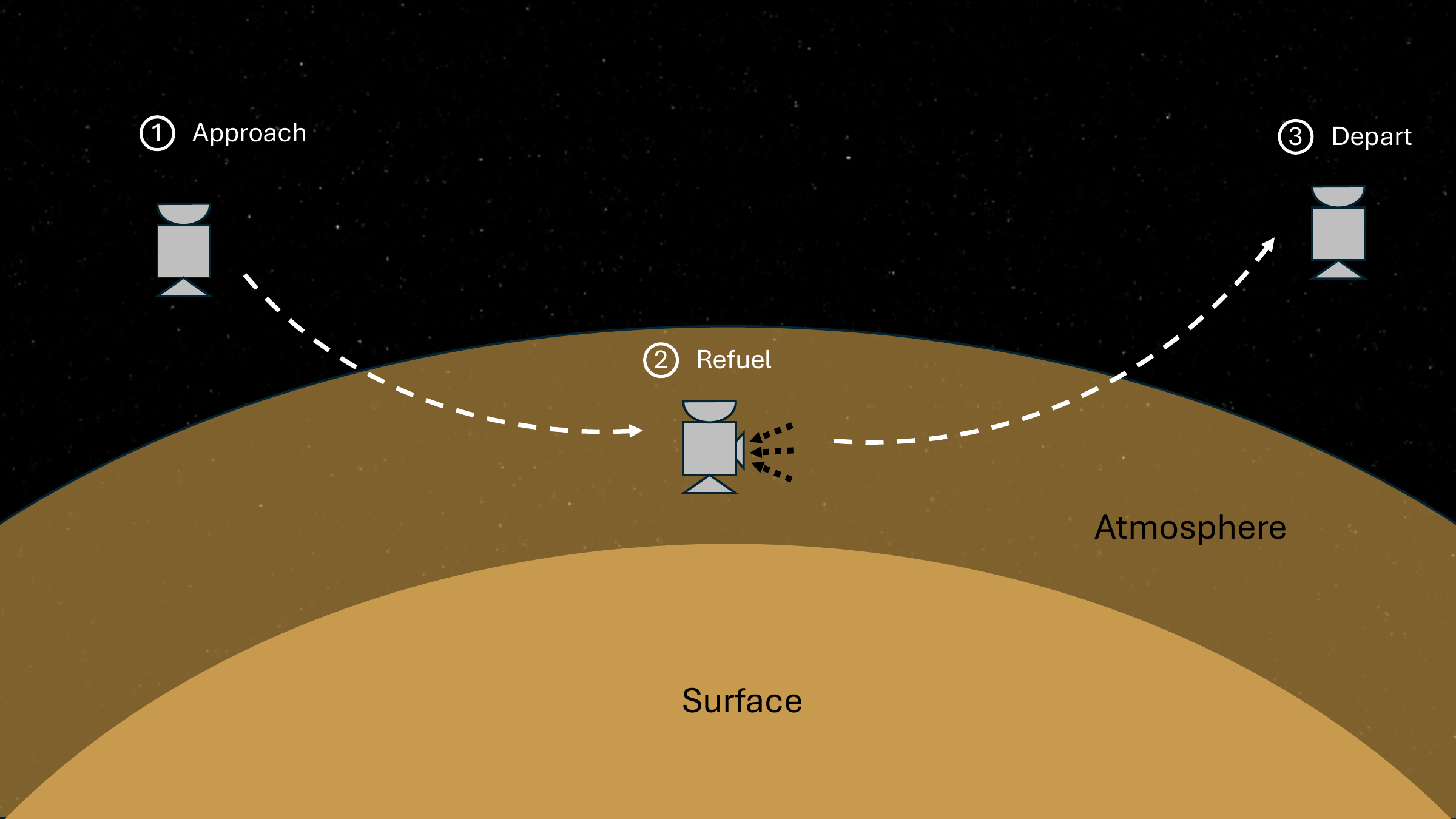}
    \caption{Titan flyby refueling mission (mission type `FB').}
    \label{fig:flyby}
\end{figure}

Type (i) would likely have several disadvantages, including only a single fuel collection pass, and the high encounter velocity that would lead to high temperatures and reaction in a collection vessel. Therefore types (ii) and (iii) seem the most favorable. Flyby missions (Fig.~\ref{fig:flyby}) have several advantages over any type of landing, including a much simpler architecture with fewer elements and therefore fewer complications and risks. They also avoid the fuel expense in retrieving fuel from the surface. Disadvantages may include a longer time (significant number of passes) to collect sufficient fuel, and the possibility of reactions occurring during fuel collection, depending on the encounter velocity. However, there exists uncertainty as to whether a net positive energy gain would be realized, between fuel energy collected versus energy loss due to atmospheric drag. 

\subsubsection{Stopovers}
\label{sect:stopovers}

\begin{figure}
    \centering
    \includegraphics[width=0.8\linewidth]{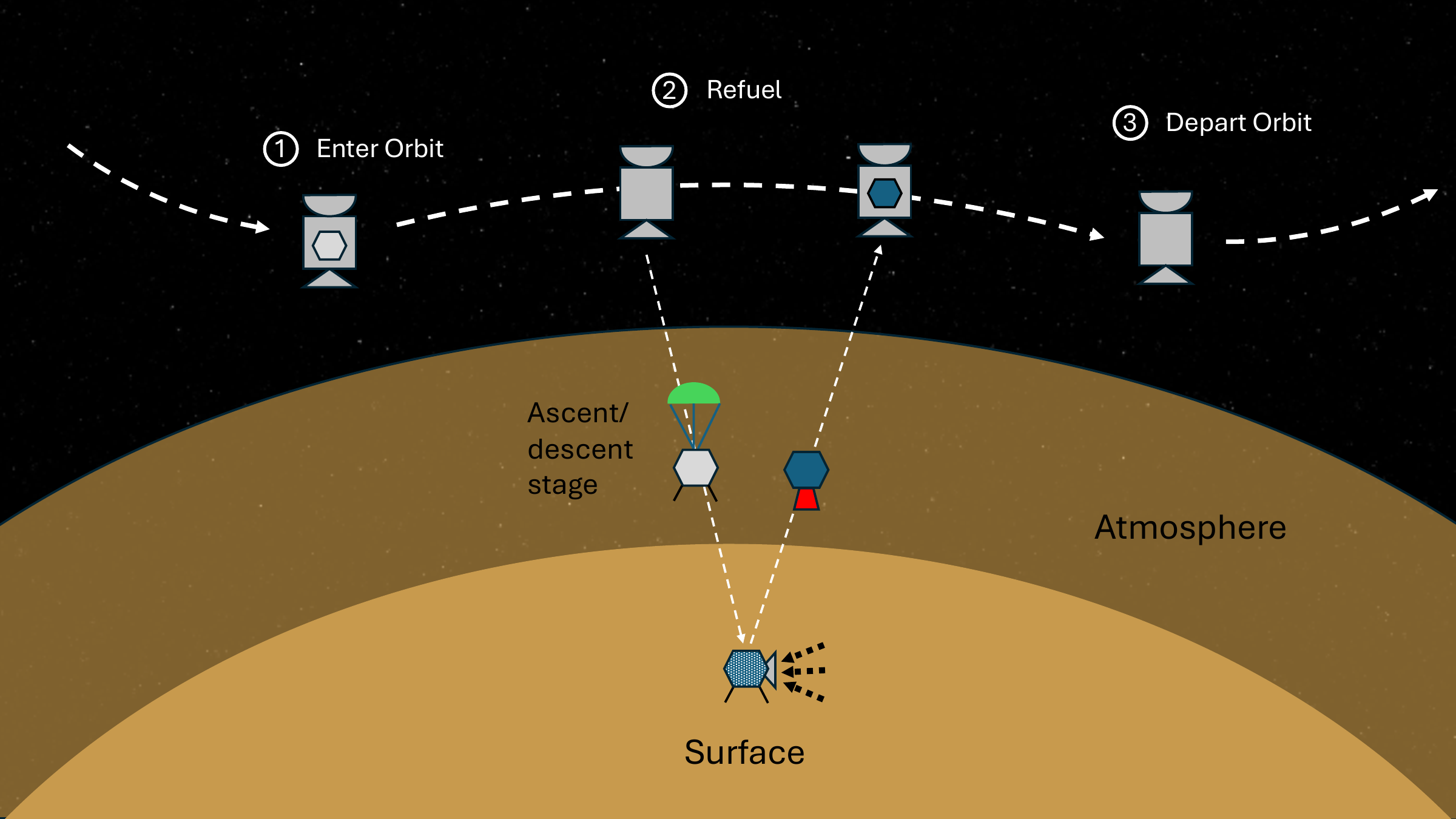}
    \caption{Titan stopover refueling mission (mission type `OL').}
    \label{fig:stopover}
\end{figure}

In the ‘stopover’ scenario (Fig.~\ref{fig:stopover}) we include missions that enter Titan (or possibly Saturn) orbit, sending a landing vehicle down to the surface to retrieve and probably refine fuel (e.g. \cite{oleson2022titan}). In such mission scenarios there is no permanent infrastructure on the surface, so the entire mission is self-contained. Fuel and other materials are collected on the surface from the lakes/seas, atmosphere, dunes and even crust (see Section~\ref{sect:collecting}). Advantages include a longer access time to collect and refine fuels (not constrained to brief encounter passes), although the overall refueling period may proceed more quickly than in the flyby scenario, which may require many flybys to collect sufficient material. Disadvantages include that the refueling ship must be entirely self-contained for refueling, including the surface landing craft, and refining and manufacturing capability. 

\subsubsection{Titan-orbiting stations}
\label{sect:orbtiting}

\begin{figure}
    \centering
    \includegraphics[width=0.8\linewidth]{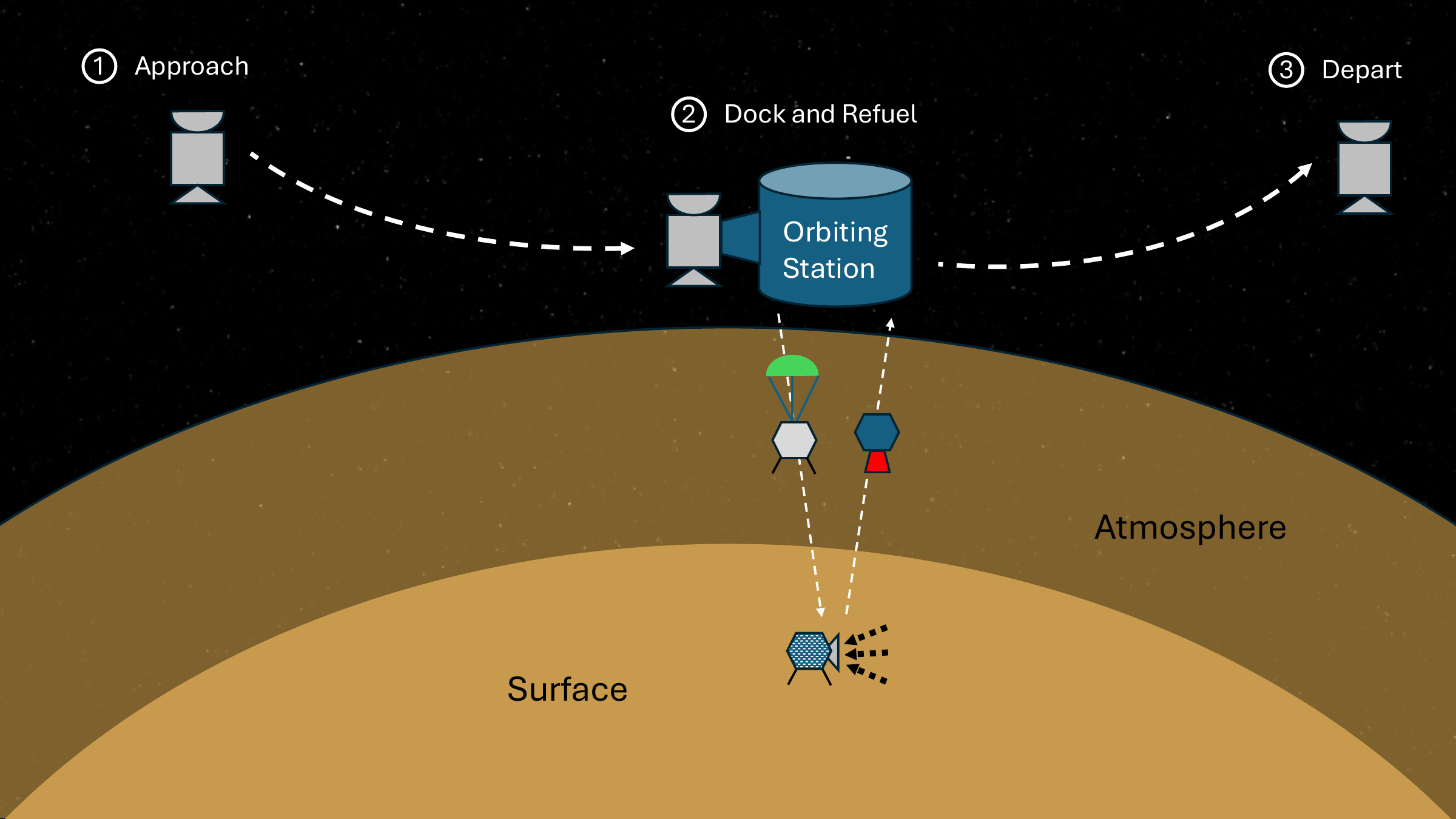}
    \caption{Titan orbital station refueling (mission type `RS').}
    \label{fig:orbital}
\end{figure}

To avoid carrying the extra load of a refueling surface landing craft (type (ii)), and refining and manufacturing equipment (types (i) and (ii)), these may be permanently stationed at Titan and available to any passing spacecraft. This scenario (Fig.~\ref{fig:orbital}) would make sense if Titan becomes a regular waystop for refueling for many spacecraft, in the case of a more developed outer solar system economy. 

In this scenario, a Titan (or perhaps Saturn) orbiting space station provides refueling and provisioning for passing spacecraft, including manufactured goods and chemicals. In turn, the station adopts the responsibility of collecting raw and refined materials from the surface, stocking these in orbit, and further processing and refining into useful high-level goods. This also allows an advantage of the station retrieving additional raw materials from other Saturn moons (e.g. ores, metals, sands etc) or being restocked from the inner solar system by a bulk cargo ship.

The advantage to the passing spacecraft is that most of the difficulties of fuel collection and refining are off-loaded onto the station, although in return the spacecraft gives up its autonomy and self-sufficiency. If for example the station was to become disabled or destroyed, a spacecraft from the inner solar system without fuel collection, refining and manufacturing capability might end up in a dire situation, running of fuel and life essentials. For this reason, having multiple backup stations, or having some amount of self-sufficiency on the passing spacecraft may be essential to protect against single-point failures.

\subsubsection{Titan Surface Depots}
\label{sect:depots}

\begin{figure}
    \centering
    \includegraphics[width=0.8\linewidth]{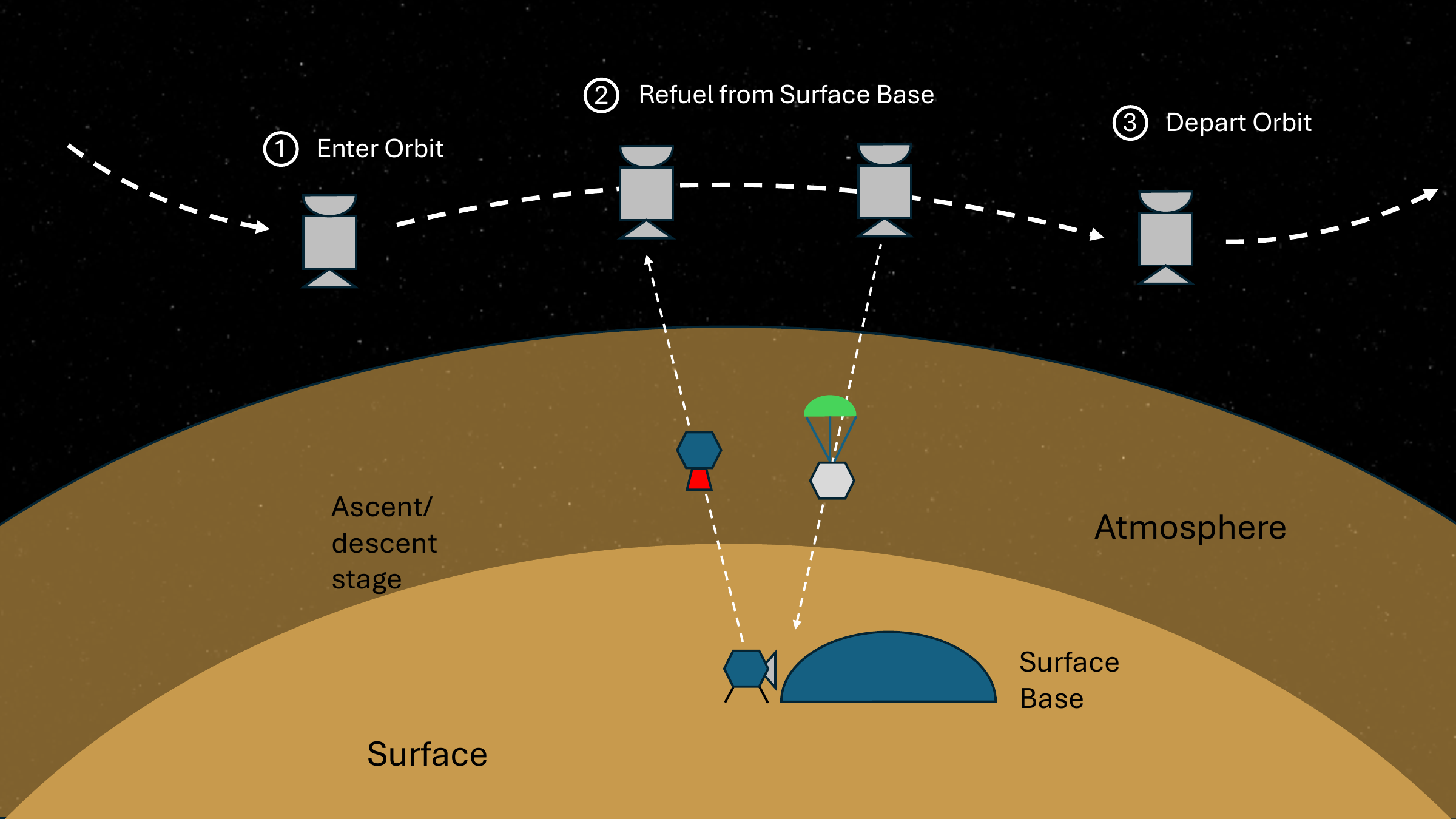}
    \caption{Titan depot refueling (mission type `AS').}
    \label{fig:depots}
\end{figure}

At some point it may become desirable to have more permanent presence on the surface of Titan itself. This would have several advantages, including allowing for longer-term exploration and even tourism, and allowing for large-scale refining and manufacturing to occur on the surface, with only the most high-value materials (refined fuels and manufactured goods) to be lifted out of Titan’s gravity well. In this scenario (Fig.~\ref{fig:depots}), a passing spacecraft could be refueled from the surface with the option of not carrying a refueling landing craft, since that could be permanently stationed on the surface and launched as needed to refuel passing spacecraft. 

Advantages of this approach again include reducing the need for passing spacecraft to carry landing craft and advanced refining and manufacturing capabilities, at the cost of less self-sufficiency. In addition, there would be the cost of establishing and maintaining a permanent base on the surface.  

\subsection{Systems Considerations}
\label{sect:systems}

\subsubsection{Propulsion System and Fuel Types}
\label{sect:propulsion}

Currently available in-space propulsion systems can be broadly categorized into three classes: chemical propulsion, electric propulsion, and propellant-less propulsion. Although electric propulsion systems offer very high propellant efficiency, their low thrust levels make them unsuitable for mission phases requiring large $\Delta V$ within short time scales, such as orbital insertion or ascent from surface. Propellant-less concepts face similar limitations in thrust magnitude, and their can be availabilities based on environment (i.e., solar sail). Therefore, chemical propulsion remains the most practical and mature option for human-class missions to Saturn and Titan. 

Table \ref{tab:propellant_summary} summarizes the common chemical propellant combinations that can be derived from elements readily available on Titan, i.e., hydrogen, carbon,  oxygen, and nitrogen. Notably, the storage temperature of oxidizer and fuel indicate the fuel storage on Titan can be strongly influenced by Titan's cryogenic surface environment at 94 K as it may fundamentally changes conventional propellant handling strategies used on Earth. Hydrocarbon fuels can be easily stored as stable liquids with no active cooling, while some cooling may still be necessary for oxygen. On the contrary, RP1 and N$_2$H$_2$ would freeze under Titan conditions and would require continuous heating to maintain suitable storage conditions. While all fuel types can be valid options, hydrocarbon-based fuels would be favored for fuel storage considerations. 

\begin{table}[h!]
\centering\footnotesize
\caption{Summary of propellant combinations: storage temperatures and performance}
\begin{tabular}{lcccc}
\hline
{Propellant Pair} & {Oxidizer Temp (K)} & {Fuel Temp (K)} & {$I_{sp}$ (s)} & {$v_e$ (m/s)} \\
\hline
O$_2$ + H$_2$      & 90.2 & 20.3   & 440--450 & 4315--4413 \\
O$_2$ + CH$_4$     & 90.2 & 111.7  & 317--370 & 3109--3628 \\
O$_2$ + RP1       & 90.2 & 273--300 & 309--330 & 3030--3236 \\
O$_2$ + N$_2$H$_4$ & 90.2 & 386.7  & 345--354 & 3383--3472 \\
N$_2$O$_4$ + N$_2$H$_4$ & 294 & 386.7 & 313--328 & 3069--3217 \\
\hline
\end{tabular}
\label{tab:propellant_summary}
\end{table}

Nuclear fission-based propulsion systems is a promising future  technology that can significantly reduce the fuel mass requirement. In particular, nuclear-power propulsion system can achieve a specific impulse of 800-1000 s, more than twice as efficient as a high-performance chemical propulsion system \cite{Kumar_2025}.

\subsubsection{Options for Launch and Spacecraft }

For deep-space human-class missions, super-heavy lift launch vehicles are required. While further technology developments are expected, the existing launch and transfer vehicle option most suitable for a human-class mission is the SpaceX Starship spacecraft with its super heavy rocket that can deliver 100+ tons of payload to Earth orbit and beyond \cite{wilken2022critical}. This system was designed for deep-space transportation, in particular establishing permanent human presence on Mars \cite{heldmann2022mission}. The Starship spacecraft also has a mid-lift-to-drag ratio hypersonic vehicle geometry that is preferred for aerocapture maneuver at Earth and Titan, enabling fuel savings for initial orbit insertion and even some orbit transfer maneuvers. 

\subsubsection{Requirement for Fuel Mass and $\Delta V$ }

\begin{figure}[h!]
    \centering
    \includegraphics[width=1\linewidth, trim={0 0 0 2cm}, clip]{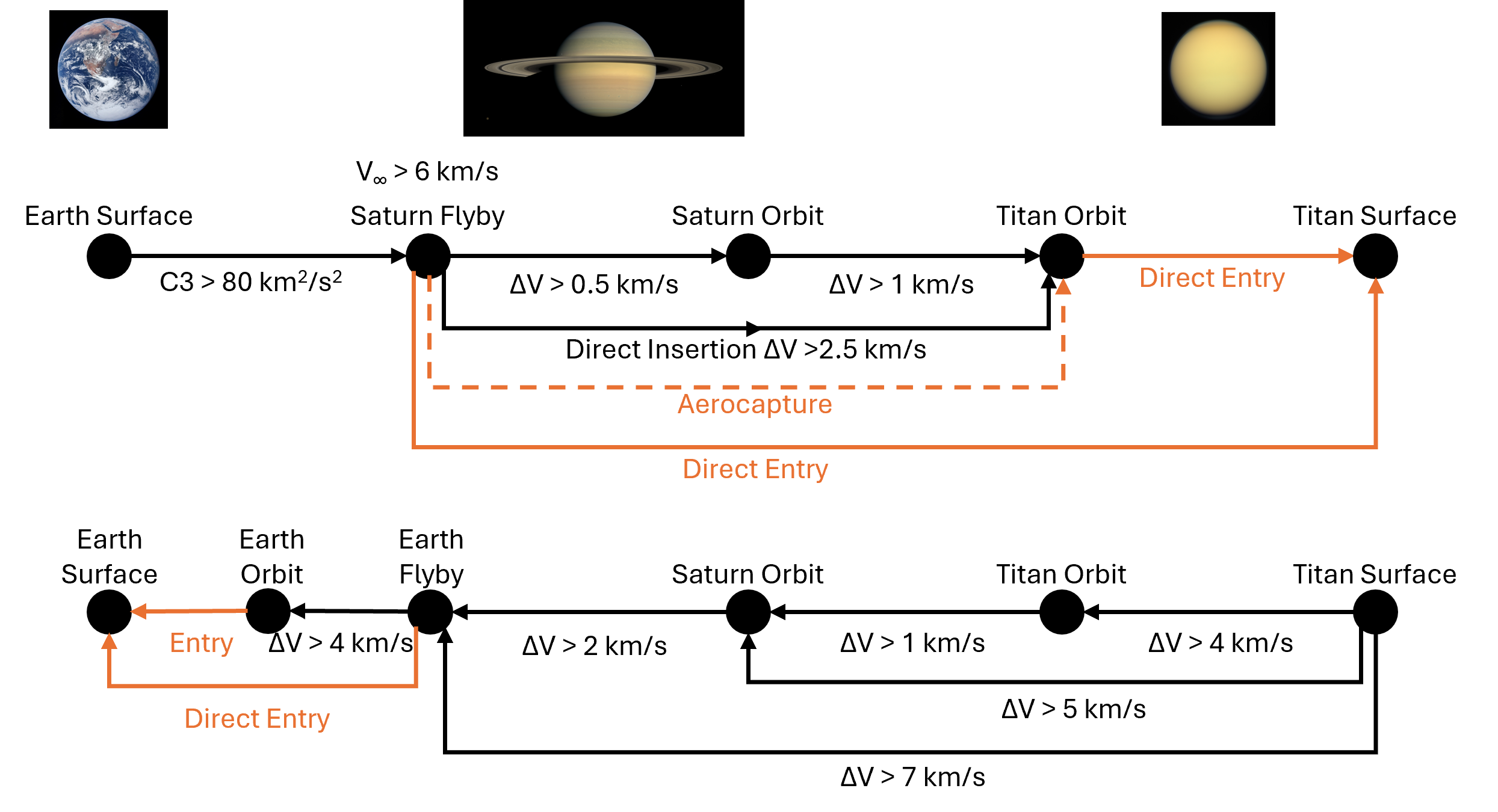}
    \caption{$\Delta V$ roadmap for transferring between Earth and Saturn and Titan. The values are representative of the typical minimums. }
    \label{fig:dvroadmap}
\end{figure}

The mass of fuel needed for a mission is tied directly to the required mission $\Delta V$ and the total mass for each mission leg. Fig.~\ref{fig:dvroadmap} summarizes the typical minimum $\Delta V$ and alternative aerocapture and entry approaches from Earth to Saturn/Titan and back. Aerocapture is a fuel-free orbit insertion maneuver and atmospheric entry allows the delivery of spacecraft to the surface without fuel-consumptions. However, both maneuvers will require significant thermal protection systems (TPS) to sustain the heating conditions during hypersonic atmospheric flight, and the addition of TPS can still save the overall mass to accommodate for payload. 




Based on the minimum $\Delta V$ requirements, assuming a modest specific impulse $I_\text{sp}$ of 370 s for Titan launch, departure, and transfers, Table~\ref{tab:dv+massfraction} shows the fuel mass requirement for selected values of $\Delta V$. 4 km/s of $\Delta V$ is representative of a Titan launch to a Titan orbit. For a representative vehicle of 100 ton dry mass (such as a StarShip), the required fuel mass will be 200 tons, bringing the total wet mass to 300 tons with a dry mass fraction of 33\%. 2 km/s of of $\Delta V$ is representative of Saturn departure maneuver to an Earth flyby.
 
\begin{table}[h!]
\centering\footnotesize
    \caption{Estimated fuel mass for the orbiting station based on the expected $\Delta V$, dry mass of 100 tons, and $I_\text{sp}$ of 370 s. All masses are in tons.}
    \begin{tabular}{lccccc}
    \hline
    $\Delta V$, km/s & Dry Mass & Fuel Mass & Total Mass & Dry Mass Fraction \\
    \hline
    1   & 100 & 31.7  & 131.7 & 75.9\% \\
    2   & 100 & 73.5   & 173.5 & 57.6\% \\
    2.5 & 100 & 99.2   & 199.2 & 50.2\% \\
    4   & 100 & 201.1  & 301.1 & 33.2\% \\
    7   & 100 & 588.3  & 688.3 & 14.5\% \\
    \hline
    \end{tabular}
    \label{tab:dv+massfraction}
\end{table}

Titan surface operation is the most viable option to produce a large quantity of fuels and parts. Table~\ref{tab:surface_fuelmass} lists the fuel mass requirement to deliver different cargo mass (fuel and parts) from Titan surface to orbit. A typical tank mass of 12\% is assumed (i.e., tank mass is 12\% of the fuel mass). Notably, fuel mass, tank mass, and total mass scales linearly with cargo mass. Meaning that the ratio among the mass values remain constant. Therefore, to deliver each ton of cargo from surface to orbit, an additional 2.7 tons of fuel has to be produced and used for Titan launch. These mass values provide a baseline fuel production requirement. 

\begin{table}[ht]
\centering\footnotesize
    \caption{Estimated fuel mass to launch cargo on surface to orbit, assuming launch $\Delta V$ of 4 km/s and $I_\text{sp}$ of 310 s. All masses are in tons}
    \begin{tabular}{cccccc}
    \hline
    Cargo Mass & Fuel Mass  & Tank Mass & Total Mass & Cargo Mass Fraction \\
    \hline
    30  & 81.8  & 9.8  & 121.6 & 24.7\% \\
    50  & 136.4 & 16.4 & 202.8 & 24.7\% \\
    100 & 272.7 & 32.7 & 405.5 & 24.7\% \\
    200 & 545.5 & 65.5 & 811.0 & 24.7\% \\
    \hline
    \end{tabular}
    \label{tab:surface_fuelmass}
\end{table}

\subsubsection{Aerocapture and Direct Atmospheric Entry}
Aerocapture maneuver and direct atmospheric entry can significantly reduce the fuel mass requirement for human-class missions. Especially, the atmospheric conditions at Titan for entry and aerocapture are very benign. Along with the nominal mid-L/D vehicle geometry, a Starship-type vehicle can be re-used multiple times for entry and aerocapture maneuvers using non-ablative thermal protection system (TPS) materials. Typical design parameters for both entry and aerocapture systems are peak g-load, peak heat rate, and total heat load. Total heat load is more meaningful when using ablative TPS material. For non-ablative TPS materials, peak g-load and peak heat rate are the two main design parameters. A general guideline for maximum g-load is 10 Earth g's for manned missions and higher for unmanned missions. Peak heat rate is limited by the capabilities of non-ablative TPS materials, for example, the space shuttle tiles are capped at 100 W/cm$^2$. 

\begin{figure}[h!]
    \centering
    \includegraphics[width=0.75\linewidth]{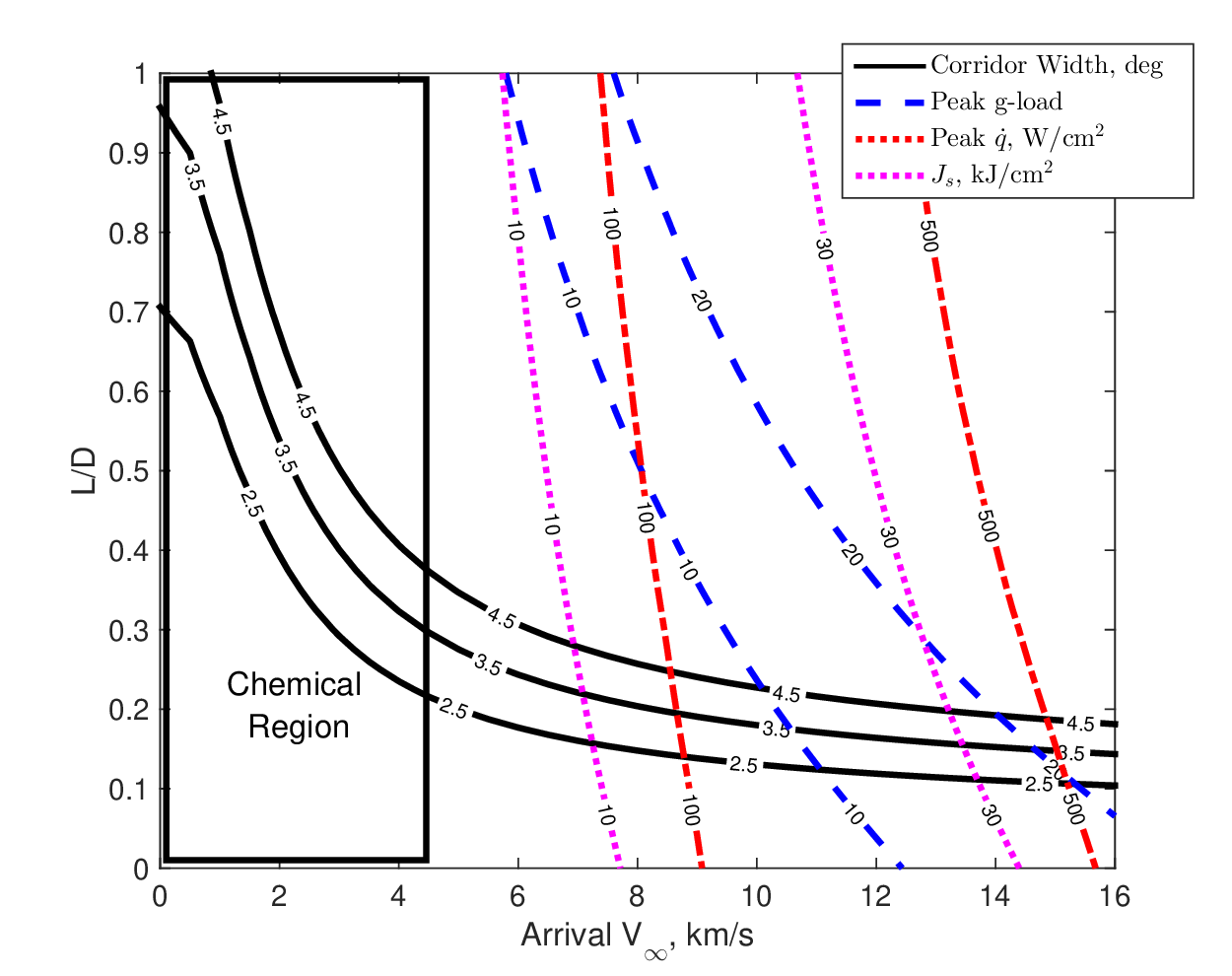}
    \caption{Aerocapture feasibility for Titan \cite{Lu2018-jc}. }
    \label{fig:aerocapture_feasibility}
\end{figure}
Prior work has established a broad feasibility of aerocapture at Titan for orbit insertion \cite{Lu2018-jc}. Fig.~\ref{fig:aerocapture_feasibility} shows the overall feasibility of Titan aerocapture, with contour lines denoting the level of g-load, peak heat rate, and total heat load.  Using a mid-L/D (i.e., L/D=0.6) vehicle geometry (i.e., SpaceX StarShip), aerocapture at Titan can maintain peak g-load to be less than 10 Earth-gs and peak heat rate less than 100 W/cm$^2$, as long as the arrival $V_\infty$ does not exceed 8 km/s, which allows for flexible mission architectures. Vehicle geometries with higher L/D will allow slightly higher arrival $V_\infty$. At lower arrival velocity, propulsive orbit insertion can be a more efficient option since no significant TPS mass is needed, the area is representatively marked as the ``chemical region''.

\begin{figure}[h!]
    \centering
    \includegraphics[width=0.75\linewidth]{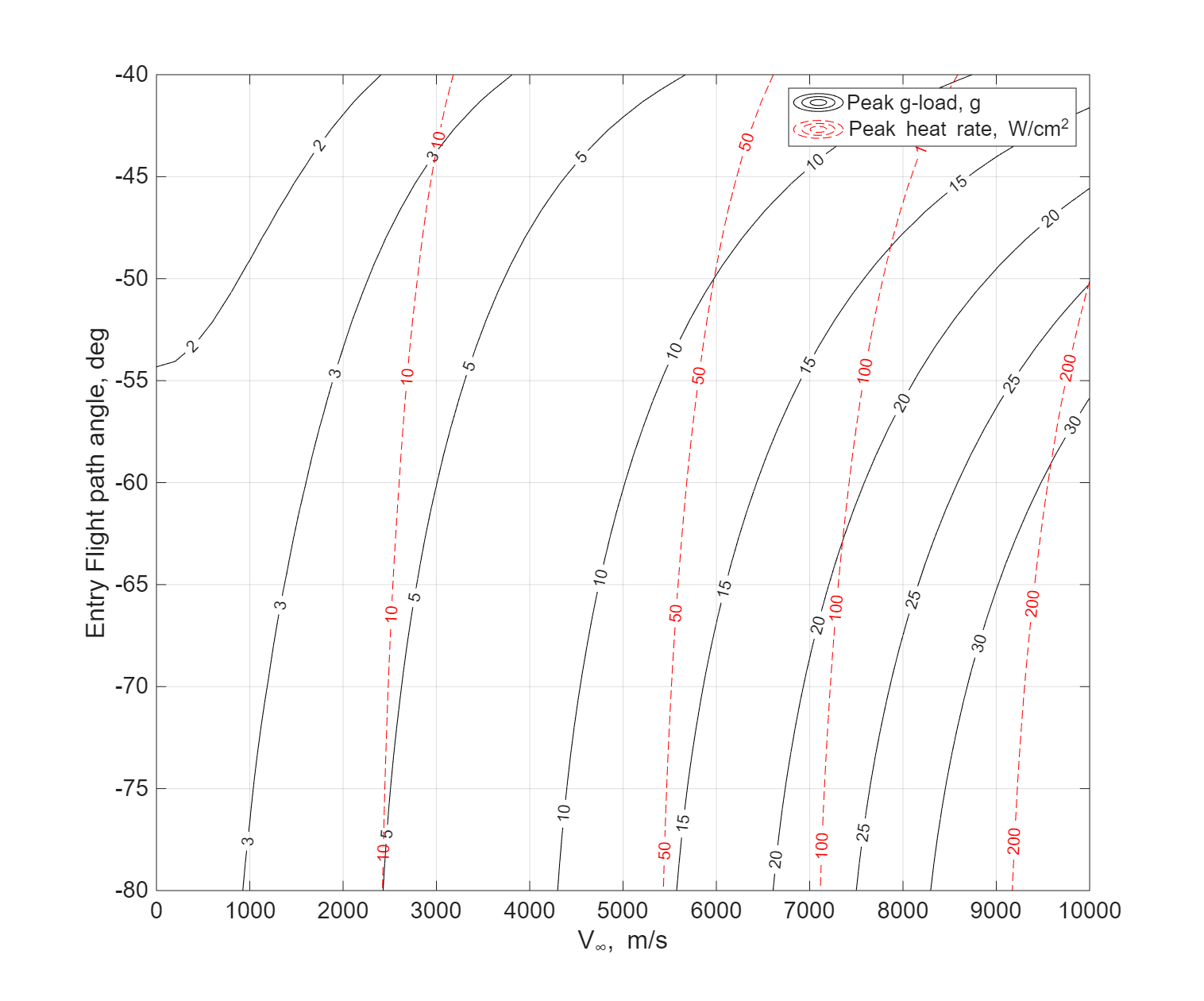}
    \caption{Titan atmospheric entry trade-off assuming ballistic entry. Black contours denote the peak g-load and red dashed contours denote peak heat rate. }
    \label{fig:entry_access}
\end{figure}

Similarly, atmospheric entry at Titan is also very benign. Fig.~\ref{fig:entry_access} shows the peak g-load and peak heat rate of Titan entry at a range of arrival $V_\infty$ and entry flight path angle. A general design guideline when using non-ablative TPS material is to use shallow entry flight path angles, which reduce both peak deceleration load and peak heat rate. As shown in Fig.~\ref{fig:entry_access}, under representative human-rating constraints (peak g-load $<10$ and peak heat rate $<100$ W/cm$^2$),  arrival $V_\infty$ > 8 km/s can be accommodated. This high arrival $V_\infty$ provides flexibility in mission architecture design. Furthermore, entry at suborbital speed ($V_\infty$<0 km/s) will also be very benign for manned missions.


\section{Resource Collection and Refining}
\label{sect:collectingrefining}


Once a mission reaches Titan, whether robotic or human, Titan's basic resources must be collected, purified, separated and processed to be ready for use as raw ingredients for the manufacture of more complex chemicals and materials. In this section we provide a high-level outline of how raw solids, liquids and gases may be processed, while acknowledging that many details remain to be filled in by future studies.


\subsection{Resource Collection}
\label{sect:collecting}

There are two principal sites of raw material collection from Titan: from flyby, and on the surface. Flyby collection includes scooping of rarefied atmosphere at high velocity, which may require multiple passes to collect sufficient raw material. In addition, gases may react during the collection process as has been documented in the Cassini mass spectrometer (`wall reactions', e.g. \cite{teolis2015revised}), making this an uncertain approach at present.

Solids, liquids and gases may also be collected from the surface. Surface locations for materials extraction are broadly divided between the organic dunes at low latitudes, and the hydrocarbon lakes and seas at higher latitudes, with many other intermediate terrains also existing \cite{lopes2016nature, lopes2020global}. The composition of the dunes is thought to be made of solid hydrocarbons overlying a bedrock of water ice \cite{bonnefoy2016compositional, brossier2018geological}. 

\begin{figure}
    \centering
    \includegraphics[width=0.95\linewidth]{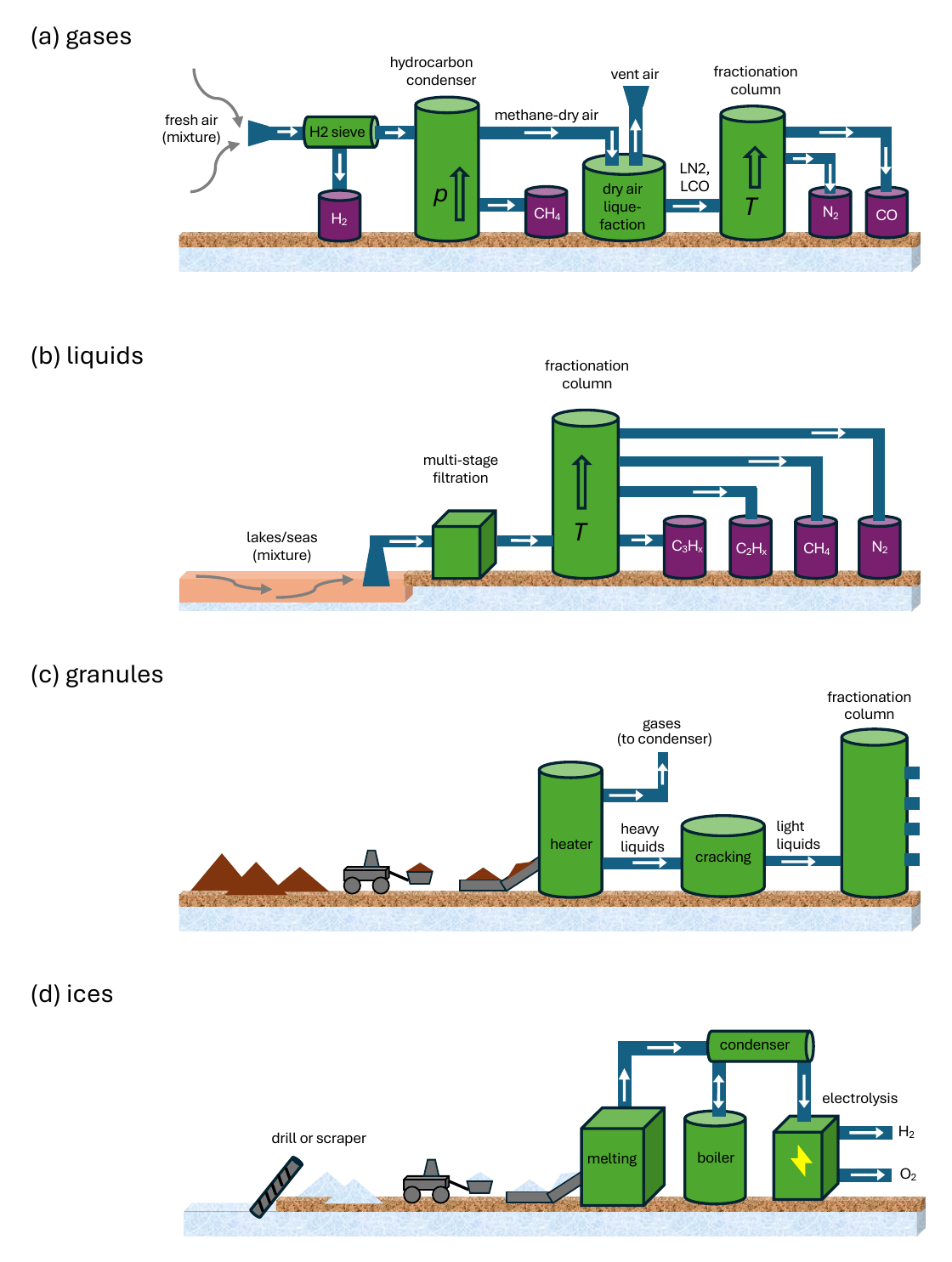}
    \caption{Mining and refining of Titan resources: (a) condensation and separation of atmospheric gases; (FIX) (b) fractionation of liquids from lakes ans seas; (c) heating and separation of solids; (d) electrolysis of ices. }
    \label{fig:mining}
\end{figure}

\subsubsection{Atmospheric gases}
\label{sect:gases}

Collection of gaseous CH$_4$ on Titan's surface was discussed in the 2022 sample return mission study \cite{oleson2022titan}. In this method, methane was condensed from Titan air, which leverages the cold ambient temperatures and pressures, where methane is already near its triple point at Titan's surface conditions. Pumps are used to continuously draw in atmosphere to keep methane saturated, and then condensation is driven by compression. Methane-dry air, composed mostly of N$_2$ which does not condense in this scenario, is continuously vented as replacement methane-rich air is drawn in.

While this method is sufficient for production of methane for fuel, a more permanent settlement will also require pure nitrogen for chemical processes, including production of fertilizer (Section 6). The lesser atmospheric components \ce{H2} (1000 ppm) and CO (50 ppm) are also desirable for chemical production, although they may be obtained by other means including from water (Eq.~\ref{eq:electrolysis}) and by partial oxidation of methane.

A potential four-gas plant could operate as follows (Fig.~\ref{fig:mining}(a)). As a first stage, hydrogen is sieved from the initial gas mixture to prevent hydrogen embrittlement \cite{cotterill1961hydrogen}, which can cause failure of later pumping stages - a well known problem with cryogenic engines \cite{dhital2014review}. Hydrogen molecular sieves come in multiple types, requiring an optimization study. Variants include zeolite-based \cite{li2010zeolitic}, carbon tubules \cite{lei2021carbon, hou2023high, sazali2020comprehensive}, metalic \cite{li2010molecular} and others.

Second, methane is removed as envisaged in the Oleson et al. study - this is equivalent to drying air on Earth prior to liquefaction. This stage will also remove other minor hydrocarbons such as ethane and acetylene, already with very small partial pressures at 93~K.

The final stages proceed similarly to liquefaction of air on Earth \cite{castle2002air}, which is routinely used to produce liquid nitrogen (LN2) and liquid oxygen (LOX). The now-dry air is again compressed (causing heating), cooled, and then expanded, exploiting the Joule-Thomson effect to produce a liquid which is mostly LN2 mixed with some liquid CO (LCO). The liquid LN2/LCO mixture can be further purified by fractional distillation.

\subsubsection{Surface liquids}
\label{sect:liquids}

Liquids may also be collected from Titan's surface, from the large lakes and seas that are found predominantly in polar regions \cite{stofan2007lakes}. While their composition is not known precisely, Cassini measurements and modeling indicate that they are mostly methane, with significant dissolved amounts of ethane and nitrogen (\cite{mastrogiuseppe2014bathymetry, cordier2009estimate}, Table~\ref{tab:resources}), and heavier organics that may be in the solid state on the lake bed. Fig.~\ref{fig:mining}(b) shows a concept where lake liquids are pumped and then separated by fractionation, where the liquid mixture is heated and different fractions are collected at different temperatures (discussed further in Section \ref{sect:refining}).

When collecting liquids from Titan's lakes, multi-stage filtration is required. Solid particulates can cause issues with mechanisms, particularly in fluid lines. Filtration of solid matter from oil (a process known as `desolidification') is already an established technique for terrestrial oil \cite{gong2022analysis}. Since filtration often relies on consumable filters, and replacement filters would be difficult to supply regularly to the outer solar system, strategies for filters that are reusable, long-lifetime, or 3D printed are essential. Reusable filters could be removed, cleaned with solvent, and reinstalled. 


\subsubsection{Surface solids}
\label{sect:solids}

Solids may be collected from the organic dunes (Fig.~\ref{fig:mining}(c)) and by `mining' crustal water ice (Fig.~\ref{fig:mining}(d)). 

\begin{figure}
    \centering
    \includegraphics[width=0.95\linewidth]{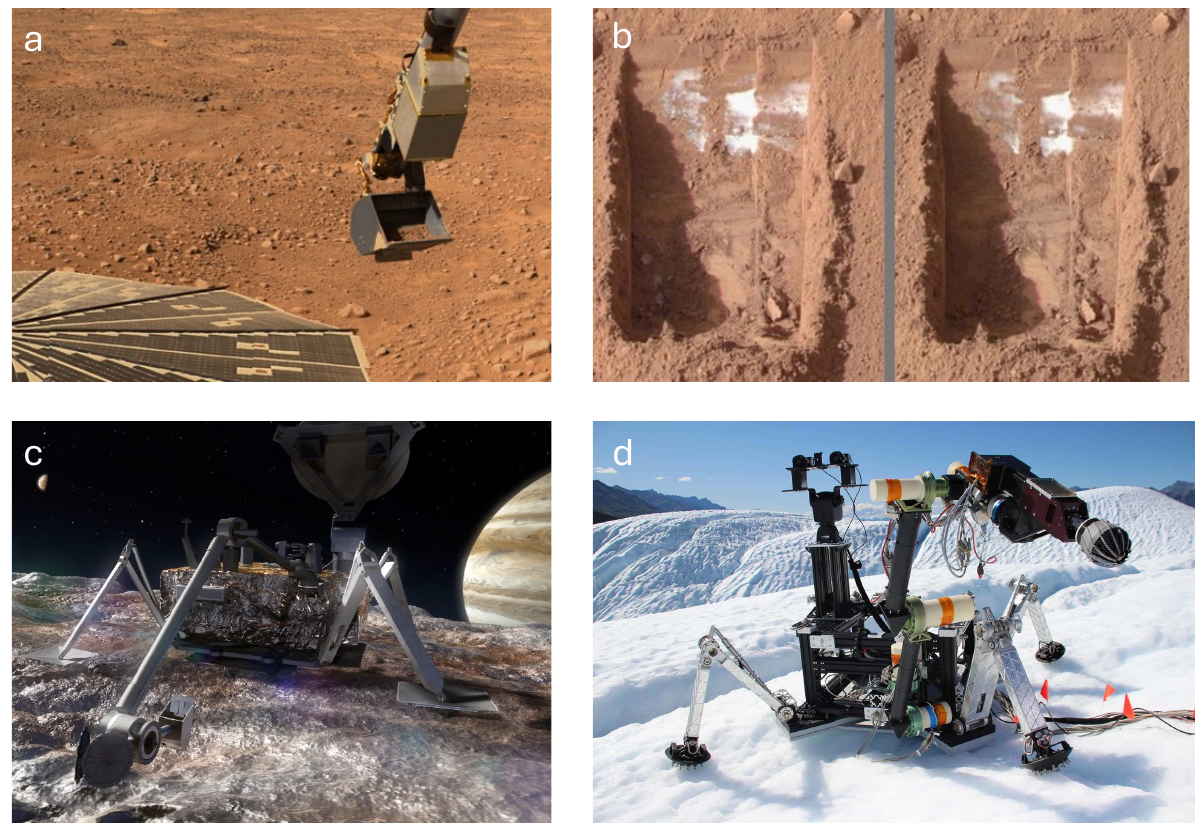}
    \caption{Recent planetary surface sampling technologies: (a) the Mars Phoenix scoop; (b) Phoenix scoop marks on Mars, showing sub-surface ice; (c) concept of the Europa lander showing instrumented arm; (d) field test of Europa Lander prototype in Alaska. Images: NASA/JPL. }
    \label{fig:sampling}
\end{figure}

For crustal water ice, sample acquisition techniques could include varieties of drilling, grinding and rasping, as already used by the Mars Phoenix Lander \cite{arvidson2009results}, and planned for the Europa Lander \cite{badescu2019sampling, badescu2021dual} (Fig.~\ref{fig:sampling}). 

Careful consideration of cohesive and sticking properties and experimentation on simulated samples at appropriate temperatures will be necessary to ensure smooth operation. Initial processing may include comminution of large pieces (grinding, milling, etc into smaller sizes). This will be followed by size separation, perhaps `washing' in liquid methane to remove organics, and initial storage of material ready for later electrolysis (see Section \ref{sect:oxygen}), after which the gases \ce{H2} and \ce{O2} can be stored. 

The nature of Titan's organic regolith remains largely unknown. Therefore, while significant progress has been made on utilization of lunar regolith for construction \cite{azami_comprehensive_2024}, much of this work will not apply to Titan's organic solids. 
While the detailed composition is not known at present, the DrAMS instrument on the Dragonfly mission should help to elucidate the chemical makeup \cite{trainer2021development, stern2023development}. 

Therefore, further discussion of refining and utilization will focus on material from the atmosphere and methane lakes where the composition is better known at present \cite{niemann2010composition, cordier2009estimate}. 

\subsubsection{Storage}
\label{sect:gasstorage}

Gases are most easily stored in pressurized gas cylinders. These should be standardized to facilitate ease of robotic handling, interfacing, and storage location flexibility. On Earth, over-pressurization is typically accomplished by relief valves, which could also be used on Titan's surface, but on-orbit, these could contribute to unwanted changes in velocity. 

\subsection{Refining}
\label{sect:refining}

\begin{figure}
    \centering
    \includegraphics[width=0.5\linewidth]{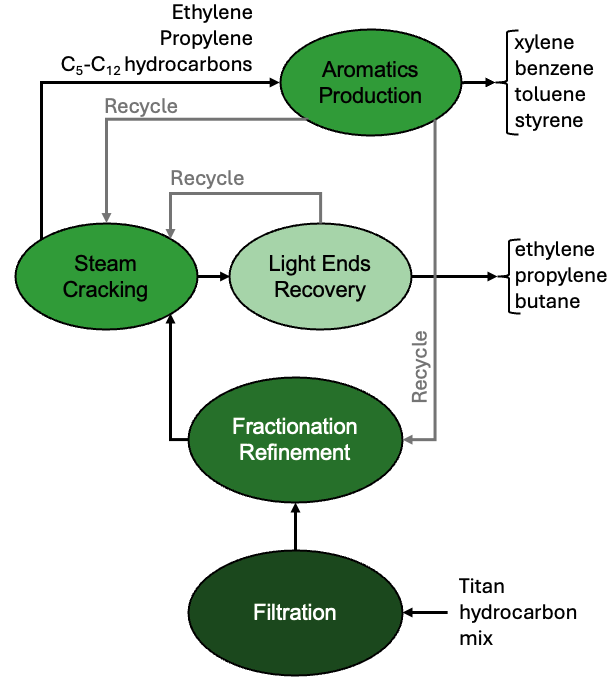}
    \caption{Concept diagram of hydrocarbon refining.}
    \label{fig:refining}
\end{figure}

After initial collection and storage, liquids and gases can be further refined, following a series of steps well-established from terrestrial refining (see Fig.~\ref{fig:refining}): 

\begin{enumerate}
    \item  {\em Separation} (distillation)
    \item  {\em Conversion} (cracking and reforming)
    \item  {\em Treatment} and purification 
\end{enumerate}

Heating and cooling are important components of refining steps. Note that on-orbit cooling of collected gases will be difficult, because radiative heat transport is the only method due to the low pressures. 

\subsubsection{Separation by fractionation}
\label{sect:fractionation}

The first step after collection is initial separation of fractions, which is typically achieved through atmospheric fractionation for lighter (more volatile) fractions. 

\begin{figure}
    \centering
    \includegraphics[width=1.0\linewidth]{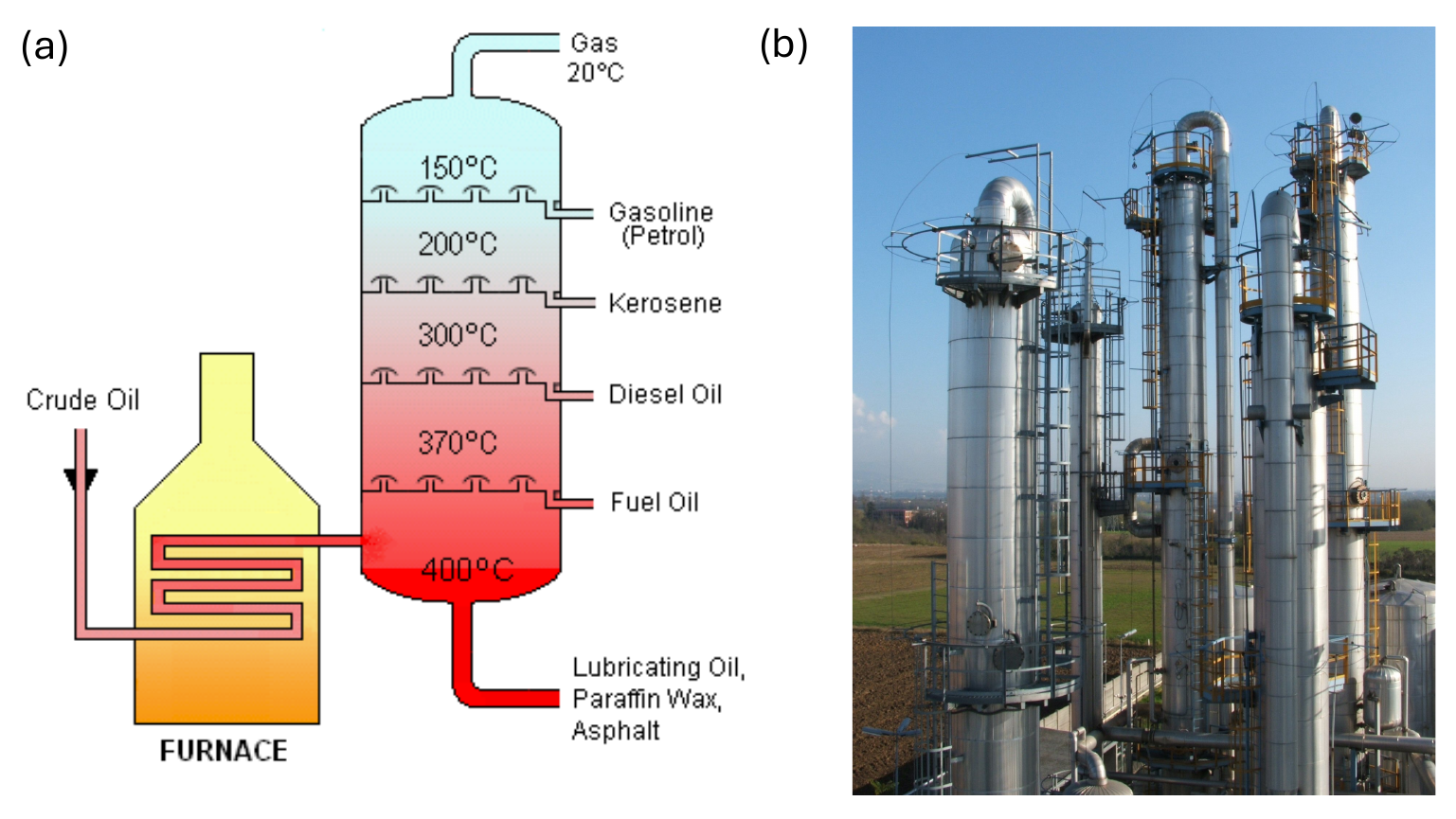}
    \caption{(a) Fractionation schematic. (b) Industrial fractionation column. (wikimedia)}
    \label{fig:fractionation}
\end{figure}

In atmospheric fractionation, a mixed `crude' is heated to a vapor, and enters a column through the bottom. As vapors rise, they cool, and different molecules condense successively from the heaviest fractions at the bottom, through to the lightest fractions at the top \cite{pripakhaylo2025separation}. See Fig.~\ref{fig:fractionation}.

Fractionation columns require careful design depending on the input mixture.
In terrestrial designs, a returned portion of condensed product at the top of the column (known as `reflux') provides cooling. The optimal number of stages (`trays') where the uprising vapors encounter the downward reflux must be carefully designed. Columns to separate simple, two-component mixtures may be designed exactly through a McCabe-Thiele diagram \cite{mccabe1925graphical}, while more complex mixtures will require modeling \cite{ibrahim2014design}.

Any unseparated, heavier fractions emerging from the bottom of the column can be separated in vacuum, where they can boil at lower temperatures \cite{zhou2010recovery}.

\subsubsection{Conversion}
\label{sect:conversion}

\begin{figure}
    \centering
    \includegraphics[width=0.95\linewidth]{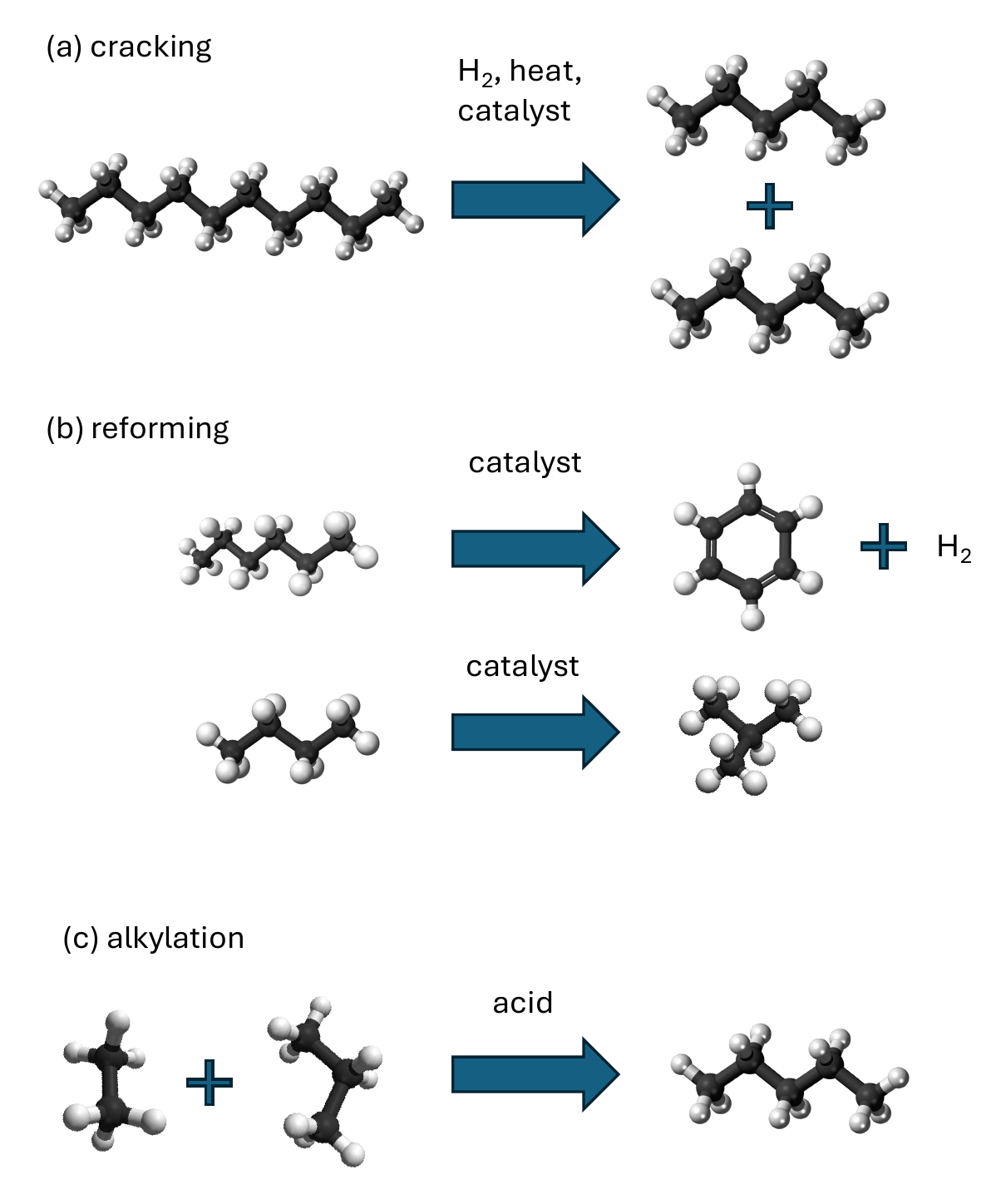}
    \caption{Types of hydrocarbon conversion: (a) cracking; (b) reforming; (c) alkylation.}
    \label{fig:conversion}
\end{figure}

Some fractions of distillate will typically be more valuable for later use than others. Therefore conversion is an important process, which allows for less useful fractions to be converted to more useful ones. There are three main types of conversion (see Fig.~\ref{fig:conversion}):

\begin{itemize}
    \item {\em Cracking}: uses heat, pressure and catalysts to break up large molecules into smaller ones \cite{greensfelder1949catalytic}.
    \item {\em Conversion or reforming}: re-arranges the chemical structure of a molecule into a more useful type \cite{pines2012chemistry, zaera2009regio}.
    \item {\em Alkylation}: the opposite of cracking, alkylation converts lighter substances to heavier ones \cite{corma1993chemistry}.
\end{itemize}

Many different types of cracking are used on Earth. Direct thermal cracking (application of high temperatures) is relatively uncontrolled, and may produce unwanted products \cite{sadrameli2015thermal, wang2023comprehensive}. It may also be less feasible on Titan where energy is scarce. Steam cracking is another high-temperature method \cite{gholami2021review}, where hydrogen is added from water steam at temperatures around 850\degC to produce smaller hydrocarbons. Fluid catalytic cracking uses an added powdered catalyst (e.g. silica-alumina) in a heated oil mixture to reduce temperatures needed to break apart heavy fractions \cite{vogt2015fluid}. Finally hydrogen cracking (`hydrocracking') adds hydrogen to produce lighter fractions from heavier ones \cite{weitkamp2012catalytic} (Fig.~\ref{fig:conversion}(a)).

Catalytic reforming (Fig.~\ref{fig:conversion}(b)) can also change the structure of hydrocarbons without changing their size \cite{sinfelt2022catalytic}. In particular, useful techniques are isomerization \cite{akhmedov2007recent}, typically converting straight-chain parafins (alkanes) into branched chains, and dehydrogenation-aromatization \cite{guisnet1992aromatization}, where linear or branched molecules are converted to cyclic molecules, losing hydrogen in the process. This produces very valuable chemical feedstocks such as benzene, xylene, tolene and styrene. Most reforming processes use platinum-based catalysts, which would be a rare and valuable material on Titan. Final products can be separated by fractionation.

Alkyation (Fig.~\ref{fig:conversion}(c)) is also an important process, used on Earth to join together several very light fractions (e.g. propene and butene) that would otherwise be gaseous at STP, into more useful heavier liquids \cite{albright2009present}. The process generally makes use of an acid such as sulfuric acid (\ce{H2SO4}) \cite{sun2013alkylation} or hydrofluoric acid (HF) \cite{simpson2007hydrofluoric} which remove hydrogen, but are themselves consumed in the process. Availability of elements S and F on Titan is hence of critical importance to the feasibility of this process in its current form.

\subsubsection{Purification}
\label{sect:purification}

The final step of refining is purification, which removes impurities such oxygen, nitrogen and sulfur from hydrocarbons \cite{ancheyta2016hydrotreating, robinson2006hydrotreating, qin2025molecular}. While fuels and fluid for cooling loops can tolerate >60 percent impurities, polymers require over 99 percent purity to hit reasonable yield and nominal materials properties. Mixtures with less purity during polymerization result in early-termination of polymerization and impurity inclusions can reduce overall polymer strength and can introduce brittle behavior into filament and final parts \cite{alo20223d}. Evaluation of purity tolerances to maintain minimum materials properties are needed. The ability to use `dirty' polymers minimizes the need for extensive purification setups. 

Hydrotreating removes impurities such as sulfurs to protect down-stream catalysts \cite{ancheyta2016hydrotreating, robinson2006hydrotreating, qin2025molecular}. This process also selectively saturates hydrocarbons, removing alkenes to improve separations \cite{bowen1950removal}. This is particularly important for acetylene and other `problematic' alkenes which contribute to separations and hardware failure if not saturated early on. The process leaves desired alkenes such as ethylene and propylene to be used for polymerization, and it requires hydrogen gas, a metal catalyst, and temperatures around 300-400\degC. While hydrogen is available on Titan, the catalyst must be brought from Earth or refined in space, and replacements made available as catalytic properties decline over use. This also implies that ease of robotic replacement of these kinds of consumables must be part of hardware design requirements. 

A possible \textit{in situ} catalyst and molecular sieve for purification are zeolites, which have been found in many natural environments including possibly on Mars \cite{ruff2004spectral, mousis2016martian}.
On Earth, these structures hydrate readily and so on Titan they would need to be heated (<500\degC) to remove water or organic solvents \cite{Coombs_1998}. They are notoriously difficult to distinguish between each other with FTIR/Raman spectroscopy \cite{sadrara2021rapid}, and may require onsite XRD for analysis \cite{norby2006situ} or on-board chemical analysis, such as Sample Analysis on Mars (SAM) \cite{mahaffy2012sample}.

By-products of any purification process at Titan should be considered for collection and reuse. The by-products largely depend on the initial composition as well as choice of purification technique, which should be assessed during feasibility and project scoping.

\subsection{Materials Characterization}
\label{sect:characterization}

Analysis of intermediate and final products during refining is a requirement for ensuring that that processes are successful and final products are usable \cite{riazi2005characterization}. Any mission intending to perform ISRU must include appropriate characterization instrumentation in addition to mining, refining and manufacturing components. 


Non-destructive techniques are preferable, such as FTIR \cite{rakhmatullin2018application, dutta2013molecular}, Raman spectroscopy \cite{kostina2023use, sharma2005portable}, UV/Vis spectroscopy \cite{banda2016crude, evdokimov2007potential}, and visual inspection. For sealed chambers, an optically transparent window is needed, which can complicate designs. Sensors inside the chamber with sealed electrical pass-throughs may be easier for the design, but severely limit characterization approaches.

Analysis techniques appropriate for volatile liquids and gases used in industry are gas chromatography with mass spectrometry (GC-MS) \cite{herod2007characterization}. For liquid polymers, high-performance liquid chromatography (HPLC) \cite{kim2015combination}, and other chromatography techniques can be used. These are used in later stages of processing for high-purity assessment. 

On Earth, these instruments require significant periodic maintenance and have several consumable parts. For example, a column in a GC-MS apparatus facilitates species separation prior to being identified by the mass spectrometer. This specialized and engineered part degrades over time and requires regular replacement, which is problematic for outer solar system use.
In addition to traditional launch and in-space operations considerations (power, volume, mass), other considerations include consumables used and required post-launch calibration or ability to do remote method development prior to materials characterization. 


\subsection{Final Storage}
\label{sect:refinedstorage}

Once desired alkenes are isolated, they are further separated based on the desired final polymers. The monomers can be stored separately, or polymerization can be initiated for any thermoplastic to be stored as discussed below. Components for thermosets, such as two-part epoxy or rubbers, must be stored separately until ready to be used.

\section{Chemical Production}
\label{sect:chemicals}

Various low melting point materials: plastics, rubbers and similar synthetic materials can be made from hydrocarbons and nitriles. In addition, a versatile range of useful materials with higher melting points can be made from carbon, especially: diamond, graphite, graphene (carbon nano-tubes), fullerenes (buckyballs). If silicon can be found on Titan or collected nearby, that expands the materials roster to include silicon carbide, as well as pure silicon for circuity. 

Solvents and other useful chemicals for paints, glues, cleaning products etc can also be produced from basic CHON elements such as benzene, cyclohexane, formaldehyde, acetone, acetic acid, ammonium hydroxide. More complex materials such as proteins, amino acids, sugars (ingredients for foods), food colorings and preservatives, and medicines can be produced solely or largely from the same four basic elements. 

Note that, for many of the processes described below, diverse catalysts are needed. Therefore wherever possible finding creative ways to distill as many needed chemicals as possible from Titan's liquids without further processing will be important.

\subsection{Fuels}
\label{sect:fuelproduction}

Fuels are produced for two purposes: for rocket propellant, and for electricity generation. There is some overlap in fuels used between these two classes, e.g. \ce{O2} as an oxidizer. 

Rocket propellants are often classified as {\em bi-propellants} (or {\em `biprops'}) - fuels such as \ce{H2} and \ce{CH4} that require an oxidizer; vs {\em mono-propellants} (or {\em `monoprops'}) - fuels that can be exothermically dissociated by passing over a catalyst to release energetic exhaust gases, hydrazine (\ce{N2H4}) being the most common, but others include \ce{H2O2} and \ce{C2H4O}. 

Other common fuel distinctions include between {\em cryogens} vs {\em non-cryogens} - meaning fuels that require liquefaction from a gaseous phase to a liquid phase for storage and use. Finally {\em hypergolics} are fuels that spontaneously ignite in the presence of an oxidizer, the most common being monomethylhydrazine (MMH, formula \ce{CH3NHNH2}) and unsymmetrical dimethylhydrazine (UDMH, formula \ce{(CH3)2NNH2}).

\subsubsection{Hydrogen peroxide}

Hydrogen peroxide (\ce{H2O2}) is a valuable and useful chemical, with applications from propulsion (as either a monopropellant or an oxidizer for biprop) to antiseptic/disinfectant. 

On Earth, it is produced almost exclusively by the anthraquinone process (Riedl–Pfleiderer process \cite{ingle2022progress}), which makes use of an anthraquinone intermediary in the presence of a palladium catalyst. The two step reaction first reduces anthraquinone to make anthrahydroquinone:


\begin{equation}
    \ce{C14H8O2} + \ce{H2} \rightarrow \ce{C14H10O2}
\end{equation}

\noindent
followed by auto-oxidization in the presence of \ce{O2}:

\begin{equation}
    \ce{C14H10O2} + \ce{O2} \rightarrow \ce{C14H8O2} + \ce{H2O2}
\end{equation}

\noindent
Alternative methods of production from \ce{H2} and \ce{O2} directly have been researched for possible implementation on the space station \cite{vijapur2019situ}, however on Titan with plentiful organic carbon the traditional process may be the most viable if sufficent palladium is available.

\subsubsection{Hydrazine}

On Earth, hydrazine is typically produced by one of two processes: (1) the peroxide process, involving oxidation of ammonia via intermediates; (2) the Olin Rashig process, which uses sodium hypochlorite (NaOCl). Given that both sodium and chlorine are of unknown abundances on Titan, the peroxide process (also known as the ketazine process, and by other names) appears to be more practical for Titan. This proceeds through multiple stages, having a net reaction of:

\begin{equation}
    2 \ce{NH3} + \ce{H2O2} \rightarrow \ce{N2H4} + 2 \ce{H2O}
\end{equation}

\noindent
Reagents required are: \ce{NH3}, \ce{H2O2}, as well as butanone (aka methyl ethyl ketone, \ce{CH3C(O)CH2CH3}). 

\subsubsection{Mono and dimethyl hydrazine}

Monomethyl hydrazine (MMH) can be produced by reaction of ammonia with sodium hypochlorite to form chloramine:

\begin{equation}
\ce{NH3} + \ce{NaOCl} \rightarrow \ce{NH2Cl} + \ce{NaOH}
\end{equation}

\noindent
followed by reaction with methylamine in a modification of the Olin Raschig reaction to form MMH:

\begin{equation}
\ce{NH2Cl} + \ce{CH3NH2} \rightarrow \ce{CH3N2H3} + \ce{HCl}
\end{equation}

Unsymmetrical dimethyl hydrazine (UDMH) can similarly be produced from dimethylamine and monochloramine:

\begin{equation}
\ce{(CH3)2NH} + \ce{NH2Cl} \rightarrow \ce{(CH3)2NNH2 . HCl}
\end{equation}

\noindent
resulting in 1,1-dimethylhydrazinium chloride. Other than chlorine, the other needed elements for this synthesis are widely available on Titan.

\subsection{Plastics and rubbers} 
\label{sect:plasticsrubbers}



Plastics and rubbers remain amongst the most straightforward applications of native Titan ingredients, being formed from polymers of simple hydrocarbons already widely found on Titan 
(see Table~\ref{tab:synthetics})
The monomers required to make plastics and rubbers may be refined from liquids in Titan lakes/seas, or perhaps created from methane.

\begin{table}[]
    \centering
    \begin{tabular}{lll}
    \hline
         polymer & monomer name & monomer formula \\
         \hline
        polyethylene & ethylene & \ce{C2H4} \\
        polypropylene & propylene & \ce{C3H6} \\
        nitrile rubber & acrylonitrile \& & \ce{C2H3CN} \\
             & butadiene & \ce{C4H6} \\
        butyl rubber & isobutylene \& & \ce{C4H8} \\
             & isoprene & \ce{C5H8} \\
    \end{tabular}
    \caption{Simple polymers and monomer ingredients.}
    \label{tab:synthetics}
\end{table}

Many different mechanisms of polymerization reactions are possible, broadly divided into (i) chain-growth polymerization and (ii) step-growth polymerization, which is further divided into polyaddition and polycondensation. Additional reaction-enabling ingredients needed for different types of polymerization can include acids (including Lewis acids such as \ce{SnCl4}, \ce{AlCl3}, \ce{BF3} and TiCl4), alcohols and esters. 

On Titan therefore, the limiting factors for such reactions are likely to be the reaction enabling ingredients rather than the raw monomer units. Before proceeding to develop chemical reactors, careful research is required into availability of these substances (see Section \ref{sect:furtherwork}).


\subsection{Glues and solvents}
\label{sect:solventsglues}

Solvents are needed for a wide variety of purposes, including cleaning, dilution of paints, detergents, and as needed materials for many types of chemical synthesis.
Solvents, which are typically liquids, may be either polar (having an electric dipole) or non-polar. Polar solvents maybe further categorized as protic (having a labile proton, H$^+$, often bonded to a nitrogen in NH2 or an oxygen in OH), or aprotic.

Examples include:

\begin{itemize}
    \item Non-polar solvents: benzene (\ce{C6H6}), toluene (\ce{C7H8}), hezane (\ce{C6H12})
    \item Polar, protic: ammonia (\ce{NH3}), water (\ce{H2O}), acetic acid (\ce{CH3COOH}), methanol (\ce{CH3OH})
    \item Polar, aprotic: acetonitrile (\ce{CH3CN}), acetone (\ce{CH3COCH3}), dichloromethane (\ce{CH2Cl2})
\end{itemize}

Glues or adhesives are a wide range of compounds that stick to two surfaces, binding them together. All glues must have three properties: wetting -  being able to maintain contact with a surface; thixotrophy - strengthening after application; transmission of load or force. The action of adhesives comes from several different mechanisms, one or more of which may cause the action of any particular glue:

\begin{itemize}
    \item Chemical bonding
    \item Van der Waals forces
    \item Electrostatic forces
    \item Diffusion into the substrate
\end{itemize}


While the chemical production of the wide variety of glues and solvents is beyond  the scope of this article, the chemistry of some basic precursors to glues and solvents is now briefly discussed.

\subsubsection{Aldehydes and ketones}

Acetaldehyde (\ce{CH3CHO}) is manufactured on Earth via the Wacker (or Hoechst-Wacker) process, which is oxidation of ethylene in the presence of a catalyst:

\begin{equation}
    2\ce{C2H4} + \ce{O2} \rightarrow \ce{C2H4}
\end{equation}

\noindent
A similar process can be used to generate acetone \ce{(CH3)2CO} from propene:

\begin{equation}
    2\ce{C3H6} + \ce{O2} \rightarrow \ce{C3H6O}
\end{equation}

\noindent
However, the catalysts required (aqueous palladium(II) chloride and copper(II) chloride) may be difficult to come by on Titan, requiring them to be either brought from Earth or mined from a more rocky body. 

Another process more commonly used to produce acetone is the cumene process, whereby propene is first added to benzene, forming cumene: 

\begin{equation}
    \ce{C3H6} + \ce{C6H6} \rightarrow \ce{C6H5CH(CH3)2}
\end{equation}

\noindent
in the presence of a catalyst (\ce{H3PO4}/\ce{SiO2}), and then oxidized using a further catalyst (e.g. \ce{H2SO4}) to form acetone and phenol:

\begin{equation}
     \ce{C6H5CH(CH3)2} + \ce{O2} \rightarrow \ce{(CH3)2CO} + \ce{C6H5OH}
\end{equation}

Acetone may then be thermally decomposed to form ketene (ethanone) by heating to 600-700\degC in the Schmidlin ketene synthesis process:

\begin{flalign}
    && 
    \ce{CH3COCH3} \rightarrow \ce{CH2=C=O + CH4}
    && \text{(Thermolysis)}
\end{flalign}

\subsubsection{Alcohols and Ethers}

A necessary ingredient to produce these species is CO, which may either be distilled from Titan air (Section \ref{sect:collecting} or by steam reforming of methane:

\begin{flalign}
    && \ce{CH4 + H2O} \rightarrow \ce{CO + 3 H2 } && \text{(Steam reforming)}
\end{flalign}

\noindent
requiring 206 kJ/mol (endothermic). 
These ingredients can then be combined to create methanol via water and \ce{CO2} intermediates:

\begin{align}
\ce{CO2 + 3 H2} &\ce{-> CH3OH + H2O} && \text{(Hydrogenation)} \label{eq:hydrogenation} \\
\ce{CO + H2O} &\ce{-> CO2 + H2} && \text{(Water gas shift)} \label{eq:wgshift} \\
\cline{1-2}
\ce{CO + 2 H2} &\ce{-> CH3OH} && \text{(Net)} \label{eq:net} 
\end{align}

\noindent
Typical reaction conditions are: 5–10 MPa (50–100 atm) and 250\degC\ (482\degF).

Dimethyl ether (DME) can then be produced by dehydration of methanol in the presence of sulfuric acid:

\begin{flalign}
    && \ce{2 CH3OH} \rightarrow \ce{(CH3)2O + H2O} && \text{(Dehydration)}
\end{flalign}

An alternative production mechanism for DME is direct synthesis from syngas (CO + \ce{H2}), as for methanol, using a two-stage catalytic process: (1) creation of methanol (\ce{CuO}-ZnO-\ce{Al2O3} catalyst) and (2) dehydration to DME over $\gamma$-\ce{Al2O3}.

On Titan, careful consideration of available catalysts and reaction intermediaries would be required to optimize such processes.


\subsection{Fertilizers and agrichemicals}
\label{sect:agrichemicals}


For food production on long-duration missions to the outer solar system, including colonies on Titan, fertilizer production is sure to be important. 
Fertilizer contains three major macronutrients: nitrogen (N), phosphorus (P) and potassium (K), which are all essential for plant growth. Other elements needed in smaller quantities are termed micronutrients, especially copper (Cu), iron (Fe), manganese (Mn), molybdenum (Mo), zinc (Zn), and boron (B).

For years, animal waste was the major source of fertilizer for farming. During the 19th century, the discovery of major deposits of nitrogen and phosphorus-rich guano (bird excrement) on islands around the world spurred a major industry to extract and import this, leading to a boom in intensive agriculture. However it was clear that this resource was in a fixed and dwindling supply, leading to intensive efforts to find chemical means to produce fertilizer. This finally came with the 1910 breakthrough of Fritz Haber and Carl Bosch in synthetic production of ammonia from nitrogen in air - mimicking the process of nitrogen fixation from air in the root nodules of plants.

\subsubsection{Ammonia}

It may be that in future ammonia will be found to exist in usable quantities on Titan, however, photochecmical models predict very low levels in the atmosphere, and hence on the surface.

The Haber-Bosch process may be used to convert Titan's nitrogen and hydrogen to \ce{NH3}:

\begin{equation}
{\displaystyle {\mathrm {N} {\vphantom {A}}_{\smash[{t}]{2}}{}+{}3\,\mathrm {H} {\vphantom {A}}_{\smash[{t}]{2}}{}\mathrel {\rightleftharpoons } {}2\,\mathrm {NH} {\vphantom {A}}_{\smash[{t}]{3}}}\qquad {\Delta H_{\mathrm {298~K} }^{\circ }=-92.28~{\text{kJ per mole of }}{\mathrm {N} {\vphantom {A}}_{\smash[{t}]{2}}}}}
\end{equation}

\noindent
The reaction is exothermic, but uncatalyzed it requires high temperatures and pressures to overcome the strong nitrogen triple bond. For that reason, catalysts including ferrite iron are important to reduce the reaction temperature. Typical catalyzed reactions occur at 250 to 350 bar and 450-550\degC.

\ce{N2} is readily accessible from Titan's air, however \ce{H2} is less readily available, as either 0.1\% of the atmosphere, or from the afore-mentioned electrolysis of water - an energy intensive process but one that will doubtless be needed to produce breathable oxygen. 

In a modern Haber-Bosch reactor, hydrogen is generated from methane using water:

\begin{equation}
    {\displaystyle {{\mathrm {CH_{4(g)}~} } + \mathrm {H} 
    {\vphantom {A}}_{\smash[{t}] {2}}\mathrm {O} 
    {\vphantom {A}}_{\smash[{t}] {{\mskip 1mu}(\mathrm {g} )}} 
    \mathrel {\longrightarrow } 
    {\mathrm {CO_{(g)}~} }{}+{}3\,\mathrm {H} 
    {\vphantom {A}}_{\smash[{t}]{2}} 
    {\vphantom {A}}_{\smash[{t}]{{\mskip 1mu}(\mathrm {g} )}}}\qquad 
    {\Delta H^{\circ }=+206\ {\mathrm {kJ} /\mathrm {mol} }}}
\end{equation}

and subsequently also oxygen:

\begin{equation}
    {\displaystyle {{\mathrm {2 {\mskip 2mu} CH_{4(g)}~} }{}+{}\mathrm {O} 
    {\vphantom {A}}_{\smash[{t}] {2}}
    {\vphantom {A}}_{\smash[{t}] {{\mskip 1mu} (\mathrm {g} )}}
    \mathrel {\longrightarrow } 
    {\mathrm {2CO_{(g)}~} } + 4\,\mathrm {H} {\vphantom {A}}_{\smash[{t}]{2}}{\vphantom {A}}_{\smash[{t}]{{\mskip 1mu}(\mathrm {g} )}}}\qquad 
    {\Delta H^{\circ }=-71\ {\mathrm {kJ} /\mathrm {mol} }}}
\end{equation}

\begin{equation}
    {\displaystyle {{\mathrm {CO_{(g)}~} }{}+{}\mathrm {H} 
    {\vphantom {A}}_{\smash[{t}]{2}}\mathrm {O_{{\mskip 1mu}(g)}} 
    \mathrel {\longrightarrow } 
    {\mathrm {CO_{2(g)}~} }+ \mathrm {H} 
    {\vphantom {A}}_{\smash[{t}]{2}}
    {\vphantom {A}}_{\smash[{t}]{{\mskip 1mu}(\mathrm {{\mskip 1mu}g} )}}}\qquad 
    {\Delta H^{\circ }=-41\ {\mathrm {kJ} /\mathrm {mol} }}}
\end{equation}

\noindent
Carbon dioxide is then scrubbed from the hydrogen and nitrogen mixture. 

\subsubsection{Phosphorus and potassium}

On the Earth, phosphorus is mainly derived from phosphate rock, such as fluoroapatite (\ce{Ca5(PO4)3F}) and hydroxyapatite (\ce{Ca5(PO4)3OH}), which are converted to phosphate salts by treatment with acids. Meanwhile potassium is also extracted from a variety of potassium-rich rocks, collectively known as `potash', usually a mxiture of potassium chloride, potassium sulfate, potassium carbonate, and/or potassium nitrate.

At present, since quantities of phosphorus and potassium are unknown, and deposits may be small, they will have to be provided from other sources, such as from nearby moons and asteroids, or brought from Earth.

\subsection{Colorings and dyes}
\label{sect:dyescolors}

Coloring and dyes are important chemicals, used now only for decorative purposes, but also for safety, printing, painting and marking, and in cell biology for staining samples, important for medical purposes.

The range of possible chemicals used for colorings (including food colorings) and dyes is vast, and beyond the scope of this article. While many require elements beyond simply C-H-O-N for their chemistry, nonetheless many are composed primarily of organic molecules, especially carbon cycles. Therefore, much of their structure can be created from abundant hydrocarbons and nitriles on Titan.

\begin{itemize}
    \item 
An important class as aniline (\ce{C6H5(NH2)}) based dyes, such as the first synthetic dye, mauveine, discovered in 1856. Mauveine is a complex derived from aniline along with toluidine (\ce{C6H4(NH2)(CH3)}).
\item 
Anthraquinone based dyes feature a tri-cyclic carbon ring (anthracene), with two oxygen atoms bonded to the middle ring, and form red to blue dyes.
\item 
Azo dyes contain the N=N double bond, along with aryl (C6H5 cycles) or  alkenyl (C=C) groups forming the larger structure.
\item 
Carbonyl dyes such as indigoids contain at least two conjugated C=O bonds. The archetype of this dye class is the naturally occuring indigo (Vat Blue 1), used in the dying of jeans.
\end{itemize}

\begin{figure}
    \centering
    \includegraphics[width=1.0\linewidth]{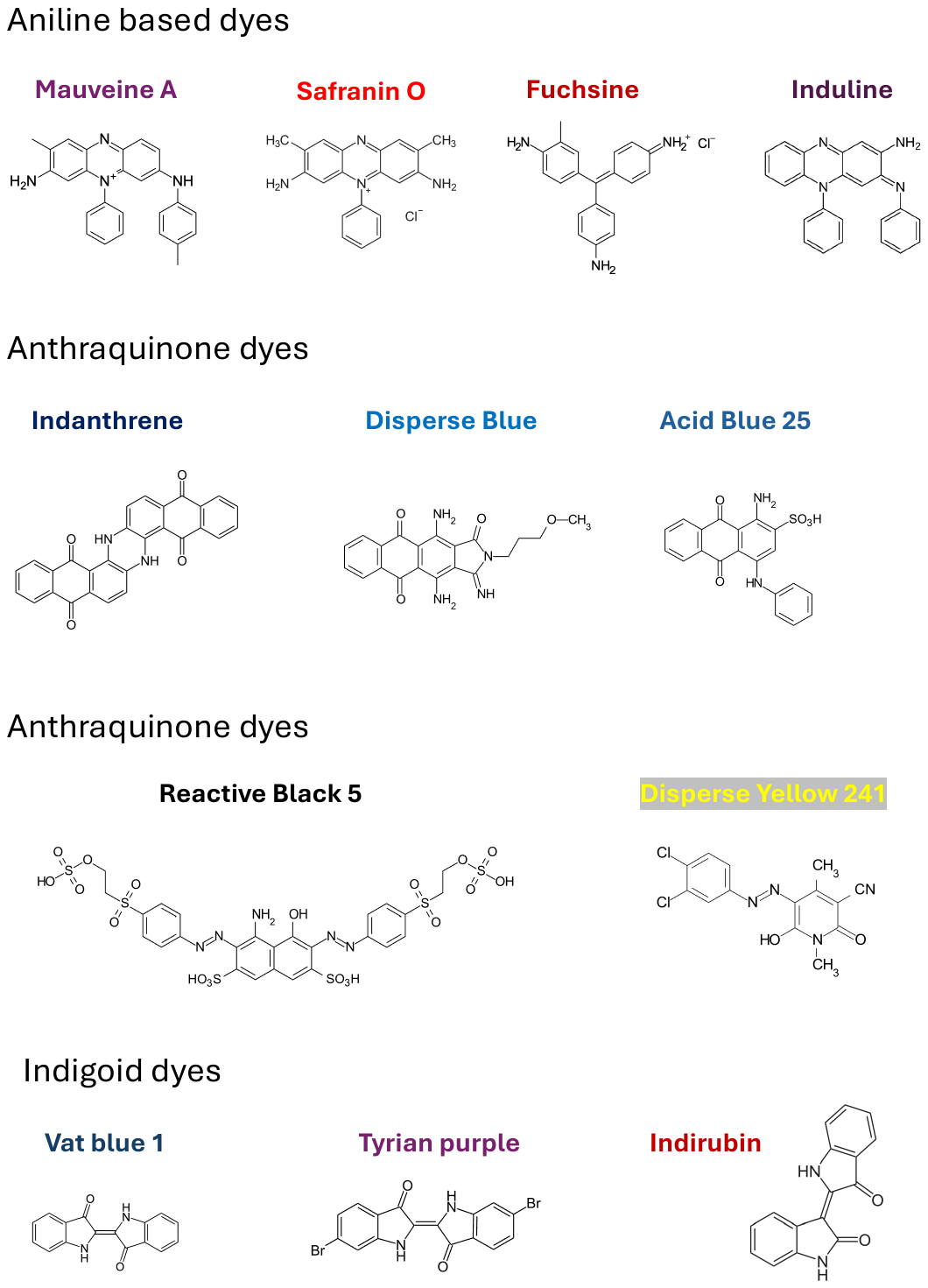}
    \caption{Examples of dyes that can be created moslty from C, H, O and N elements found abundantly on Titan.}
    \label{fig:dyes}
\end{figure}

Some examples of chemical structures are shown in Fig.~\ref{fig:dyes}.

\subsection{Textiles}
\label{sect:textiles}


Knitted or woven textiles needed for a wide variety of uses (clothing, bandages etc) are made from fibers, which may be categorized into the natural (silk, cotton, wool, linen), synthetic (nylon, polyester), or hybrid (e.g. rayon). On Titan and in other space locations it may eventually be possible to grow natural fibers, which would provide a wider variety of useful materials.

\begin{figure}
    \centering
    \includegraphics[width=1.0\linewidth]{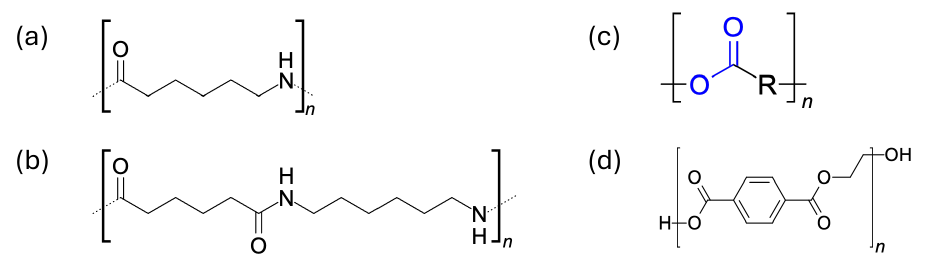}
    \caption{Monomer building blocks of synthetic fibers: (a) nylon-6; (b) nylon 6,6; (c) generic polyester; (d) polyethylene terephthalate (PET), a type of polyester. }
    \label{fig:fibers}
\end{figure}

In the nearer term it may be more practical to manufacture any needed fibers entirely synthetically, as have been done on Earth for around a century, beginning with the widespread usage of nylon in the 1940s. Synthetic fibers are typically made from the four light CHON elements widely available on Titan, through polymerization of a wide variety of basic monomer units (see examples in Fig.~\ref{fig:fibers}) that are themselves produced in a series of steps from more basic organic chemicals such as described earlier in this section \cite{morgan1981brief}.


\subsection{Storage}
\label{sect:chemstorage}

\begin{figure}[ht]
    \centering
    \includegraphics[width=1\linewidth]{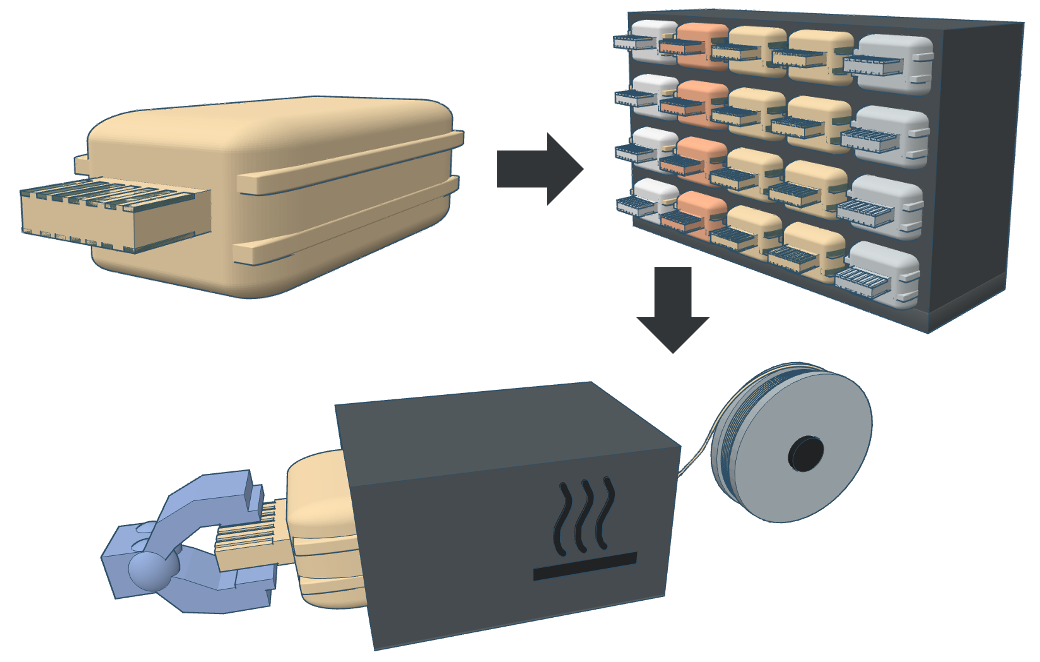}
    \caption{Standardized storage form factors for thermoset polymers facilitate ease of storage and automation, and this illustration depicts a version of this process which results in filament for use in additive manufacturing.}
    \label{fig:FormFactorToExtruder}
\end{figure}

Polymers intended for use in 3D printing or injection molding are thermoplastics, so they can be stored in any form factor prior to manufacturing use. Assuming future missions will rely on robotic systems, forms with flat surfaces will be the easiest for robotic training, calibration, and performance, and so uniform bars should be the primary storage form factor (Figure \ref{fig:FormFactorToExtruder}. Additional considerations to this should take into consideration the intended manufacturing process.

\section{Manufacturing}
\label{sect:manufacturing}

NASA programs such as On-orbit Servicing Assembly and Manufacturing (OSAM, \cite{arney2021orbit}) have  explored avenues for in-space manufacturing, including projects such as On-Demand Manufacturing of Metal Components (ODMM) and On-Demand Manufacturing of Electronics (ODME) \cite{prater2020space, roberts2022game}. Programs like this have laid the groundwork for long-term human presence in the outer solar system. 


Limitations due to availability of heavy elements will constrain the types of materials that can be manufactured. It is important to also note that many polymer parts do not perform well in a vacuum environment \cite{deutsch2007polymers}, therefore parts manufactured using polymers should be used inside of spacecraft or terrestrial stations. Additionally, performance properties of materials manufactured at Titan will need be measured and confirmed. Common mechanical and thermal properties used to characterize materials are tensile strength, Young's modulus, elongation, indentation hardness (Shore), compression set, and thermal expansion.

\begin{figure}
    \centering
    \includegraphics[width=0.8\linewidth]{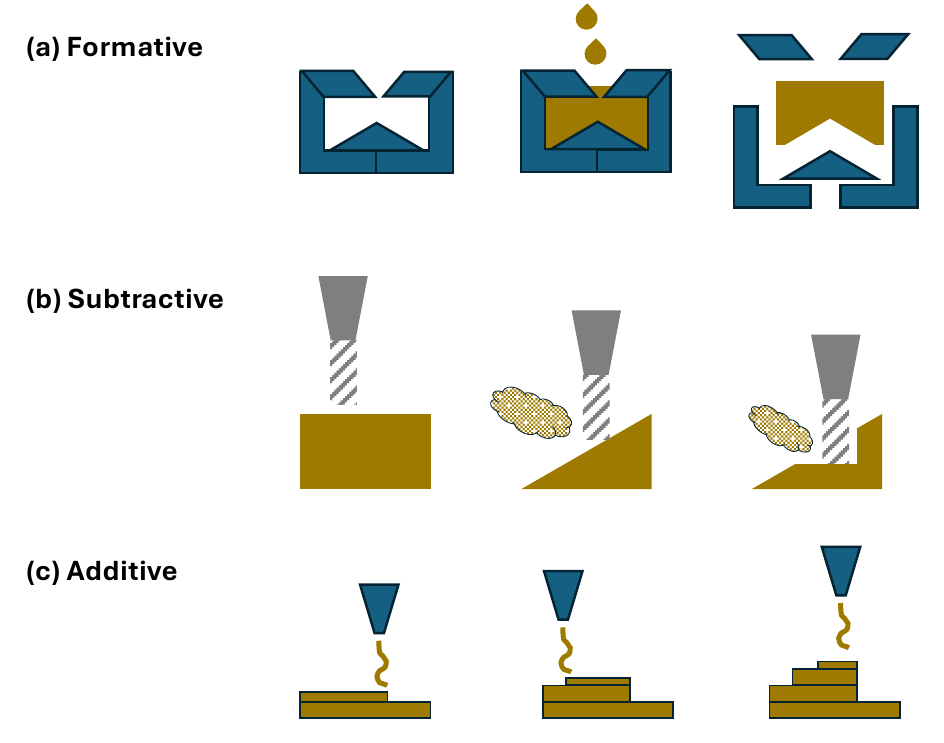}
    \caption{Types of manufacturing.}
    \label{fig:manufacturing}
\end{figure}

For the manufacture of useful parts from raw materials, there are three main approaches:

\begin{enumerate}

\item
{\bf Forming} (Fig.~\ref{fig:manufacturing}a) involves flowing a liquid or powder into a pre-defined mold with material-specific temperature and pressure requirements. 
\item 
{\bf Subtractive } processes (Fig.~\ref{fig:manufacturing}b) involve machining bulk material into the desired part. 
\item {\bf Additive} processes (Fig.~\ref{fig:manufacturing}c) are best visualized through 3D printing, where 2D layers are stacked in sequence to form the part. 
\end{enumerate}

All three of these approaches have requirements for the form-factor of the starting material, and each has limitations on the kind of parts/objects they can make in varying gravitation, pressure, and temperatures.

\subsection{Additive Manufacturing}
\label{sect:additive}

Additive manufacturing (AM) is a production process that builds objects layer by layer from digital designs, and spans techniques such as extrusion 3D printing, stereolithography, and selective laser sintering \cite{wong2012review}. AM can be performed with a wide variety of materials and has found applications beyond hobbyists in aerospace, healthcare, automotive, and more \cite{haleem2019additive, javaid2018additive}. 

For in-space manufacturing, digital design files can be remotely uploaded to planetary bodies or stored locally. Part size is limited to the size of the print volume of the printer, and polymer choice drives printing configuration. For 3D printing on-orbit, gravity-agnostic methods must be used. Example on-orbit considerations are the use of bulk-solids for the feedstock rather than powders, and forced- over gravity- extrusion \cite{paek2022composites}.

\begin{table}[]
\footnotesize
    \begin{centering} 
     \caption{3D printers that have operated successfully on the ISS} \label{tab:3dprinters}
    \begin{tabular}{lllll}
    \hline
Name & Material & Year & Printable  & Temp. 
\\
 & & & Size (cm) & ($^\circ$C)  \\
    \hline
        3D Printer$^*$ & ABS & 2014 & $6{\times}6{\times}12$ & 170-200  \\
        Additive Manufactur.   & ABS, Ultem 9085, & 2017  & $10{\times}10{\times}14$ & 180-260 \\
        \hspace{4mm} Facility$^*$ & \hspace{4mm} HDPE  & &  & \\
        ReFabricator$^{**}$ & Scrap plastic & 2018 & $6{\times}6{\times}12$ & 300-360 \\
        Regolith Printer$^\ddagger$ & Regolith simulant & 2021& $14{\times}10{\times}10$ & 1050-1300  \\
         3D Metal Printer$^\dagger$ & Stainless steel & 2024 & $9{\times}5$ & 1200  \\
    \hline
    \end{tabular}
    \end{centering}
    \\
        $^*$Made In Space $^{**}$Tethers Unlimited $^\dagger$ESA/Airbus/AddUp $^\ddagger$Redwire\\
    \normalsize   
\end{table}

AM has been demonstrated successfully on the ISS for multiple materials (Table~\ref{tab:3dprinters}) \cite{hoffmann2023space}). The first 3D printer, designed by the company Made In Space, was shipped to the ISS in 2014 and successfully printed small demonstration parts using the plastic material ABS (Acrylonitrile Butadiene Styrene) \cite{werkheiser20143d, bean2015international}.

After successful demonstration, a space station 3D printer was used to print a needed spare part - the Oxygen Generator System (OGS) Velocicalc Adapter  \cite{o2018turn}. A decade later in 2024, an ESA-contributed experiment successfully demonstrated 3D printing for metal using stainless steel \cite{makaya2023towards}. Metal printing is much more challenging than plastic printing, with temperatures exceeding 1200\degC rather than 180-300\degC as for plastics.

\begin{figure}
    \centering
    \includegraphics[width=1.0\linewidth]{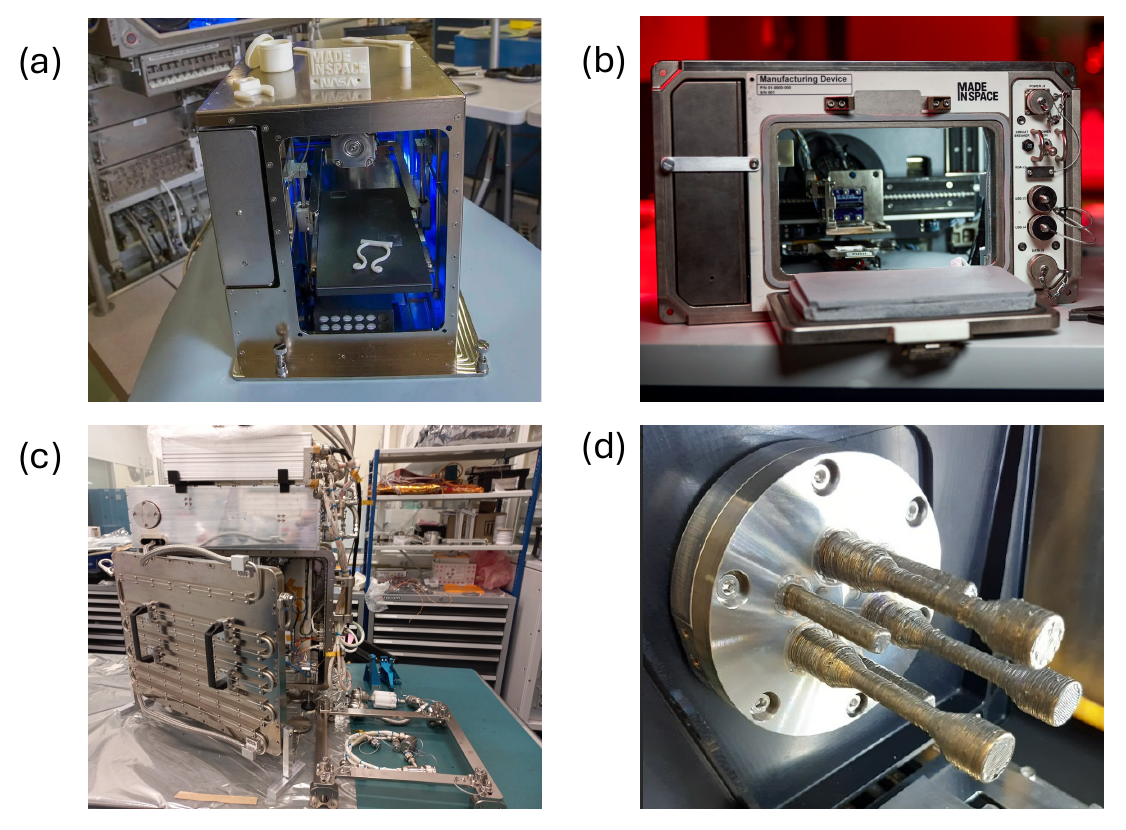}
    \caption{Examples of additive manufacturing in space: (a) (b) Redwire Regolith Printer (RRP) (c) ESA/Airbus 3D Metal Printer (d) Finished stainless steel test sample. Image credits: NASA/ESA}
    \label{fig:3dprinting}
\end{figure}

In the class of stereolithography, Turbine Ceramic Manufacturing Module (T-CMM) was the first demonstration of stereolithography ceramic fabrication in space, using a ceramic resin and UV radiation to generate a single-piece turbine blade \cite{nasa_inspa_awards}. 

Yet another type of printing was the Redwire Regolith Printer (RRP) was a technology demonstration mission in 2021, developed in partnership with NASA’s Marshall Space Flight Center. This unit (Fig.~\ref{fig:3dprinting}b) demonstrated 3D printing using simulated lunar regolith material in a type of printing called deposition modeling printing \cite{redwire_lunar_additive_iss}. 

In lunar-based manufacturing and construction, Azami et al. review numerous processes that have been experimentally demonstrated on Earth using lunar simulant These include laser powder bed fusion, solar sintering, microwave sintering, binder jetting, fused filament fabrication (FFF) of polymer/regolith composites, and molten regolith extrusion \cite{azami_comprehensive_2024}. The authors identify SLA as the most mature for precise indoor fabrication, contour crafting (architectural-size FFF) as most promising for large-scale construction, and FFF as advantageous for their simplicity and low energy consumption. 

NASA’s 3D-Printed Habitat Challenge (2015-2019) was a Centennial Challenges Program competition designed to advance additive construction technology needed to create sustainable housing solutions for Earth, the Moon, Mars and beyond \cite{nasa_3dhabitat}. Large-scale additive construction systems were solicited that were capable of fabricating structures from \textit{in situ} materials and mission recyclables, progressing to physical habitat fabrication. The challenge winner, Marsha by SpaceFactory, was 15 feet tall.

\subsection{Forming}
\label{sect:forming}

\begin{figure}[h]
    \centering
    \includegraphics[width=1\linewidth]{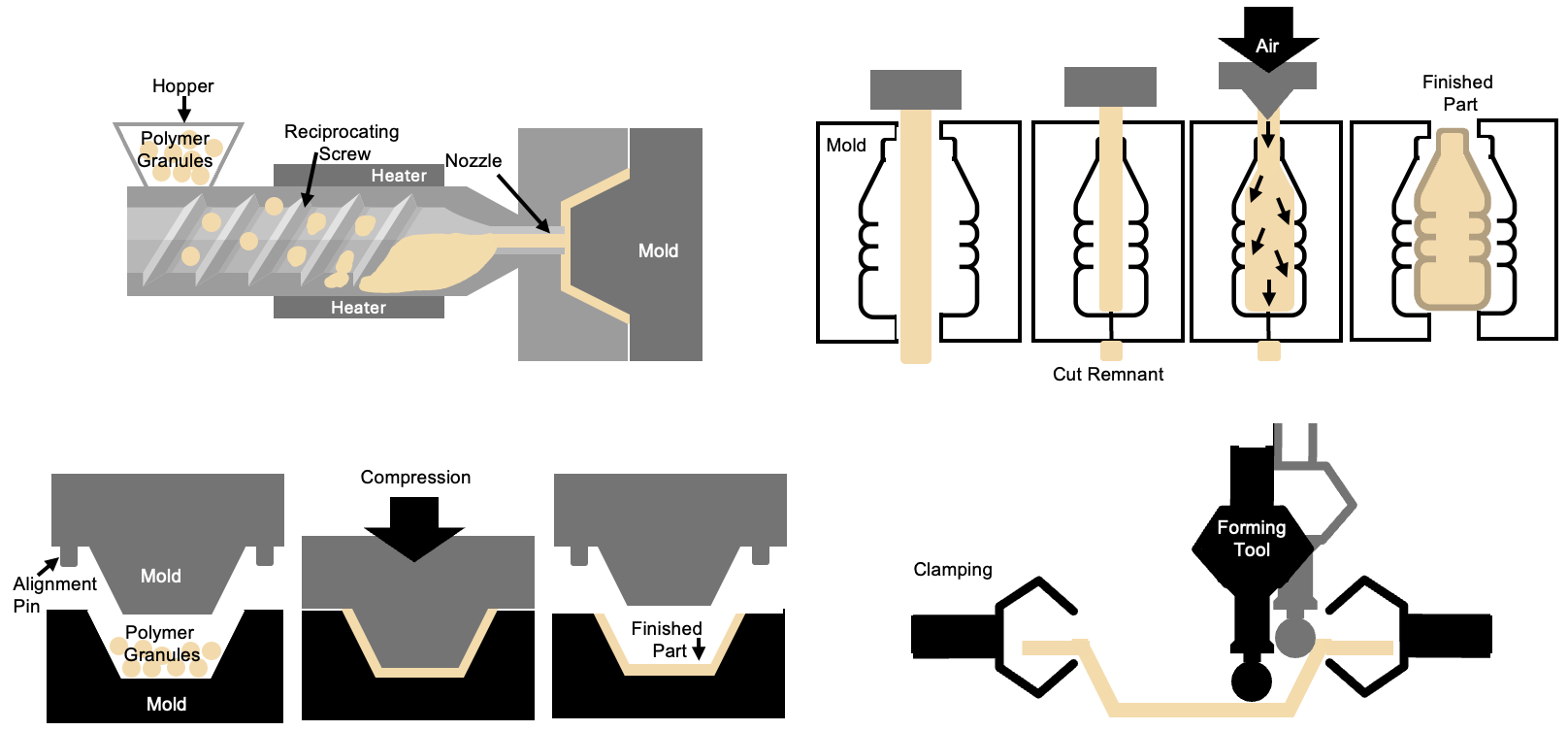}
    \caption{Simple diagrams of forming methods. From top left to bottom right: injection molding, blow molding, compression molding, and single-point incremental forming (SPIF).}
    \label{fig:forming-methods}
\end{figure}

Forming alters a material's shape through applied force without removing material, and can be used in most solids. Forces are applied through compression, tensile, or bending, and tools such as dies, molds, or rollers are employed at elevated temperatures. Common methods when manufacturing with polymers are injection molding (high volume parts, Fig.~\ref{fig:forming-methods}), compression molding (high strength parts), and blow molding (hollow parts) (Figure \ref{fig:forming-methods}. 

The choice of method is driven by end use of the part, with polymer strength or elastic properties strongly dependent on manufacturing strategy \cite{Mejia2021Effect}. For example, compression molding used with polymers, uses high pressures to eliminate voids which increases part strength. \cite{TATARA2024389} This method requires engineered molds (similar to injection molding) and high pressure equipment. The major use case for this approach is envisaged for storage of polymerized and ready-to-use thermoset materials, as discussed in the automation section.

Modern approaches to forming leverage the flexibility of programmable systems seen in CNCs (computer numerical control) and 3D printer gantries, such as in single-point incremental forming (SPIF) \cite{Hassan2021Progress}. Relatively new to the polymer manufacturing landscape, SPIF transforms a sheet of polymer using incremental applied force delivered by a movable tool head (Fig.~\ref{fig:forming-methods}). This approach allows for highly customized parts without engineered molds, creates a path for remote operations. This approach also aligns with recommendations throughout this paper for standardization of form factors for raw material storage, where polymer sheets are easy to store and handle robotically. Forming methods have not yet been tried on orbit or on other rocky bodies in our solar system.

\subsection{Subtractive Processes}
\label{sect:subtractive}

Subtractive manufacturing involves removing material from a solid bulk workpiece to achieve a desired shape, using processes such as CNC milling, turning, drilling, grinding, and laser cutting. This approach is widely used in machine shops and is valued for its high accuracy and precision, smooth surface finishes, and compatibility with a broad range of materials. It has also seen development for specialized applications such as on-orbit manufacturing. In 2018, the Metal Advanced manufacturing Bot-assisted Assembly (MAMBA) from Tethers Unlimited, Inc.  demonstrated automated CNC milling for use on-orbit \cite{nasa_mamba_2018}, however it never flew.

\subsection{Complex Manufacturing}


Establishing standards and procedures for foundational additive, formative, and subtractive methods should be done prior to complex manufacturing methods. However, there are several unique opportunities Titan may offer which are worth discussing. 

3D printing of basic circuit boards has been demonstrated on Earth \cite{dong2018sustainable}, though the results remain relatively crude compared to conventionally manufactured electronics. The primary approaches include conductive filament printing (extrusion of metal-impregnated polymer), electron beam melting of metal wires, and physical vapor deposition of atomized metals. In a space context, these are feasible in principle but require metallic feedstock brought from Earth, and the resulting boards are limited to simple, low-density circuitry. Complex multilayer boards with fine feature sizes remain beyond current in-space AM capabilities.

Given that metals will be scarce on Titan, \textit{in situ} fabricated organic optoelectronics should be explored if \textit{in situ} electronics are needed. They rely on $\pi$-conjugated small molecules and polymers that can transport charge and absorb or emit light. Many materials are shared across organic light emitting diodes (OLEDs), organic solar cells (OSCs), and organic semiconductors due to all devices neededing good charge mobility and tunable bandgap. Commonly used are conjugated polymers, featuring alternating single and double bonds, such as Poly(p-phenylene vinylene) (PPV) and its derivatives, as well as oligomers and small molecules \cite{polym12112627}.

3D printing organ-like tissue has been demonstrated on the ISS via the BioFabrication Facility (BFF) \cite{moroni2022can}. This program serves as a stepping stone toward the long-term goal of manufacturing whole human organs in space. This process is difficult in Earth's gravity, which causes soft biological structures to collapse under their own weight. Milestones to date include the first successful 3D bioprinting of a human knee meniscus in orbit, announced in September 2023 \cite{klarmann20243d}, as well as ongoing cardiac tissue printing investigations aimed at developing patches to replace damaged heart tissue. While printing organs at Titan isn't an immediate or typical need, far future humans traveling in the outer solar system may benefit from this technology under development today.





\subsection{Foodstuffs}
\label{sect:foods}

Traditional agriculture should be possible on Titan, as has been prototyped in space on the ISS \cite{massa2016growth}, provided that the full range of needed minerals can be supplied in addition to the basics such as \ce{CO2}, light, water and growth substrate.

3D printing of food may also be attractive, and is a growth area in terrestrial applications today. 3D printing of foods provides a number of potential advantages over traditional cuisine:

\begin{itemize}
    \item flexibility of combinations of ingredients, including personalized nutrition;
    \item fine-scale control over design, including layering and structure;
    \item use of novel ingredients;
    \item reduction of waste through on-demand production of items at time of consumption.
\end{itemize}

Many types of 3D printing techniques are possible, with material extrusion being the most widely used for food printing applications \cite{enfield2023future}.
For space, extensive use applications have been proposed \cite{terfansky20133d, santhoshkumar20243d}, although usage remains limited to date. A notable success was the 2019 growth and printing of meat on the ISS, using a combination of a bioreactor from Aleph Farms and a 3D food printer from 3D Bioprinter Solutions.

\subsection{Product Life Cycle}
\label{sect:lifecycle}

Given that material collection and refinement may have costly power or process requirements, reusing previously manufactured material should be considered. Selection of polymers to produce at Titan is primarily determined by the local chemistry and ease of processing, but additional consideration should be made for recyclability. Thermoplastics are re-formable, but often pick up contaminants through use. Thermosets, such as rubbers and resins, cannot be re-formed once polymerized and would rely on comminution and use of the particlized solids, possibly by adding a thermoplastic binder to make bricks. Recycling of metals, brought from elsewhere, will be particularly important given their scarcity at Titan. 

Future development is focused on expanding the range of recyclable materials, including metals (e.g. the MAMBA ground demonstration), reducing processing power requirements, and understanding how recycled materials perform over long-duration missions. The first demonstration of this concept on the ISS was the ReFabricator, installed in 2019, which was intended to recycle 3D-printed polymer parts back into filament feedstock, though an anomaly prevented full operation \cite{nasa_isam_2023}. Another example, through a NASA SBIR, the Cornerstone Research Group developed a reversible copolymer material which can be used for additive manufacturing \cite{snyder_2017_thermoset}. Earth-based mechanical testing of recycled mission materials should be performed and cataloged now as the next generation of projects are funded.




\section{Further work needed}
\label{sect:furtherwork}

In this section we examine some of the future directions, both in scientific study of Titan and also technology development, needed to fully realize the potential of using Titan's resources {\em in situ}.

\subsection{Titan Science}
\label{sect:scienceneeds}

{\em Atmosphere:}
Around two dozen chemicals have currently been identified in Titan's atmosphere through remote sensing spectroscopy \cite{nixon2024composition}, and more have been inferred to be present from molecular mass sampling of the upper atmosphere using Cassini's INMS instrument  \cite{vuitton2009composition, westlake2012titan}. However, photochemical models require a much vaster array of possible chemicals to be produced considering chemistry of elements C, H, O and N \cite{loison2015neutral, lara2014time, vuitton2019simulating} and P \cite{hickson2014evolution}. Evidence for further chemical complexity is present in the heavy ion spectrum of Cassini's CAPS instrument \cite{waite2007process}. Examples of some small molecular species likely to be present in the atmosphere (and condensed on the surface) are listed in Table~\ref{tab:needed}. 

Further work to detect and measure abundances of these is therefore important to more fully comprehend the availability of resources at Titan. Currently, the only means for detection available are astronomical techniques, which have proved capable for identifying new hydrocarbons \cite{coustenis2003titan, nixon2013detection, lombardo2019detection, nixon2020detection, nixon2025atmosphere} and nitrile species \cite{moreno2011first, cordiner2015ethyl, palmer2017alma, thelen2020detection}. Current facilities with powerful molecular detection capabilities (i.e. high resolution spectroscopy) include the Atacama Large Millimeter-sub-millimeter Array (ALMA) \cite{moreno2014alma}, the James Webb Space Telescope (JWST) \cite{nixon2016titan}, and the NASA InfraRed Telescope Facility (IRTF) \cite{nixon2024high}. 

\begin{table}[]
\small
    \caption{Examples of important chemicals predicted to be present on Titan}
        \label{tab:needed}
    \centering
    \begin{tabular}{lrl}
\hline
Name & Formula & Industrial uses \\
    \hline
    & & \\
    \multicolumn{3}{l}{{\em Hydrocarbons:}} \\
    Cyclopropane & c-\ce{C3H6} &  Solvent; pharmaceuticals \\
       Butadiene & \ce{C4H6} &  ABS plastic; SBR rubber \\
      Butene(s) & \ce{C4H8} &  Refrigerant; aerosols; rubber \\
      Butane(s) & \ce{C4H10} & Fuel; refrigerant; aerosols \\
      Isoprene & \ce{C5H8} & Rubbers; adhesives; glues; sealants \\
          Toluene & c-\ce{C7H8}  & Paints; coatings; inks; adhesives \\
      Styrene  & c-\ce{C8H6} &  ABS and SBR; polystyrene\\
          & &  \\
    \multicolumn{3}{l}{{\em Nitrogen species:}} \\
    Ammonia & \ce{NH3} &  Refrigerant; disinfectant; fertilizer \\
    Hydrazine & \ce{N2H4}  & Propellent; pharmaceuticals; agrichemicals \\
    Methanamine & \ce{CH3NH2} & Herbicides/fungicides; pharmaceuticals \\
    Cyanobenzene & \ce{C6H5CN} & Solvent; polymers; electroplating \\
              & &  \\
    \multicolumn{3}{l}{{\em Oxygen species:}} \\
     Formaldehyde & \ce{CH2O} &  Resins; preservatives; textiles; vaccines \\
     Methanol  & \ce{CH3OH} &  Paints; plastics; adhesives; synthetic fibers\\
     Acetaldehyde  & \ce{CH3CHO}  & Flavorings; pharmaceuticals; resins; dyes\\
     Dimethyl ether & \ce{(CH3)2O}  & Fuel; aerosol propellant; refrigerant \\
     Acetone & \ce{(CH3)2CO} & Solvent; cleaner; cosmetics \\
            & &  \\
    \multicolumn{3}{l}{{\em Others:}} \\
    Hydrogen sulfide & \ce{H2S} &  Sulfuric acid; rayon; refining \\
    Phosphine & \ce{PH3} &  Semiconductor (doping); fumigant \\
\hline
    \end{tabular}
\end{table}

{\em Surface:} With few exceptions (\ce{H2}, \ce{C2H4}, CO), molecules produced in the atmosphere inevitability condense onto Titan's surface, where vast seas and lakes of organic molecules are found \cite{stofan2007lakes} and also dunes of solid organic particulates \cite{lorenz2006sand}. At present, further investigation of their physical and chemical properties is mostly via theoretical \cite{tokano2009limnological, cordier2009estimate, cordier2012titan, steckloff2020stratification} and laboratory techniques \cite{cable2021titan, luspay2015experimental, burr2015titan, yu2017direct, yu2018does, yu2020single, benkoski2023effects, hirai2023rapid, he2015nmr, malaska2017laboratory}, although some compositional insights have been obtained from Cassini near-infrared and radar data \cite{brown2008identification, mastrogiuseppe2014bathymetry, mastrogiuseppe2019deep}.

Future {\em in situ} exploring missions are therefore vital to further characterize these materials. the {\em Dragonfly} rotorcraft mission scheduled to arrive at Titan in 2034 \cite{barnes2021science, turtle2024dragonfly} will sample dune material \cite{trainer2021development, freissinet2024unveiling}, provided the first detailed compositional analysis. Future missions to Titan's seas \cite{oleson2015titan, stofan2013time, rodriguez2022science} would provide important information on their depths, composition and layering.

Near-term laboratory work in this area and future chemical analysis of Titan lakes and seas will be needed (e.g. \cite{stofan2013time, oleson2015titan, rodriguez2022science}) before specialized planning of surface collection missions.

\subsection{Spacecraft and mission design to reach Titan}
\label{sect:aerospaceneeds}

\begin{figure}
    \centering
    \includegraphics[width=1.0\linewidth]{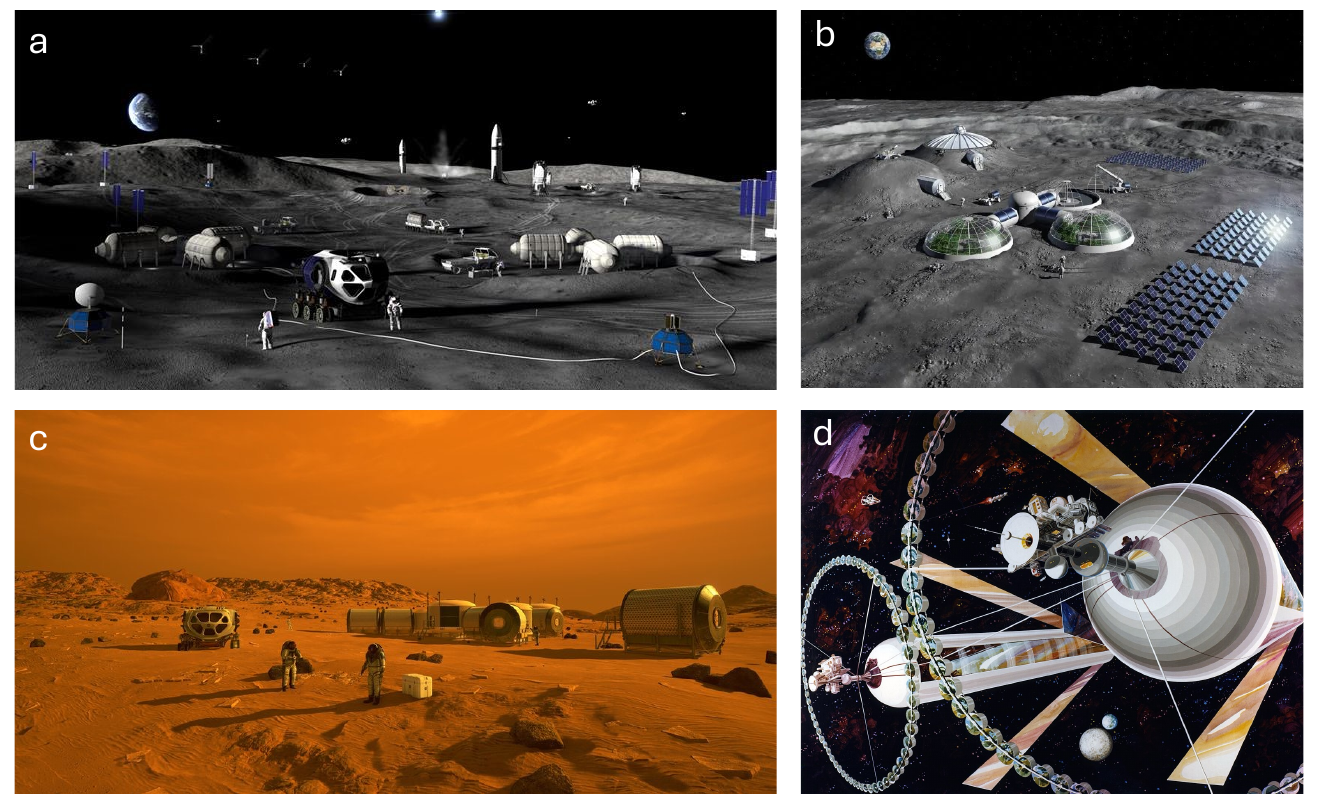}
    \caption{Artists concepts of future colonies on the Moon (a,b), Mars (c) and deep space O'Neill type habitats (d). Image credits: NASA/ESA.}
    \label{fig:colonies}
\end{figure}

Studies for future space habitats and colonization have tended to focus heavily on the Moon and Mars \cite{al2004elements, gavert2006lunar, snyder2008lunar, nair2008strategic, levchenko2021mars, neukart2024towards, zubrin1991mars, zubrin1992mars, zubrin1995economic, zubrin2018economic, dastagiri2017theory} (Fig~\ref{fig:colonies}(a,b,c)), being our closest large neighboring worlds other than the very challenging Venus \cite{landis2003colonization}. However, moving outwards from not-unreasonable colonization of Mars, to the much more distant Titan may not happen without intervening steps. 

These may include establishing outposts on main belt asteroids \cite{kecskes2002scenarios, joyce2013human, joyce2014technologies} - perhaps Ceres or Vesta \cite{swartz2017paraterraforming} - and amongst the moons of Jupiter \cite{steklov2018europe, kerwick2012colonizing}, before proceeding onwards to Titan despite its obvious resources. Another likely option will be establish large, rotating space stations as envisaged by Gerard O'Neill \cite{oneill1974colonization, oneill2000high} (Fig~\ref{fig:colonies}(d)) that can be placed at strategic locations throughout the solar system - or even have propulsion allowing them to function as large spacecraft. 

Much work remains to be done to complete these steps, including the capability for space stations to be self-sufficient for long durations. Establishing close-cycle habitats has proved difficult both on the ISS \cite{gentry2015international, ichimura2025assessment, ridley2024international} and on the ground, e.g. Biosphere 2 \cite{nelson1993using, nelson2018pushing, nelson2018some}. Major systems work or long duration missions to be further developed includes: (a) closed cycle air and water recycling \cite{jones2017would}; (b) recycling of metals and plastics \cite{hall2021recycling, mariappan2019conceptual, mariappan2021theoretical, schroeder2023space, kamal2021integrated, yang2025resource}; (c) food growth \cite{seedhouse2020growing, meinen2018growing, shaw2022farm, musgrave2008growing, langhans1988challenges}; (d) ability to make internal and external repairs, likely with the assistance of robotics \cite{pernigoni2023self, davis2019orbit, hastings2006orbit, burger2018caesar}; (e) in-space medical procedures \cite{stewart2007emergency}; (f) health and wellness in different gravities \cite{barratt2008principles}, including psychological \cite{kanas1990psychological, kanas2012psychology}; and more.

Developing the capability to mine, refine and utilize materials from asteroids will also prove important \cite{grandl2023asteroid}. While it may not be necessary to mine metallic ores and other useful substances (e.g. ices) from nearby asteroids for return to Earth (even if commercially exciting) \cite{grandl2013near}, such a capability could be vital for missions far from Earth in the outer solar system. Therefore developing such a capability could prove another stepping stone to reaching Titan with a human presence. Alongside the engineering development, scientific knowledge of the mineralogy of asteroids across the solar must reach a higher level \cite{papagiannis1983importance}, and we must have rapid tools for assaying any new bodies encountered \cite{elvis2014many}.


\subsection{Power generation on Titan}
\label{sect:powerneeds]}

\begin{figure}
    \centering
    \includegraphics[width=1.0\linewidth]{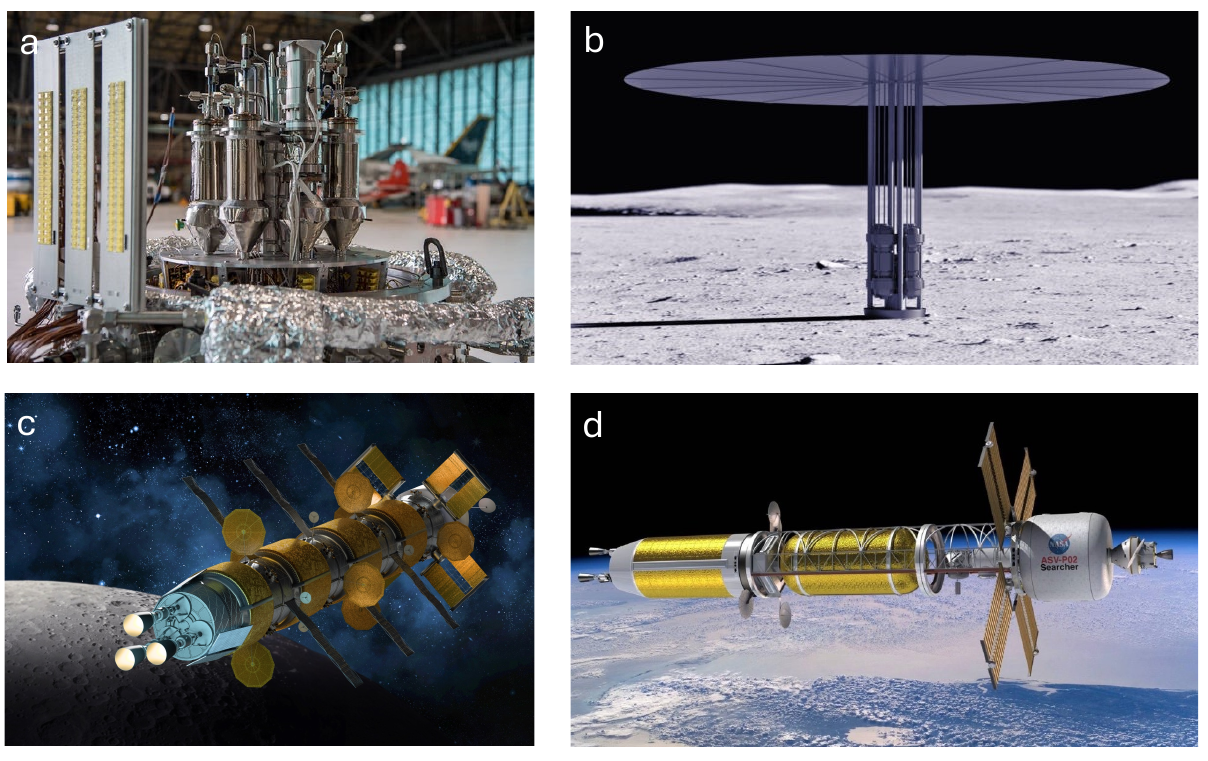}
    \caption{Nuclear power in space: (a) Kilopower test reactor (DOE/NASA); (b) NASA Artemis program reactor concept on lunar surface; (c) Nuclear electric spacecraft concept. (d) NASA Copernicus Nuclear Rocket concept. Images: NASA. }
    \label{fig:nuclear}
\end{figure}

As mentioned in Section~\ref{sect:power}, reasonable levels of power generation on Titan can only be expected in the near term to come from small fission reactors, similar to those currently used on nuclear submarines \cite{holbert2025review} and as envisaged for lunar and martian colonies \cite{nikitaeva2022power, nikitaeva2022analysis, buden1985nuclear,  balint2004nuclear, bushman2004martian} and for spacecraft propulsion \cite{robinson2018isru, gilbert2026nuclear} (Fig.~\ref{fig:nuclear}). 

Areas of power generation requiring further study include low-maintenance reactors \cite{schubert2021ultra, boccelli2024mass}, extraction and enrichment of uranium from worlds other than Earth \cite{schubert2019nuclear, karkadakattil2025laser, schubert2020nuclear}, transportation of ore or enriched fuel safety to Titan \cite{barry2024evaluation}, and disposal of spent fuel \cite{blessing2006design, huff2025taking}.

\subsection{Materials processing and manufacturing}
\label{sect:materialsneeds}

{\em Sample acquisition:}
As discussed in Section~\ref{sect:collectingrefining}, materials collection may come from Titan air, liquids or solids. Before relying on apparently simple pumping, drilling and scraping techniques, any tools to be sent to Titan must be carefully tested under realistic conditions in a simulation chamber on Earth and/or in a field environment to ensure functioning. This has proved invaluable for mechanisms sent to Mars \cite{sobrado2014mimicking, cozzolino2020martian, motamedi2015design, mavsek2017thermo, malin2017mars} and is currently underway for missions such as Europa Lander and Dragonfly \cite{bowkett2025autonomous, wagner2024demonstrating, mortensen2024development, hines2022designing}.

{\em Chemical refining and processing:}
Many industrial chemical processes on Earth make heavy use of metals: for containers (e.g. pressure vessels); pipes; mechanisms such as pistons and valves; and importantly for catalysts of diverse types. However since metals will be unlikely to be found on Titan in useful quantities they will need to be imported at great cost, therefore requiring careful use and reuse and the imaginative use of alternative materials wherever possible.

For pipes and pressure vessels carbon fiber, made from carbon extracted from readily available methane, may prove valuable. However carbon fiber has different characteristics from the metal it be replacing \cite{ahmad2020review}. While fives times stronger and two times stiffer than steel, it is also more brittle, failing catastrophically rather than bending or deforming like a metal, so usage must be carefully planned to avoid disasters such as befell the Titan submersible \cite{weijermars2025comprehensive}. Silicon carbide could also prove a useful replacement for cutting tools and structures, as is already used in space applications \cite{krenkel2005c, robichaud2012silicon}.

For molecular separation (sieving), variants of carbon nanotubes and graphene will be important, and these have already been demonstrated to be useful for separations of \ce{O2}, \ce{N2}, \ce{CH4}, \ce{H2O}, ce{CO2} and Ar \cite{nejad2016investigation, safavi2010single, tseng2009preparation} and even He \cite{li2015efficient, niechcial2021separation} and \ce{H2} \cite{sharma2009aligned}.

{\em Manufacturing:}
Section~\ref{sect:manufacturing} of this paper has already focused on the use of available materials on Titan for many purposes. Key work leading up to such deployments must include demonstrations of end-to-end, systems level processes (from extraction to refinement to manufacturing to extended use) in closed systems, and under relevant conditions of low or no gravity \cite{antony2025space, zocca2022challenges, ishfaq2022opportunities, mehta2017impact, patel2025space}.

\subsection{Robotics, AI and autonomy}
\label{sect:aineeds}

Human missions to the outer solar system will be multiple years to even decades in length, given current technologies, and therefore require extensive planning and preparation. Such missions will need to be largely self-sufficient, excepting for resources that can be obtained e.g. at Titan. Inevitably, human missions will be preceded by extensive robotic reconnaissance and - most likely - placement of foundational infra-structure such as habitats, spare parts and fuel depots to reduce risk.

{\em Studies of long-duration missions:}
For missions to Titan, a `terrestrial planet' world, extensive previous studies of human long-duration missions to the Moon and Mars will prove valuable sources of ideas and eventually experience. A description of historical lunar and martian colony studies is beyond the scope of this paper, however some key ideas are worth noting. In particular, the `Mars Direct' (hereafter MD) plan advocated by Zubrin and colleagues through the 1990s \cite{zubrin1991mars, zubrin1992mars} was a pivotal change in astronautical thinking about how to handle the risks and logistics of a lengthy (2-year) mission far from Earth remedial support. 

In the MD plan, a key change from earlier missions was the use of {\em In Situ Resource Utilization} (ISRU) - in the form of production of fuel on the surface to save on the large mass requirement of landed fuel from Earth. Specifically MD produced methane and oxygen from atmospheric \ce{CO2} via use of the Sabatier reaction (Eg.~\ref{eq:sabatier}, Fig.~\ref{fig:cho-cycle}) for \ce{CH4} and electrolysis of the resulting water (Section~\ref{sect:oxygen})

{\em AI and Robotics:}
At the time of writing, very rapid progress is being made in artificial intelligence, with Agentic AI being on the cutting edge of technology \cite{acharya2025agentic}. These technologies are already powerful and many predict that they will soon be to practice self-improvement without human intervention, leading towards a self-aware `Artificial General Intelligence' (AGI) in the not-distant future \cite{yenduri2025artificial}.

While in-silico developments are developing very rapidly, mostly unhindered by physical constraints (except power and processor cycles), the development of highly capable humanoid robots is also developing apace, as evidenced by the recent success of a robotic `winning' a human footrace and setting a new half-marathon record for a biped \cite{sparkes2026humanoid}. As well as biomechanical developments in walking, running, jumping etc \cite{guizzo2019leaps}, progress is also rapid in sensory areas such as touch, taste and smell \cite{de2015touch, dahiya2012robotic, ciui2018chemical, tian2025robotic, france2025artificial, ishida2012chemical}; while vision and hearing are already highly advanced (e.g. \cite{sonka2013image, lyon2017human}).

By the time of sending long-duration, self-sustaining missions to the outer planets, whether with humans or without, robots and AI will undoubtedly play a huge part. Work is therefore in needed in areas such as: (a) robot repair and maintenance by self or other robots \cite{fallahiarezoodar2025review}; (b) robot tool usage \cite{tee2018towards}; (c) AI autonomy and decision making \cite{nesnas2021autonomy}, including spacecraft guidance, hazard avoidance etc; (d) on-surface operations \cite{fink2011robotic, weisbin2002nasa}; (e) interactions with human crew, many of which are already areas of active development \cite{hambuchen2021review, wedler2018single}.


\section{Summary and Conclusions}
\label{sect:summary}

\begin{figure}
    \centering
    \includegraphics[width=0.9\linewidth]{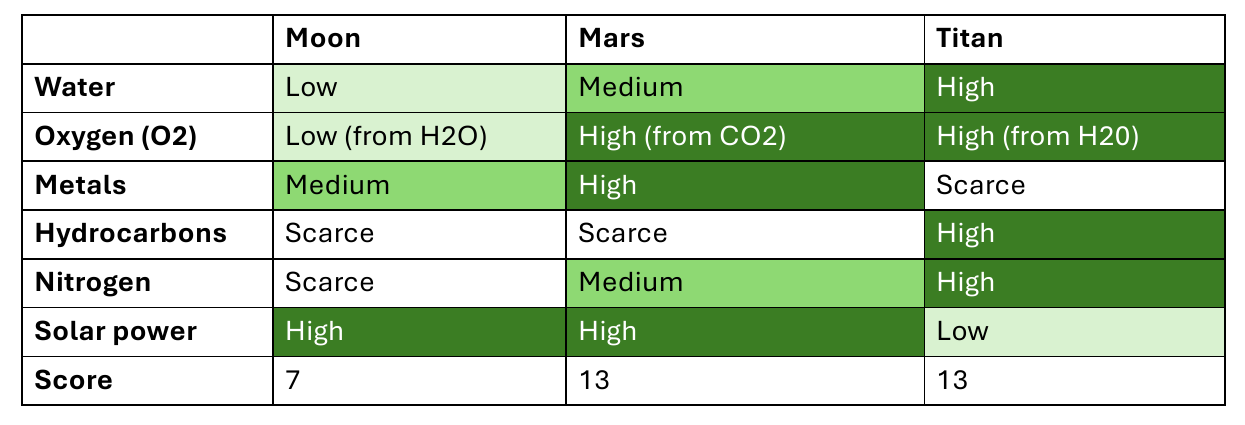}
    \caption{Comparison of resources readily available on Earth, Moon and Mars. Scoring is: Scarce (0); Low availability (1); Medium availability (2); High availability (3). The current scoring is not intended to rigorous: prevalence of some resources e.g. organics on Mars is still uncertain and requires further characterization.}
    \label{fig:comparison}
\end{figure}

This paper has focused on two topics: the availability of useful resources on Titan, and the use of those resources to facilitate deep space exploration, such as human missions or stations on Titan. the resources available on Titan are extensive, although different in nature from those available on the Moon and Mars (Section~\ref{sect:intro}) (Table~\ref{fig:comparison}). This is as expected for a world in the outer solar system `beyond the snow line' (Titan) compared to inner solar system worlds more depleted in light volatile elements and molecules \cite{grasset2017water}.

While current and future planned robotic missions do not require any resources from Titan (e.g. Dragonfly), eventually more ambitious missions will require use of local resources. An example is the recently studied Titan sample return \cite{oleson2022titan} that used refueling on Titan to attain mission closure. Beyond these missions, we have discussed much more ambitious, generalized mission architectures such as permanent orbital or surface outposts that would enable much longer term presence. While such visions are speculative for the time being the unique resources available on Titan in the outer solar system imply that eventually missions will be be developed to take advantage of them. 

In the nearer term, significant research can be undertaken to help make these dreams closer to reality, including: scientific characterization of Titans dunes, lakes and seas; development of low-g and zero-g refining and manufacturing; and ability to process and refine hydrocarbons at cold temperatures. 

    









\clearpage

\bibliographystyle{elsarticle-num} 
\bibliography{titan.bib}

\end{document}